%% file: main.tex
\documentclass[sigconf,balance=false]{acmart}
\usepackage{popets}
\usepackage[T1]{fontenc}
\setcopyright{popets}
\copyrightyear{YYYY}

\acmYear{YYYY}
\acmVolume{YYYY}
\acmNumber{X}
\acmDOI{XXXXXXX.XXXXXXX}
\acmISBN{}
\acmConference{Proceedings on Privacy Enhancing Technologies}
\usepackage{url}
\usepackage{xspace}
\usepackage{rotating}
\usepackage{framed}
\usepackage{verbatim}
\usepackage[inline]{enumitem}
\usepackage{xcolor}
\usepackage{cleveref}
\usepackage{booktabs}
\usepackage{longtable}
\usepackage{makecell}
\usepackage{tabularx}
\usepackage{multicol}

\usepackage{tikz}
\newcommand*\circled[1]{\tikz[baseline=(char.base)]{
            \node[shape=circle,draw,inner sep=2pt] (char) {#1};}}

\usepackage{macros}

\begin{document}

\title[Compact and Divisible E-Cash with Threshold Issuance]{Compact and Divisible E-Cash with Threshold Issuance}

\author{Alfredo Rial}
\email{alfredo@nymtech.net}
\affiliation{%
  \institution{Nym Technologies}
  \country{}
}
\author{Ania M. Piotrowska}
\email{ania@nymtech.net}
\affiliation{%
  \institution{Nym Technologies}
  \country{}
}

\begin{abstract}
  \input{0Abstract.tex}

\end{abstract}


\maketitle

\input{1Introduction.tex}

\input{3offlineecash.tex}
\input{4IdealFunctionality.tex}
\input{10ConstructionEcash.tex}
\input{5InstantiationOfThresholdOfflineEcash.tex}

\input{6EfficiencyComparison.tex}
\input{11Conclusion.tex}

\bibliographystyle{ACM-Reference-Format}
\bibliography{main}

\appendix
\input{ASecurityDefinitionsBuildingBlocks.tex}

\input{7UCSecurity.tex}

\input{9DefinitionsIdealFunctionalities.tex}
\input{BSecurityAnalysisCompact.tex}

\input{CSecurityAnalysisDivisible.tex}

\input{DCompactECashRangeProof.tex}

\input{EDivisibleECashSpendProof.tex}

\end{document}

%% file: 0Abstract.tex
Decentralized, offline, and privacy-preserving e-cash could fulfil the need for both scalable and byzantine fault-resistant payment systems. 
Existing offline anonymous e-cash schemes are unsuitable for distributed environments due to a central bank. 
We construct a distributed offline anonymous e-cash scheme, in which the role of the bank is performed by a quorum of authorities, and present its two instantiations. Our first scheme is compact, i.e. the cost of the issuance protocol and the size of a wallet are independent of the number of coins issued, but the cost of payment grows linearly with the number of coins spent. Our second scheme is divisible and thus the cost of payments is also independent of the number of coins spent, but the verification of deposits is more costly. We provide formal security proof of both schemes and compare the efficiency of their implementations. 

%% file: 1Introduction.tex
\section{Introduction}
\label{sec:introduction}


At the present moment, there is a pressing need for private electronic cash (e-cash), as shown by growing interest in blockchain-based systems and Centrally-Banked Digital Currencies (CBDCs), but no privacy-enhanced and decentralized system exists that scales to the requirements of this application scenario. To address this problem, we introduce the first decentralized offline e-cash scheme with provable security and full implementation, that can efficiently support electronic payments. In contrast to the current privacy-enhanced blockchain, systems do not require a constant online presence of the payees. The issuance of coins is non-interactive, i.e., the authorities do not need to synchronise, as we do not rely on MPC protocols.
Moreover, our scheme maps to distributed payment systems such as CBDCs, a technology that urgently requires further attention in terms of privacy.
The idea of distributing issuance among a quorum of authorities has been so far explored in the context 
of attribute-based credentials~\cite{DBLP:conf/ndss/SonninoABMD19, cryptoeprint:2022:011} and online anonymous e-cash schemes~\cite{DBLP:journals/iacr/BaudetSKD22}.
\newline\noindent \textit{\textbf{Contributions:}}
Our paper makes the following contributions:
\begin{itemize}[leftmargin=*, noitemsep,topsep=0pt]
    \item We introduce a construction $\mathrm{\Pi}_{\CEC}$ for threshold issuance offline e-cash ($\CEC$). To the best of our knowledge, this is the first offline e-cash scheme with threshold issuance. To this end, we define the system model and the security properties in the ideal-world/real-world paradigm~\cite{DBLP:conf/focs/Canetti01} by proposing an ideal functionality $\Functionality_{\CEC}$ for $\CEC$ (Sections \S\ref{sec:offlineecash}, \S\ref{sec:constructionCEC}).
    \item We propose two instantiations of  $\mathrm{\Pi}_{\CEC}$ based on compact~\cite{DBLP:conf/eurocrypt/CamenischHL05} and divisible~\cite{DBLP:conf/pkc/PointchevalST17} e-cash (Section \S\ref{sec:instantiation}). Our schemes are more efficient than~\cite{DBLP:conf/eurocrypt/CamenischHL05, DBLP:conf/pkc/PointchevalST17} thanks to the use of Pointcheval-Sanders signatures in the random oracle model and to decreasing the number of coin secrets. 
    We formally prove that our construction $\mathrm{\Pi}_{\CEC}$ realizes $\Functionality_{\CEC}$ when instantiated with the algorithms of our compact and divisible e-cash schemes (Sections \S\ref{sec:securityProofCompact},  \S\ref{sec:securityProofDivisible}). 
    \item We provide an open-source Rust implementation of both schemes and present an extensive evaluation of their performance and trade-offs (Section \S\ref{sec:efficiencycomparison}). To the best of our knowledge, this is first such practical comparison.
    \item We conclude by outlining how our schemes can be integrated with a blockchain-based bulletin board (Section \S\ref{sec:integration}). This would allow our scheme to fulfil the requirements for a distributed privacy-enhanced CBDC~\cite{digitaleuro} and even provide scalability via offline transactions for existing blockchain systems like ZCash~\cite{zcash}. 
\end{itemize}

\section{Background and Motivation}
\label{background}
Anonymous e-cash was originally proposed by Chaum as a digital analog of regular cash, which also allows for private payments~\cite{DBLP:conf/crypto/Chaum82}. 
Unlike physical cash, e-coins are easy to duplicate, hence e-cash schemes must prevent double-spending and typically that is done by having a centralized bank~\cite{DBLP:conf/eurocrypt/CamenischHL05,DBLP:conf/pairing/BelenkiyCKL09,DBLP:conf/eurocrypt/CanardG07,DBLP:conf/acns/CanardPST15,DBLP:conf/pkc/CanardPST15,DBLP:conf/pkc/PointchevalST17,DBLP:conf/asiacrypt/BoursePS19,DBLP:conf/crypto/OkamotoO89,DBLP:conf/crypto/OkamotoO91,DBLP:conf/pkc/BaldimtsiCFK15,DBLP:conf/pkc/BauerFQ21}. Thus, there has been a revival of interest in adopting privacy to decentralized blockchain systems, in which coins are authenticated by proving in ZK that they belong to a public list of valid coins maintained on the blockchain, thus they do not require a central bank to prevent double-spending~\cite{DBLP:conf/sp/MiersG0R13}. 

Although blockchain-based privacy-enhanced systems such as ZCash and Monero have users~\cite{zcash, cryptonote,DBLP:journals/ledger/NoetherM16}, such 
decentralized e-cash systems require being online to check the status of payments, which simply does not scale to the speed of transactions needed by real-world payment systems or support the reality of transactions that need to be made without internet access, so the usage of blockchains for payments remains small in proportion to traditional payments. Also, as could happen to any other low-transaction fee blockchain that advertises high throughput, the ZCash blockchain has suffered an  attack of \emph{`spam'} transactions that have increased the blockchain size to such an extent that ZCash has suffered from what is effectively a \emph{denial-of-service} attack that has collapsed its throughput (i.e., simply downloading the blockchain becomes nearly impossible)~\cite{zcashattack1}. Similarly, `layer 2' solutions based on blockchain technologies such as zero-knowledge roll-ups, which increase transaction speed and (in some cases) privacy, are also vulnerable to these attacks~\cite{zamyatin2021sok}.  By virtue of not requiring a merchant to be online all the time but only needing eventual settlement over regular epochs, our system avoids these issues while maintaining the advantages of decentralization. Hence, it is to be expected that current privacy-enhanced blockchain systems that require an online setting will evolve into decentralized offline e-cash systems similar to the one presented in this paper. While proposals exist to enable offline payments in blockchain-based cryptocurrencies
such as payment channel networks~\cite{DBLP:conf/ccs/0001M17}, 
they typically do not offer strong privacy protection as only users who share a channel can transact with each other and so users who make a payment are not anonymous, similarly as in~\cite{10.1145/3052973.3052980}, and cashing out payments still requires online blockchain interactions. Even privacy-preserving offline payment channels effectively restrict payment transactions and the network topology of payment channels can lead participants to be effectively de-anoymized~\cite{ DBLP:conf/fc/KapposYPKDMM21, sharma2022anonymity}. 

On the other side of the spectrum, centralized CBDCs have gained increasing interest in over a hundred countries and will soon be deployed in Europe and China~\cite{digitaleuro, xu2022developments}. CBDC systems typically sacrifice user transaction privacy from the settlement layer and can lead to the possibility of dystopian surveillance of user financial transactions~\cite{danezis2015centrally}. Although financial transaction data could be considered a matter of national security, MIT and the Federal Reserve of Massachusetts in the United States have begun exploring blockchain systems for its CBDC efforts without transaction unlinkability~\cite{lovejoy2022high}. In stark contrast, Switzerland's Central Bank  has put forward a centralized online privacy-preserving scheme called `eCash 2.0', but offline payments would require special hardware and are currently not supported. The European Central Bank (ECB) recently published a list of requirements for the Digital Euro, which include privacy of user transactions from the settlement layer and offline transactions~\cite{digitaleuro}. Our scheme would fulfil those criteria, and we present how it could be integrated with a blockchain.
Furthermore, distributing the issuing power would be practical for emerging multi-nation economic proposals for joint issuance of currencies such as the South American joint reserve currency SUR in the `Banco del Sur' recently endorsed by Brazil and Argentina~\cite{marshall2009financing}. Even more importantly, practically preventing byzantine faults in centralized CBDCs requires it to be managed by a set of distributed parties, similar to Facebook's Diem project's consensus protocol based on HotStuff~\cite{yin2019hotstuff}. In this manner, our system unifies the objectives of blockchain research for decentralization while maintaining compatibility with the requirements for privacy-preserving CBDC by decentralizing e-cash.


\vspace{-5pt}
\paragraph{Online vs Offline Ecash}
To revisit the original e-cash proposal, the solution to double-spending is to associate each coin with a unique \emph{serial number}, which is used to detect double-spending by dishonest users and prevent dishonest providers from depositing a payment twice~\cite{DBLP:conf/crypto/Chaum82}.
In \emph{online e-cash} schemes~\cite{DBLP:conf/crypto/Chaum82,DBLP:conf/ndss/SonninoABMD19}, providers are constantly connected to the bank
and can then check if a coin has been double-spent before accepting a payment. An alternative and more realistic solution space is given by \emph{offline e-cash} schemes~\cite{DBLP:conf/eurocrypt/CamenischHL05,DBLP:conf/pairing/BelenkiyCKL09,DBLP:conf/eurocrypt/CanardG07,DBLP:conf/acns/CanardPST15,DBLP:conf/pkc/CanardPST15,DBLP:conf/pkc/PointchevalST17,DBLP:conf/asiacrypt/BoursePS19}, which do not require a permanent online connection between a provider and the bank.
The provider can accept payment and deposit it at a later settlement stage as there is a guarantee that users who double-spend  will be identified by the bank.

An important issue in the design of anonymous e-cash schemes is paying the exact amount as users cannot receive change,
since the providers are not anonymous towards the bank. If a user receives change, the user would in fact become a provider and lose anonymity. 
In online e-cash, the user can contact the bank in order to exchange a coin for lower denominations in order to make the payment. In offline e-cash, contacting the bank is not allowed during the spending phase.
\emph{Transferable e-cash} schemes~\cite{DBLP:conf/crypto/OkamotoO89,DBLP:conf/eurocrypt/ChaumP92,DBLP:conf/pkc/BaldimtsiCFK15} are one solution to this problem,
allowing a user to further spent a previously received coin without interacting with the bank. Hence, providers can return the change to the users that paid.
Although transferable e-cash is appealing, state-of-the-art schemes~\cite{DBLP:conf/pkc/BauerFQ21} are much less efficient than non-transferable ones.

An alternative solution to preserve user unlinkability is to use coins of the smallest denomination. However, the large number of coins that may need to be spent easily yields an inefficient scheme. To solve this problem, researchers have focused on designing offline e-cash protocols whose complexity does not depend on the number of coins withdrawn or spent.
In \emph{anonymous compact e-cash} schemes~\cite{DBLP:conf/eurocrypt/CamenischHL05,DBLP:conf/pairing/BelenkiyCKL09}, the cost of storing a wallet of $N$ coins and the cost of withdrawing $N$ coins is independent of $N$. However, the cost of spending $n \leq N$ coins grows linearly with $n$.
\emph{Anonymous divisible e-cash} schemes~\cite{DBLP:conf/crypto/OkamotoO91,DBLP:conf/eurocrypt/CanardG07,DBLP:conf/acns/CanardPST15,DBLP:conf/pkc/CanardPST15,DBLP:conf/pkc/PointchevalST17,DBLP:conf/asiacrypt/BoursePS19} improve the efficiency of compact e-cash and allow the user to spend $n \leq N$ coins with cost independent of $n$. Therefore, in the last decade, research has focused solely on divisible e-cash. However, divisible schemes achieve constant spending cost at the expense of much more expensive deposit and identification phases, and to our knowledge, efficiency analysis of compact e-cash schemes with multiple denominations has not been conducted, as well as a fair comparison between practical implementations of divisible and compact e-cash has never been done. Therefore, it is unclear which scheme is better for the use-case of offline e-cash as required by privacy-enhanced CBDCs~\cite{digitaleuro} and blockchain-based scalability. 
Our analysis shows that compact e-cash with multiple denominations is preferable for small payments, which would naturally compose the majority of offline e-cash transactions in application scenarios such as a user-facing blockchain or CBDC where practical deployment concerns would necessitate distributed authorities.
In order to address the urgent scalability issues of blockchain systems and possibly dangerous centralization of CBDCs without privacy, a formal treatment of offline e-cash is needed, including a fair comparison of compact and divisible e-cash.

%% file: 3offlineecash.tex
\section{Threshold Issuance Offline E-Cash}
\label{sec:offlineecash}

In this section, we introduce an offline e-cash scheme with threshold issuance ($\CEC$). First, we outline its system model and informally define the security properties. 
To define formally the security properties of $\CEC$, we construct the ideal functionality $\Functionality_{\CEC}$ and explain how it guarantees those properties.


\subsection{System model}

An $\CEC$ scheme involves $n$ authorities $(\fcecAuthority_1, \allowbreak \ldots, \allowbreak \fcecAuthority_n)$, any number of users $\fcecUser_j$ and any number of providers $\fcecProvider_k$.
The interaction between those parties takes place through a \emph{setup}, \emph{withdrawal}, \emph{spend} and \emph{deposit} phase. Users withdraw wallets containing one or more electronic coins from the authorities and spend them with providers, who then deposit them back with the authorities.  

In the \textbf{setup} phase, a trusted third party 
generates the public parameters $\cecparams$. Next, the public verification key  
$\spk$ is generated, alongside key pairs $(\ssk_{\fcecAuthority_i}, \allowbreak \spk_{\fcecAuthority_i})_{i\in[1,n]}$ for each of the authorities $\fcecAuthority_i$. The keys are generated in such a way that a user needs to receive a withdrawal from at least $t$ authorities in order to create a wallet.
The key generation can be executed by a trusted third party or run in a distributed way~\cite{DBLP:journals/iacr/KateHG12, cryptoeprint:2021:339}. 
Finally, each user $\fcecUser_j$ generates a key pair $(\ssk_{\fcecUser_j}, \allowbreak \spk_{\fcecUser_j})$. 

To obtain a wallet with $L$ coins, where $L$ is a public parameter of the scheme, a user $\fcecUser_j$ engages in the \textbf{withdrawal} protocol. To this end, $\fcecUser_j$ sends a request $\cecrequest$ to a set of $t$ different authorities. 
Each $\fcecAuthority_i$ verifies  $\cecrequest$ and using its secret key $\ssk_{\fcecAuthority_i}$ issues back a response $\cecresponse$. 
$\fcecUser_j$ verifies $\cecresponse$ and extracts from it a partial wallet $\cecwallet_i$. Once $\fcecUser_j$ has completed the protocol with at least $t$ authorities, 
and collected the threshold number of partial wallets,  
$\fcecUser_j$ aggregates them to form
a single consolidated wallet $\cecwallet$ with $L$ coins of the same monetary value. To \textbf{spend} $V$ coins  with a provider $\fcecProvider_k$ the user generates a payment $\cecpayment$ using her wallet $\cecwallet$ and payment information $\cecpaymentinfo$. $\cecpaymentinfo$ contains the provider's identity and other information about the payment and must be unique for each payment. $\fcecUser_j$ sends $\cecpayment$ to $\fcecProvider_k$, who verifies it. 
To \textbf{deposit} the payment $\cecpayment$, provider $\fcecProvider_k$ sends $\cecpayment$ to a bulletin board $\BB$. Each authority reads $\cecpayment$ from $\BB$, verifies it and checks it against all the payments previously written on $\BB$, in order to rule out double spending and double depositing. 
The double spending detection mechanism reveals the public key of the user $\fcecUser_j$ if the user double-spent any coin, while double depositing detection reveals that the payment is deposited twice if two payments contain the same payment information $\cecpaymentinfo$. Otherwise, the payment is deposited successfully.

\subsection{Security Properties}
\label{sec:securityoffec}
As defined in~\cite{DBLP:conf/asiacrypt/BoursePS19}, secure anonymous offline e-cash schemes should satisfy four properties. We describe them informally, taking into account that, in our schemes, the bank is replaced by a number of authorities.

\begin{description}[leftmargin=*, noitemsep,topsep=0pt]

    \item[Traceability:] guarantees that no more coins can be deposited than those that have been withdrawn. In particular, adversarial parties are not able to forge wallets. It also guarantees that an honest authority is able to identify a user who double-spends a coin. Double-depositing is also detected by the authority.

    \item[Unlinkability:] ensures that no coalition of dishonest authorities, users and providers is able to link the withdrawal of the wallet with the spending of its coins. It also guarantees that multiple spendings performed by the same user cannot be linked with each other.



    \item[Exculpability:] requires that an honest user cannot be found guilty of double-spending.

    \item[Clearance:] ensures that only the provider that receives a payment is able to deposit it.

\end{description}

%% file: 4IdealFunctionality.tex
\subsection{Ideal Functionality}\label{sec:idealfun}\label{sec:idealfunctionalityCEC}

We define the security properties of offline e-cash with threshold issuance in the ideal-world/real-world paradigm. 
To this end, in Figure~\ref{fig:idealfun} we define the ideal functionality 
$\Functionality_{\CEC}$. $\Functionality_{\CEC}$ interacts with authorities $(\fcecAuthority_1, \ldots, \fcecAuthority_n)$, any number of users $\fcecUser_j$ and any number of providers $\fcecProvider_k$. $\Functionality_{\CEC}$ is parameterized by a threshold $t$, a universe of pseudonyms $\fcecunivnym$, a universe of wallet identifiers $\fcecunivwalletid$, a universe of request identifiers $\fcecunivrequestid$, a universe of payment information $\fcecunivpaymentinfo$, and a number $L$ of coins in a wallet. In~\S\ref{sec:securitydefinitionsbuildingblocks}, we define the security properties of the cryptographic primitives used in our $\CEC$ schemes. 

\paragraph{Remarks about $\Functionality_{\CEC}$.} When describing ideal functionalities, we use the conventions introduced in~\cite{DBLP:conf/crypto/CamenischDR16} and summarised in~\S\ref{sec:securitymodel}. 
\begin{description}[leftmargin=*, noitemsep,topsep=0pt]

\item[Aborts.] When invoked by any party, $\Functionality_{\CEC}$ first checks the correctness of the input. $\Functionality_{\CEC}$ aborts if any of the inputs does not belong to the correct domain. $\Functionality_{\CEC}$ also aborts if an interface is invoked at an incorrect moment in the protocol. For example, an authority $\fcecAuthority_i$ cannot invoke the $\fcecissue$ interface on input a request identifier $\fcecrequestid$ if that authority did not receive a request associated with $\fcecrequestid$. Similar abortion conditions are listed when $\Functionality_{\CEC}$ receives a message from the simulator $\Simulator$.

\item[Session identifier.] The session identifier $\sid$ has the structure $(\fcecAuthority_1, \allowbreak \ldots, \allowbreak \fcecAuthority_n, \allowbreak \sid')$. This allows any authorities $(\fcecAuthority_1, \allowbreak \ldots, \allowbreak \fcecAuthority_n)$ to create an instance of $\Functionality_{\CEC}$. After the first invocation of $\Functionality_{\CEC}$, $\Functionality_{\CEC}$ implicitly checks that the session identifier in a message is equal to the one received in the first invocation.

\item[Query identifiers.] Before $\Functionality_{\CEC}$ queries $\Simulator$, $\Functionality_{\CEC}$ saves its state, which is recovered when receiving a response from $\Simulator$. If an interface, e.g. $\fcecrequest$, can be invoked by a party more than once, $\Functionality_{\CEC}$ creates a query identifier $\qid$, which allows $\Functionality_{\CEC}$ to match a query to $\Simulator$ to a response from $\Simulator$. Creating $\qid$ is not necessary if an interface, such as $\fcecsetup$, can be invoked only once by each authority, and the authority identifier is revealed to $\Simulator$.

\end{description}

\begin{figure*}
    \begin{framed}
        {\small
        \vspace*{-10pt}
        \begin{multicols}{2}
        \begin{enumerate}[leftmargin=*]
        
        \item On input $(\fcecsetupini, \allowbreak \sid)$ from an authority $\fcecAuthority_i$:
        
        \begin{itemize}[leftmargin=*]
        
            \item Abort if $\sid \neq (\fcecAuthority_1, \ldots, \fcecAuthority_n, \sid')$, or if $\fcecAuthority_i \notin \sid$, or if $n < t$, or if $(\sid, \fcecAuthority_i, 0)$ is already stored.
        
            \item Store $(\sid, \fcecAuthority_i, 0)$ and send $(\fcecsetupsim, \allowbreak \sid, \allowbreak \fcecAuthority_i)$ to $\Simulator$.
        
        \end{itemize}
        
        \item[S.] On input $(\fcecsetuprep, \allowbreak \sid, \allowbreak \fcecAuthority_i)$ from $\Simulator$:
        
        \begin{itemize}[leftmargin=*]
        
            \item Abort if $(\sid, \fcecAuthority_i, 0)$ is not stored or if $(\sid, \fcecAuthority_i, 1)$ is already stored.
        
            \item Store $(\sid, \fcecAuthority_i, 1)$ and send $(\fcecsetupend, \allowbreak \sid)$ to $\fcecAuthority_i$.
        
        \end{itemize}
        
        \item On input $(\fcecregisterini, \allowbreak \sid)$ from  user $\fcecUser_j$ or provider $\fcecProvider_k$:
        
        \begin{itemize}[leftmargin=*]
        
        \item Abort if there is a tuple $(\sid, \allowbreak \fcecUser_j, \allowbreak 0)$ stored.
        
        \item Store $(\sid, \allowbreak \fcecUser_j, \allowbreak 0)$ and send $(\fcecregistersim, \allowbreak \sid, \allowbreak \fcecUser_j)$ to $\Simulator$.
        
        \end{itemize}
        
        \item[S.] On input $(\fcecregisterrep, \allowbreak \sid, \allowbreak \fcecUser_j)$ from the simulator $\Simulator$:
        
        \begin{itemize}[leftmargin=*]
        
        \item Abort if $(\sid, \allowbreak \fcecUser_j, \allowbreak 0)$ is not stored or if $(\sid, \allowbreak \fcecUser_j, \allowbreak 1)$ is stored.
        
        \item Store $(\sid, \allowbreak \fcecUser_j, \allowbreak 1)$ and send $(\fcecregisterend, \allowbreak \sid)$ to $\fcecUser_j$.
        
        \end{itemize}
        
        \item On input $(\fcecrequestini, \allowbreak \sid, \allowbreak \fcecAuthority_i, \allowbreak \fcecrequestid, \allowbreak \fcecwalletnum)$ from user $\fcecUser_j$:
        
        \begin{itemize}[leftmargin=*]
        
            \item Abort if $\sid \allowbreak \neq \allowbreak (\fcecAuthority_1, \allowbreak \ldots, \allowbreak \fcecAuthority_n, \allowbreak \sid')$, or if $\fcecAuthority_i \allowbreak \notin \allowbreak \sid$, or if $n < t$, or if $\fcecrequestid \allowbreak \notin \allowbreak \fcecunivrequestid$.
        
            \item Abort if $(\sid, \allowbreak \fcecUser_j, \allowbreak 1)$ is not stored, or if $(\sid, \allowbreak \fcecUser'_j, \allowbreak \fcecAuthority'_i, \allowbreak \fcecrequestid', \allowbreak \fcecwalletnum, \allowbreak \mathrm{user})$ stored such that $\fcecUser'_j \allowbreak = \allowbreak \fcecUser_j$, $\fcecAuthority'_i \allowbreak = \allowbreak \fcecAuthority_i$ and $\fcecrequestid' = \fcecrequestid$.
        
            \item Store $(\sid, \allowbreak \fcecUser_j, \allowbreak \fcecAuthority_i, \allowbreak \fcecrequestid, \allowbreak \fcecwalletnum, \allowbreak \mathrm{user})$.
        
            \item If $(\sid, \allowbreak \fcecUser'_j, \allowbreak \fcecwalletcount, \allowbreak \fcecdb)$ such that $\fcecUser'_j \allowbreak = \allowbreak \fcecUser_j$ is not stored, store $(\sid, \allowbreak \fcecUser_j, \allowbreak \fcecwalletcount, \allowbreak \fcecdb)$, where $\fcecwalletcount$ is a counter of the number of wallets of $\fcecUser_j$ initialized to $1$, and $\fcecdb$ is a (initially empty) database. 
        
            \item If $\fcecwalletnum \allowbreak \notin \allowbreak [1, \allowbreak \fcecwalletcount]$, set $\fcecwalletcount \allowbreak \gets \allowbreak \fcecwalletcount + \allowbreak 1$, set $\fcecwalletnum \allowbreak \gets \allowbreak \fcecwalletcount$, pick random $\fcecwalletid \allowbreak \gets \allowbreak \fcecunivwalletid$, and store an entry $[\fcecwalletnum, \allowbreak \fcecwalletid, 0, \allowbreak \emptyset]$ in $\fcecdb$. Update $\fcecwalletcount$ and $\fcecdb$ in the tuple $(\sid, \allowbreak \fcecUser_j, \allowbreak \fcecwalletcount, \allowbreak \fcecdb)$.
        
        
            \item Create a fresh $\qid$ and store $(\qid, \allowbreak \fcecUser_j, \allowbreak \fcecAuthority_i, \allowbreak \fcecrequestid, \allowbreak \fcecwalletnum)$.
        
        
            \item If $\fcecAuthority_i$ is honest, send $(\fcecrequestsim, \allowbreak \sid, \allowbreak \qid, \allowbreak \fcecUser_j, \allowbreak \fcecAuthority_i)$ to the simulator $\Simulator$, else pick $\fcecwalletid$ from the entry $[\fcecwalletnum', \allowbreak \fcecwalletid, 0, \allowbreak \emptyset] \allowbreak \in \allowbreak \fcecdb$ such that $\fcecwalletnum' \allowbreak = \allowbreak \fcecwalletnum$ and send $(\fcecrequestsim, \allowbreak \sid, \allowbreak \qid, \allowbreak \fcecUser_j, \allowbreak \fcecAuthority_i, \allowbreak \fcecwalletid)$ to the simulator $\Simulator$.
        
        \end{itemize}
        
        \item[S.] On input $(\fcecrequestrep, \allowbreak \sid, \allowbreak \qid)$ from the simulator $\Simulator$:
        
        \begin{itemize}[leftmargin=*]
        
            \item Abort if $(\qid', \allowbreak \fcecUser_j, \allowbreak \fcecAuthority_i, \fcecrequestid, \allowbreak \fcecwalletnum)$ such that $\qid' \allowbreak = \allowbreak \qid$ is not stored, or if $(\sid, \allowbreak \fcecAuthority_i, \allowbreak 1)$ is not stored.

            \item Store $(\sid, \allowbreak \fcecUser_j, \allowbreak \fcecAuthority_i, \allowbreak \fcecrequestid, \allowbreak \fcecwalletnum, \allowbreak \mathrm{authority})$, delete $(\qid, \allowbreak \fcecUser_j, \allowbreak \fcecAuthority_i, \allowbreak \fcecrequestid, \allowbreak \fcecwalletnum)$, and send $(\fcecrequestend, \allowbreak \sid, \allowbreak \fcecUser_j, \allowbreak \fcecrequestid)$ to $\fcecAuthority_i$.
        
        \end{itemize}
        
        \item On input $(\fcecissueini, \allowbreak \sid, \allowbreak \fcecUser_j, \allowbreak \fcecrequestid)$ from an authority $\fcecAuthority_i$:
        
            \begin{itemize}[leftmargin=*]
        
                \item Abort if a tuple $(\sid, \allowbreak \fcecUser'_j, \allowbreak \fcecAuthority'_i, \allowbreak \fcecrequestid', \allowbreak \fcecwalletnum, \allowbreak \mathrm{authority})$ such that $\fcecUser'_j \allowbreak = \allowbreak \fcecUser_j$, $\fcecAuthority'_i \allowbreak = \allowbreak \fcecAuthority_i$ and $\fcecrequestid' \allowbreak = \allowbreak \fcecrequestid$ is not stored.
        
                \item Create a fresh $\qid$, store $(\qid, \allowbreak \fcecUser_j, \allowbreak \fcecAuthority_i, \allowbreak \fcecrequestid, \allowbreak \fcecwalletnum, \allowbreak \mathrm{issue})$ and delete $(\sid, \allowbreak \fcecUser_j, \allowbreak \fcecAuthority_i, \allowbreak \fcecrequestid, \allowbreak \fcecwalletnum, \allowbreak \mathrm{authority})$.
        
                \item Send $(\fcecissuesim, \allowbreak \sid, \allowbreak \qid, \allowbreak \fcecAuthority_i, \allowbreak \fcecUser_j)$ to the simulator $\Simulator$.
        
            \end{itemize}
        
            \item[S.] On input $(\fcecissuerep, \allowbreak \sid, \allowbreak \qid)$ from the simulator $\Simulator$:
        
            \begin{itemize}[leftmargin=*]
        
                \item Abort if $(\qid', \allowbreak \fcecUser_j, \allowbreak \fcecAuthority_i, \allowbreak \fcecrequestid, \allowbreak \fcecwalletnum, \allowbreak \mathrm{issue})$ such that $\qid' = \qid$ is not stored.
        
                \item If $\fcecUser_j$ or $\fcecAuthority_i$ are honest, take the tuple $(\sid, \allowbreak \fcecUser'_j, \allowbreak \fcecwalletcount, \allowbreak \fcecdb)$ such that $\fcecUser'_j \allowbreak = \allowbreak \fcecUser_j$, and replace the entry $[\fcecwalletnum', \allowbreak \fcecwalletid, \allowbreak l, \allowbreak \fcecauthset]$ in $\fcecdb$ such that $\fcecwalletnum' = \fcecwalletnum$ by $[\fcecwalletnum', \allowbreak \fcecwalletid, \allowbreak l, \allowbreak \fcecauthset \cup \{\fcecAuthority_i\}]$.
        
                \item Delete $(\qid, \allowbreak \fcecUser_j, \allowbreak \fcecAuthority_i, \allowbreak \fcecrequestid, \allowbreak \fcecwalletnum, \allowbreak \mathrm{issue})$, and delete $(\sid, \allowbreak \fcecUser'_j, \allowbreak \fcecAuthority'_i, \allowbreak \fcecrequestid', \allowbreak \fcecwalletnum, \allowbreak \mathrm{user})$ such that $\fcecUser'_j \allowbreak = \allowbreak \fcecUser_j$, $\fcecAuthority'_i \allowbreak = \allowbreak \fcecAuthority_i$ and $\fcecrequestid' = \fcecrequestid$.
        
                \item Send $(\fcecissueend, \allowbreak \sid, \allowbreak \fcecrequestid, \allowbreak \fcecAuthority_i)$ to $\fcecUser_j$.
        
            \end{itemize}
        
        \item On input $(\fcecspendini, \allowbreak \sid, \allowbreak \fcecwalletnum, \allowbreak V, \allowbreak \fcecdsset, \allowbreak \cecpaymentinfo, \allowbreak \fcecnym, \allowbreak \fcecProvider_k)$ from $\fcecUser_j$:
        
        \begin{itemize}[leftmargin=*]
        
            \item Abort if $\fcecnym \allowbreak \notin \allowbreak \fcecunivnym$, or if $\cecpaymentinfo \allowbreak \notin \allowbreak \fcecunivpaymentinfo$, or if $\cecpaymentinfo$ does not contain $\fcecProvider_k$, or if $(\sid,  \allowbreak \fcecUser'_j, \allowbreak \fcecwalletcount, \allowbreak \fcecdb)$ such that $\fcecUser'_j \allowbreak = \allowbreak \fcecUser_j$ is not stored.
        
            \item Abort if there is not an entry $[\fcecwalletnum', \allowbreak \fcecwalletid, \allowbreak l, \allowbreak \fcecauthset] \allowbreak \in \allowbreak \fcecdb$ such that $\fcecwalletnum' \allowbreak = \allowbreak \fcecwalletnum$. Else, proceed as follows:
            \begin{itemize}[leftmargin=*]
        
                \item Abort if $\fcecUser_j$ is honest and either $V \notin [1,L]$ or $l \allowbreak + \allowbreak V \allowbreak > \allowbreak L$.
        
                \item Abort if $\fcecUser_j$ is corrupt and $\fcecdsset \notin [1,L]$. Else overwrite $V \gets |\fcecdsset|$.
        
                \item Abort if $|\fcecauthset| \allowbreak < \allowbreak t'$, where $t' \allowbreak = \allowbreak t$, if $\fcecUser_j$ is honest, or $t' \allowbreak = \allowbreak t - \allowbreak \tilde{t}$, if $\fcecUser_j$ is corrupt and there are $\tilde{t}$ corrupt authorities.
        
            \end{itemize}
        
            \item If $\fcecUser_j$ is honest, update $[\fcecwalletnum, \allowbreak \fcecwalletid, \allowbreak l, \allowbreak \fcecauthset]$ to $[\fcecwalletnum, \allowbreak \fcecwalletid, \allowbreak l+V, \allowbreak \fcecauthset]$.
        
            \item Create a fresh $\qid$ and store $(\qid, \allowbreak \fcecUser_j, \allowbreak \fcecwalletnum, \allowbreak V, \allowbreak \fcecdsset, \allowbreak \cecpaymentinfo, \allowbreak \fcecnym, \allowbreak \fcecProvider_k)$.
        
            \item Send $(\fcecspendsim, \allowbreak \sid, \allowbreak \qid)$ to $\Simulator$.
        
        \end{itemize}
        
        \item[S.] On input $(\fcecspendrep, \allowbreak \sid, \allowbreak \qid)$ from $\Simulator$:
        
        \begin{itemize}[leftmargin=*]
        
            \item Abort if $(\qid', \allowbreak \fcecUser_j, \allowbreak \fcecwalletnum, \allowbreak V, \allowbreak \fcecdsset, \allowbreak \cecpaymentinfo, \allowbreak \fcecnym, \allowbreak \fcecProvider_k)$ such that $\qid' \allowbreak = \allowbreak \qid$ is not stored and if $(\sid, \allowbreak \fcecProvider'_k, \allowbreak 1)$ such that $\fcecProvider'_k \allowbreak = \allowbreak \fcecProvider_k$ is not stored.
        
            \item Create a random unique payment identifier $\fcecpaymentid$ and store $(\sid, \allowbreak \fcecpaymentid, \allowbreak \fcecUser_j, \allowbreak \fcecwalletnum, \allowbreak V, \allowbreak \fcecdsset, \allowbreak \cecpaymentinfo, \allowbreak \fcecnym, \allowbreak \fcecProvider_k, \allowbreak 0)$.
        
            \item Delete $(\qid, \allowbreak \fcecUser_j, \allowbreak \fcecwalletnum, \allowbreak V, \allowbreak \fcecdsset, \allowbreak \cecpaymentinfo, \allowbreak \fcecnym, \allowbreak \fcecProvider_k)$.
        
            \item Send $(\fcecspendend, \allowbreak \fcecpaymentid, \allowbreak V, \allowbreak \cecpaymentinfo, \allowbreak \fcecnym)$ to $\fcecProvider_k$.
        
        \end{itemize}
        
        \item On input $(\fcecdepositini, \allowbreak \sid, \allowbreak \fcecpaymentid)$ from a provider $\fcecProvider_k$:
        
        \begin{itemize}[leftmargin=*]
        
            \item Abort if a tuple $(\sid, \allowbreak \fcecpaymentid', \allowbreak \fcecUser_j, \allowbreak \fcecwalletnum, \allowbreak V, \allowbreak \fcecdsset, \allowbreak \cecpaymentinfo, \allowbreak \fcecnym, \allowbreak \fcecProvider'_k, \allowbreak b)$ such that $\fcecpaymentid' \allowbreak = \allowbreak \fcecpaymentid$, $\fcecProvider'_k \allowbreak = \allowbreak \fcecProvider'_k$ and $b \allowbreak = \allowbreak 0$ is not stored.
        
            \item Create a fresh $\qid$ and store $(\qid, \allowbreak \fcecpaymentid, \allowbreak \fcecProvider_k)$.
        
            \item If $\fcecUser_j$ is corrupt and there is at least one corrupt authority, send $(\fcecdepositsim, \allowbreak \sid, \allowbreak \qid, \allowbreak \fcecUser_j, \allowbreak \fcecwalletnum, \allowbreak V, \allowbreak \fcecdsset, \allowbreak \cecpaymentinfo, \allowbreak \fcecnym, \allowbreak \fcecProvider_k)$ to $\Simulator$. Else if $\fcecUser_j$ is honest and there is at least one corrupt authority, send $(\fcecdepositsim, \allowbreak \sid, \allowbreak \qid, \allowbreak V, \allowbreak \cecpaymentinfo)$ to $\Simulator$. Else, send $(\fcecdepositsim, \allowbreak \sid, \allowbreak \qid)$ to $\Simulator$.
        
        \end{itemize}
        
        \item[S.] On input $(\fcecdepositrep, \allowbreak \sid, \allowbreak \qid)$ from $\Simulator$:
        
        \begin{itemize}[leftmargin=*]
        
            \item Abort if a tuple $(\qid', \allowbreak \fcecpaymentid, \allowbreak \fcecProvider_k)$ such that $\qid' \allowbreak = \allowbreak \qid$ is not stored.
        
            \item Update the stored tuple $(\sid, \allowbreak \fcecpaymentid', \allowbreak \fcecUser_j, \allowbreak \fcecwalletnum, \allowbreak V, \allowbreak \fcecdsset, \allowbreak \cecpaymentinfo, \allowbreak \fcecnym, \allowbreak \fcecProvider'_k, \allowbreak b)$ such that $\fcecpaymentid' \allowbreak = \allowbreak \fcecpaymentid$, $\fcecProvider'_k \allowbreak = \allowbreak \fcecProvider'_k$ and $b \allowbreak = \allowbreak 0$ to contain $b = 1$.
        
            \item Delete $(\qid, \allowbreak \fcecpaymentid, \allowbreak \fcecProvider_k)$.
        
            \item Send $(\fcecdepositend, \allowbreak \sid, \allowbreak \fcecpaymentid)$ to $\fcecProvider_k$.
        
        \end{itemize}
        
        \item On input $(\fcecdepvfini, \allowbreak \sid, \allowbreak \fcecusers, \allowbreak \cecpaymentinfo)$ from an authority $\fcecAuthority_i$:
        
        \begin{itemize}[leftmargin=*]
        
            \item Abort if $(\sid, \fcecAuthority'_i, 1)$ such that $\fcecAuthority'_i \allowbreak = \allowbreak \fcecAuthority_i$ is not stored or if there exists $\fcecUser_j \allowbreak \in \allowbreak \fcecusers$ such that a tuple $(\sid, \allowbreak \fcecUser_j, \allowbreak 1)$ is not stored.
        
            \item Create a fresh $\qid$ and store $(\qid, \allowbreak \fcecusers, \allowbreak \cecpaymentinfo, \allowbreak \fcecAuthority_i)$.
        
            \item Send $(\fcecdepvfsim, \allowbreak \sid, \allowbreak \qid, \allowbreak \fcecusers', \allowbreak \fcecnumdeposits)$ to $\Simulator$. $\fcecusers'$ is the set of users $\fcecUser_j \allowbreak \in \allowbreak \fcecusers$ such that $\fcecUser_j$ did not send a request to $\fcecAuthority_i$ and $\fcecUser_j$ was not received previously as input by $\fcecAuthority_i$ in another set $\fcecusers$. $\fcecnumdeposits$ is the number of payments deposited since $\fcecdepvf$ was last invoked by $\fcecAuthority_i$.
        
        \end{itemize}
        
        \item[S.] On input $(\fcecdepvfrep, \allowbreak \sid, \allowbreak \qid)$ from $\Simulator$:
        
        \begin{itemize}[leftmargin=*]
        
            \item Abort if $(\qid', \allowbreak \fcecusers, \allowbreak \cecpaymentinfo, \allowbreak \fcecAuthority_i)$ such that $\qid' \allowbreak = \allowbreak \qid$ is not stored.
        
            \item If there is no tuple $(\sid, \allowbreak \fcecpaymentid, \allowbreak \fcecUser_j, \allowbreak \fcecwalletnum, \allowbreak V, \allowbreak \fcecdsset, \allowbreak \cecpaymentinfo', \allowbreak \fcecnym, \allowbreak \fcecProvider_k, \allowbreak b)$ such that $\cecpaymentinfo' = \cecpaymentinfo$ and $b = 1$, set $c \allowbreak \gets \allowbreak 0$.
        
            \item If there are $K > 1$ tuples $(\sid, \allowbreak \fcecpaymentid, \allowbreak \fcecUser_j, \allowbreak \fcecwalletnum, \allowbreak V, \allowbreak \fcecdsset, \allowbreak \cecpaymentinfo', \allowbreak \fcecnym, \allowbreak \fcecProvider_k, \allowbreak b)$ such that $\cecpaymentinfo' = \cecpaymentinfo$ and $b = 1$, set $c \allowbreak \gets \allowbreak (\fcecProvider_1,\ldots,\fcecProvider_K)$, where $(\fcecProvider_1,\ldots,\fcecProvider_K)$ are the provider identities such that either $\fcecProvider_k$ is not included in $\cecpaymentinfo$, or there are two or more tuples deposited by $\fcecProvider_k$.
        
            \item If there is 1 tuple $(\sid, \allowbreak \fcecpaymentid, \allowbreak \fcecUser_j, \allowbreak \fcecwalletnum, \allowbreak V, \allowbreak \fcecdsset, \allowbreak \cecpaymentinfo', \allowbreak \fcecnym, \allowbreak \fcecProvider_k, \allowbreak b)$ such that $\cecpaymentinfo' = \cecpaymentinfo$ and $b = 1$, proceed as follows:
        
            \begin{itemize}[leftmargin=*]
        
                \item If $\fcecProvider_k$ is not in $\cecpaymentinfo$, output $c \allowbreak \gets \allowbreak \fcecProvider_k$.
        
                \item Else, if $\fcecUser_j$ is honest, set $c \allowbreak \gets \allowbreak 1$.
        
                \item Else, if $\fcecUser_j$ is corrupt, check if there are other tuples $(\sid, \allowbreak \fcecpaymentid, \allowbreak \fcecUser'_j, \allowbreak \fcecwalletnum', \allowbreak V, \allowbreak \fcecdsset', \allowbreak \cecpaymentinfo, \allowbreak \fcecnym, \allowbreak \fcecProvider_k, \allowbreak b)$ such that $\fcecUser'_j \allowbreak = \allowbreak \fcecUser_j$, $\fcecwalletnum' \allowbreak = \allowbreak \fcecwalletnum'$ and $b \allowbreak = \allowbreak 1$. For all such tuples, check if $\fcecdsset' \cap \fcecdsset = \emptyset$. If that is the case for all tuples, set $c \allowbreak \gets \allowbreak 1$. Else, if $\fcecUser_j \allowbreak \in \allowbreak \fcecusers$, set $c \allowbreak \gets \allowbreak \fcecUser_j$, else set $c \allowbreak \gets \allowbreak \bot$.
        
            \end{itemize}
        
            \item Delete $(\qid, \allowbreak \fcecusers, \allowbreak \cecpaymentinfo, \allowbreak \fcecAuthority_i)$.
        
            \item Send $(\fcecdepvfend, \allowbreak \sid, \allowbreak \cecpaymentinfo, \allowbreak c)$ to $\fcecAuthority_i$.
        
        \end{itemize}
        
        \end{enumerate}
        \end{multicols}
        }
   \end{framed}
\vspace*{-3mm}
\caption{Ideal Functionality $\Functionality_{\CEC}$}
\label{fig:idealfun}
\end{figure*}

\paragraph{Description of $\Functionality_{\CEC}$.} In the following, we explain how each of the interfaces of $\Functionality_{\CEC}$ works:

\circled{1} An authority $\fcecAuthority_i$ uses the $\fcecsetup$ interface to set up $\Functionality_{\CEC}$. $\Functionality_{\CEC}$ stores the fact that $\fcecAuthority_i$ has run the setup interface and enforces that each authority runs the setup interface only once. The simulator $\Simulator$ is allowed to learn that $\fcecAuthority_i$ has run the setup interface. $\Functionality_{\CEC}$ allows each authority $\fcecAuthority_i$ to run the setup interface independently of other authorities, i.e.\ the execution of the setup interface for one authority can be finalized without the involvement of other authorities. Therefore, $\Functionality_{\CEC}$ is realizable by protocols where authorities run the setup interface independently of each other. For example, protocols where each authority creates its own keys, or protocols where authorities obtain their keys from a trusted third party that generates them. $\Functionality_{\CEC}$ can be modified so that it is realizable by protocols in which the setup interface requires interaction between authorities, e.g. protocols that use a distributed key generation protocol as a building block.

\circled{2} A user $\fcecUser_j$ or a provider $\fcecProvider_k$ use the $\fcecregister$ interface to register. $\Functionality_{\CEC}$ stores the fact that $\fcecUser_j$ or $\fcecProvider_k$ has registered, and enforces that each user or provider runs the registration interface only once. The simulator $\Simulator$ is allowed to learn that $\fcecUser_j$ or $\fcecProvider_k$ has run the setup interface. 

\circled{3} 
A user $\fcecUser_j$ runs the $\fcecrequest$ interface given an authority identifier $\fcecAuthority_i$, a request identifier $\fcecrequestid$ and a wallet number $\fcecwalletnum$. The request identifier $\fcecrequestid$ is used to bind a request to its subsequent issuance, while the wallet number $\fcecwalletnum$ is used to associate the request to a wallet. $\Functionality_{\CEC}$ checks that the user has run the registration interface, and that there is not a request pending from $\fcecUser_j$ to $\fcecAuthority_i$ with the same identifier. Then, if a database for $\fcecUser_j$ is not stored, $\Functionality_{\CEC}$ stores a tuple $(\sid, \allowbreak \fcecUser_j, \allowbreak \fcecwalletcount, \allowbreak \fcecdb)$, where $\fcecwalletcount$ is a counter of the number of wallets of $\fcecUser_j$ initialized to $1$, and $\fcecdb$ is a (initially empty) database. $\fcecdb$ has entries of the form $[\fcecwalletnum, \allowbreak \fcecwalletid, \allowbreak l, \allowbreak \fcecauthset]$, where $\fcecwalletnum$ is the wallet number, $\fcecwalletid$ is a wallet identifier, $l$ is the number of coins spent from the wallet $\fcecwalletnum$, and $\fcecauthset$ is the set of authorities that issued requests for the creation of this wallet.
If $\fcecwalletnum$ received as input is such that $\fcecwalletnum \allowbreak \notin \allowbreak [1, \fcecwalletcount]$, then $\Functionality_{\CEC}$ increments $\fcecwalletcount$, sets $\fcecwalletnum \allowbreak \gets \allowbreak \fcecwalletcount$, picks a random wallet identifier $\fcecwalletid$ and stores a new entry $[\fcecwalletnum, \allowbreak \fcecwalletid, \allowbreak 0, \allowbreak \emptyset]$ in the database for $\fcecUser_j$. The reason for creating $\fcecwalletid$ is that, in our e-cash schemes, requests for the same wallet can be linked to each other if authorities communicate with each other. Therefore, when $\fcecAuthority_i$ is dishonest, $\Functionality_{\CEC}$ leaks $\fcecwalletid$ to the simulator $\Simulator$. $\Functionality_{\CEC}$ does not leak $\fcecwalletnum$ because dishonest authorities do not necessarily learn how many wallets a user requests.

After being prompted by the simulator $\Simulator$, $\Functionality_{\CEC}$ checks that $\fcecAuthority_i$ has run the setup interface, records that $\fcecUser_j$ sends a request to $\fcecAuthority_i$ with identifier $\fcecrequestid$ for wallet $\fcecwalletnum$, and sends $\fcecUser_j$ and $\fcecrequestid$ to $\fcecAuthority_i$. 

\circled{4} An authority $\fcecAuthority_i$ runs the $\fcecissue$ interface given a user identifier $\fcecUser_j$ and a request identifier $\fcecrequestid$. $\Functionality_{\CEC}$ checks whether there is a request identifier $\fcecrequestid$ pending for a request from $\fcecUser_j$ to $\fcecAuthority_i$. In that case, after being prompted by the simulator $\Simulator$, $\Functionality_{\CEC}$ records that $\fcecAuthority_i$ has run the issuance for the request $\fcecrequestid$. Concretely, $\Functionality_{\CEC}$ updates the entry $[\fcecwalletnum, \allowbreak \fcecwalletid, \allowbreak l, \allowbreak \fcecauthset]$ in $\fcecdb$ such that $\fcecwalletnum$ is the wallet number associated with $\fcecrequestid$ to contain $\fcecAuthority_i$ in the set $\fcecauthset$. Finally, $\Functionality_{\CEC}$ informs $\fcecUser_j$ that $\fcecAuthority_i$ has run the issuance for the request $\fcecrequestid$.

\circled{5} A user $\fcecUser_j$ runs the $\fcecspend$ interface given a wallet number $\fcecwalletnum$, a number $V$ of coins to be spent, a set of coin indices $\fcecdsset$, payment information $\cecpaymentinfo$, a pseudonym $\fcecnym$ and a provider identifier $\fcecProvider_k$. 
The payment information $\cecpaymentinfo$ should be unique for each spending, but $\Functionality_{\CEC}$ checks that later when verifying a deposit. If $\fcecUser_j$ is honest, $\Functionality_{\CEC}$ checks that there are enough non-spent coins related to $\fcecwalletnum$ ($l+V \leq L$, where $l$ is in the entry $[\fcecwalletnum, \allowbreak \fcecwalletid, \allowbreak l, \allowbreak \fcecauthset] \allowbreak \in \allowbreak \fcecdb)$ and in that case adds $V$ to the number of spent coins. In contrast, if $\fcecUser_j$ is corrupt, $\Functionality_{\CEC}$ records that the coins with indices $\fcecdsset$ are spent, regardless of whether they were spent before or not.  $\Functionality_{\CEC}$ also checks that enough authorities have issued the wallet $\fcecwalletnum$ to $\fcecUser_j$. In that case, after being prompted by the simulator, $\Functionality_{\CEC}$ checks that $\fcecProvider_k$ has run the registration interface and creates a payment identifier $\fcecpaymentid$ to store the information related to this spending. Finally $\Functionality_{\CEC}$ sends $\fcecpaymentid$, $V$, $\cecpaymentinfo$ and $\fcecnym$ to $\fcecProvider_k$. We remark that $\fcecProvider_k$ does not learn $\fcecUser_j$, and $\fcecnym$ can be different for each spending so that spendings by $\fcecUser_j$ are unlinkable to each other and to the request phase.

\circled{6} A provider $\fcecProvider_k$ runs the $\fcecdeposit$ interface on input a payment identifier $\fcecpaymentid$. If there is a payment with identifier $\fcecpaymentid$ related to $\fcecProvider_k$ that is not deposited, $\Functionality_{\CEC}$ proceeds to deposit it. If no authorities are corrupt, $\Functionality_{\CEC}$ does not leak to $\Simulator$ any information about the deposited payment. However, if at least one authority is corrupt and the user that computed the payment is corrupt, $\Functionality_{\CEC}$ leaks the full information about the payment. If at least one authority is corrupt but the user that computed the payment is honest, the authority leaks the number $V$ of coins spent and the payment information $\cecpaymentinfo$. After being prompted by the simulator, $\Functionality_{\CEC}$ marks that payment as deposited and informs $\fcecProvider_k$ that the payment has been deposited.

\circled{7} An honest authority $\fcecAuthority_i$ runs the $\fcecdepvf$ interface given a list of user identifiers $\fcecusers$ and payment information $\cecpaymentinfo$. $\Functionality_{\CEC}$ checks that $\fcecAuthority_i$ has run the setup interface and that all the users in $\fcecusers$ have run the registration interface.  $\Functionality_{\CEC}$ leaks to the simulator the identities of those users in $\fcecusers$ that were unknown by $\fcecAuthority_i$. This is done because, in our protocol, $\fcecAuthority_i$ needs to retrieve the public key for that user, and the adversary learns that. $\Functionality_{\CEC}$ also leaks the number of deposits that were made since the last time $\fcecAuthority_i$ run the deposit verification interface. This is done because, in our protocol, the authority needs to read the new deposits from the bulletin board, and the adversary learns that.

After being prompted by the simulator, $\Functionality_{\CEC}$ checks the deposited payment with payment information $\cecpaymentinfo$. If no such deposited payment exists, $\Functionality_{\CEC}$ sets $c \gets 0$. If there is more than one deposited payment with $\cecpaymentinfo$, $\Functionality_{\CEC}$ sets $c$ to contain the identifiers of the provider(s) that deposited those payments more than once, or just once if the identity of the provider is not in $\cecpaymentinfo$. If there is 1 such tuple,  $\Functionality_{\CEC}$ sets $c$ to the identity of the provider that deposited the payment if that identity is not in $\cecpaymentinfo$. Else, $\Functionality_{\CEC}$ sets $c \gets 1$ if the user that made a payment with $\cecpaymentinfo$ is honest, which means that there has not been a double spending. If the user is corrupt, $\Functionality_{\CEC}$ checks whether there are deposited payments where the coins spent in the payment related with $\cecpaymentinfo$ have also been spent. If that is not the case, $\Functionality_{\CEC}$ sets $c \allowbreak \gets \allowbreak 1$. Else, if the user identifier is in $\fcecusers$, $\Functionality_{\CEC}$ sets $c$ to the identifier of the user that double spent, and otherwise sets $c \allowbreak \gets \allowbreak \bot$, which indicates that double spending has been detected but the user has not been identified. $\Functionality_{\CEC}$ sends $c$ along with $\cecpaymentinfo$ to $\fcecAuthority_i$.

\paragraph{Security properties.} We now argue that $\Functionality_{\CEC}$ guarantees the security properties
defined in \S\ref{sec:securityoffec}. 
\begin{description}[leftmargin=*, noitemsep,topsep=0pt]

    \item[Traceability.] In the $\fcecspend$ interface, when a user wishes to spend coins in the wallet $\fcecwalletnum$, $\Functionality_{\CEC}$ checks that the user has been issued that wallet by at least $t-\tilde{t}$ authorities. This guarantees that users cannot forge wallets. Moreover, in the $\fcecdepvf$ interface, $\Functionality_{\CEC}$ guarantees that, if coins were double spent, the user is identified whenever the user identifier is included in the set $\fcecusers$. $\Functionality_{\CEC}$ also finds providers guilty of wrongly depositing a payment, either when they deposit a payment with the same $\cecpaymentinfo$ more than once, or when they deposit a payment with $\cecpaymentinfo$ that does not contain the provider's identity.

    \item[Unlinkability.] In the $\fcecspend$ interface, the user identity is not revealed to the provider. The provider only receives a pseudonym, which can be different at each spending. This guarantees that payments from the same user cannot be linked with each other or to withdrawals by that user.

    \item[Exculpability.] $\Functionality_{\CEC}$ never finds an honest user guilty of double spending. Therefore, any protocol that realizes $\Functionality_{\CEC}$ must guarantee that.

    \item[Clearance.] In the $\fcecdepvf$ interface,  $\Functionality_{\CEC}$ never accepts a deposit as valid if the identity of the provider that made the deposit is not contained in $\cecpaymentinfo$.

\end{description}
We remark that $\Functionality_{\CEC}$ does not take into account the case where $t$ or more authorities are corrupt. We will analyze the security of construction $\mathrm{\Pi}_{\CEC}$ under the assumption that at most $t-1$ authorities are corrupt.

%% file: 10ConstructionEcash.tex
\section{Construction $\mathrm{\Pi}_{\CEC}$}
\label{sec:constructionCEC}

\begin{figure*}
    \begin{framed}
        {\small
        \vspace*{-10pt}
        \begin{multicols}{2}
        \begin{enumerate}[leftmargin=*]

    \item On input $(\fcecsetupini, \allowbreak \sid)$, $\fcecAuthority_i$ does the following:

    \begin{itemize}[leftmargin=*]

        \item Abort if $\sid \neq (\fcecAuthority_1, \ldots, \fcecAuthority_n, \sid')$, or if $\fcecAuthority_i \allowbreak \notin \allowbreak \sid$, or if $n < t$.

        \item Abort if $(\sid, \allowbreak \cecparams, \allowbreak \spk, \allowbreak  \ssk_{\fcecAuthority_i}, \allowbreak \spk_{\fcecAuthority_i})$ is already stored.


        \item Send $(\fkggetkeyini, \allowbreak \sid)$ to $\Functionality_{\KG}$. In its first invocation, $\Functionality_{\KG}$ runs $\cecparams \allowbreak \gets \allowbreak \cecSetup(1^{\securityparameter}, \allowbreak L)$ and $(\spk, \allowbreak \langle \ssk_{\fcecAuthority_i}, \allowbreak \spk_{\fcecAuthority_i} \rangle_{i\in[1,n]}) \allowbreak \gets \allowbreak \cecKeyGenA(\cecparams, \allowbreak t, \allowbreak n)$. $\Functionality_{\KG}$ sends $(\fkggetkeyend, \allowbreak \sid, \allowbreak \cecparams, \allowbreak \spk, \allowbreak \ssk_{\fcecAuthority_i}, \allowbreak \spk_{\fcecAuthority_i})$ to $\fcecAuthority_i$.

        \item Store $(\sid, \allowbreak \cecparams, \allowbreak \spk, \allowbreak  \ssk_{\fcecAuthority_i}, \allowbreak \spk_{\fcecAuthority_i})$ and output $(\fcecsetupend, \sid)$.

    \end{itemize}

    \item On input $(\fcecregisterini, \sid)$, $\fcecUser_j$ (or  $\fcecProvider_k$) does the following:

    \begin{itemize}[leftmargin=*]


        \item Abort if $(\sid, \allowbreak \ssk_{\fcecUser_j}, \allowbreak \spk_{\fcecUser_j})$ is already stored.

        \item Send $(\fkgretrieveini, \allowbreak \sid)$ to $\Functionality_{\KG}$. $\Functionality_{\KG}$ sends $(\fkgretrieveend,  \allowbreak \sid,  \allowbreak \fkgvalue)$. If $\fkgvalue = (\cecparams, \allowbreak \spk, \allowbreak \langle  \spk_{\fcecAuthority_i} \rangle_{i\in[1,n]})$, store $(\sid, \allowbreak \cecparams, \allowbreak \spk, \allowbreak \langle  \spk_{\fcecAuthority_i} \rangle_{i\in[1,n]})$, else abort.

        \item Run $(\ssk_{\fcecUser_j}, \allowbreak \spk_{\fcecUser_j}) \gets \cecKeyGenU(\cecparams)$.

        \item Set $\sid_{\Freg} \gets (\fcecUser_j, \sid')$ and send $(\fregregisterini, \allowbreak \sid_{\Freg}, \allowbreak \spk_{\fcecUser_j})$ to $\Functionality_{\Freg}$. $\Functionality_{\Freg}$ sends $(\fregregisterend, \allowbreak \sid_{\Freg})$ to $\fcecUser_j$.

        \item Store $(\sid, \allowbreak \ssk_{\fcecUser_j}, \allowbreak \spk_{\fcecUser_j})$ and output $(\fcecregisterend, \allowbreak \sid)$.

    \end{itemize}

    \item On input $(\fcecrequestini, \allowbreak \sid, \allowbreak \fcecAuthority_i, \allowbreak \fcecrequestid, \allowbreak \fcecwalletnum)$, $\fcecUser_j$ and $\fcecAuthority_i$ do:

    \begin{itemize}[leftmargin=*]

        \item $\fcecUser_j$ aborts if $\sid \allowbreak \neq \allowbreak (\fcecAuthority_1, \allowbreak \ldots, \allowbreak \fcecAuthority_n, \allowbreak \sid')$, or if $\fcecAuthority_i \allowbreak \notin \allowbreak \sid$, or if $n \allowbreak < \allowbreak t$, or if $\fcecrequestid \allowbreak \notin \allowbreak \fcecunivrequestid$, or if $(\sid, \allowbreak \ssk_{\fcecUser_j}, \allowbreak \spk_{\fcecUser_j})$ is not stored, or if there is $(\sid, \allowbreak \fcecAuthority'_i, \allowbreak \fcecrequestid', \allowbreak \fcecwalletnum)$ such that $\fcecrequestid' \allowbreak = \allowbreak \fcecrequestid$ and $\fcecAuthority'_i \allowbreak = \allowbreak \fcecAuthority_i$.

        \item If there is not a tuple $(\sid, \allowbreak \fcecwalletcount)$, $\fcecUser_j$ stores $(\sid, \allowbreak 0)$.

        \item If $\fcecwalletnum \allowbreak \notin \allowbreak [1, \allowbreak \fcecwalletcount]$, $\fcecUser_j$ sets $\fcecwalletcount \allowbreak \gets \allowbreak \fcecwalletcount \allowbreak + \allowbreak 1$, sets $\fcecwalletnum \allowbreak \gets \allowbreak \fcecwalletcount$, runs $(\cecrequest, \allowbreak \cecrequestinfo) \allowbreak \gets \allowbreak \cecRequest(\cecparams, \allowbreak \ssk_{\fcecUser_j})$, stores $(\sid, \allowbreak \fcecwalletnum, \allowbreak \cecrequest, \allowbreak \cecrequestinfo)$ and updates $(\sid, \allowbreak \fcecwalletcount)$. $\fcecUser_j$ stores $(\sid, \allowbreak \fcecAuthority_i, \allowbreak \fcecrequestid, \allowbreak \fcecwalletnum)$.


        \item $\fcecUser_j$ sets $\sid_{\SMT} \allowbreak \gets \allowbreak (\fcecUser_j,\fcecAuthority_i,\sid')$ and sends $(\fsmtsendini, \allowbreak \sid_{\SMT}, \allowbreak \langle \fcecrequestid, \allowbreak \cecrequest \rangle)$ to $\Functionality_{\SMT}$.

        \item $\fcecAuthority_i$ receives $(\fsmtsendend, \allowbreak \sid_{\SMT}, \allowbreak \langle \fcecrequestid, \allowbreak \cecrequest \rangle)$ from $\Functionality_{\SMT}$.

        \item $\fcecAuthority_i$ aborts if $(\sid, \allowbreak \cecparams, \allowbreak \spk, \allowbreak  \ssk_{\fcecAuthority_i}, \allowbreak \spk_{\fcecAuthority_i})$ is not stored.

        \item $\fcecAuthority_i$ aborts if there is a tuple $(\sid, \allowbreak \fcecrequestid', \allowbreak \cecrequest, \allowbreak \fcecUser'_j)$ stored such that $\fcecrequestid' \allowbreak = \allowbreak \fcecrequestid$ and $\fcecUser'_j \allowbreak = \allowbreak \fcecUser_j$.

        \item $\fcecAuthority_i$ parses $\sid_{\SMT}$ as $(\fcecUser_j, \allowbreak \fcecAuthority_i, \allowbreak \sid')$. If $(\sid, \allowbreak \fcecUser'_j, \spk_{\fcecUser_j})$ such that $\fcecUser'_j = \fcecUser_j$ is not stored, $\fcecAuthority_i$ does the following:
        \begin{itemize}

            \item $\fcecAuthority_i$ sets $\sid_{\Freg} \allowbreak \gets \allowbreak (\fcecUser_j, \allowbreak \sid')$ and sends $(\fregretrieveini, \allowbreak \sid_{\Freg})$ to $\Functionality_{\Freg}$. $\Functionality_{\Freg}$ sends $(\fregretrieveend, \allowbreak \sid_{\Freg}, \allowbreak \spk_{\fcecUser_j})$ to $\fcecAuthority_i$.

            \item If $\spk_{\fcecUser_j} = \bot$, $\fcecAuthority_i$ aborts, else $\fcecAuthority_i$ stores $(\sid, \allowbreak \fcecUser_j, \spk_{\fcecUser_j})$.

        \end{itemize}

        \item $\fcecAuthority_i$ runs $b \gets \cecRequestVf(\cecparams, \allowbreak \cecrequest, \allowbreak \spk_{\fcecUser_j})$. If $b \allowbreak = \allowbreak 0$, $\fcecAuthority_i$ aborts, else $\fcecAuthority_i$ stores $(\sid, \allowbreak \fcecrequestid, \allowbreak \cecrequest, \allowbreak \fcecUser_j)$.

        \item $\fcecAuthority_i$ outputs $(\fcecrequestend, \sid, \fcecUser_j, \fcecrequestid)$.

    \end{itemize}

    \item On input $(\fcecissueini, \sid, \fcecUser_j, \fcecrequestid)$, $\fcecAuthority_i$ and $\fcecUser_j$  do the following:

    \begin{itemize}[leftmargin=*]

        \item $\fcecAuthority_i$ aborts if $(\sid, \allowbreak \fcecrequestid', \allowbreak \cecrequest, \allowbreak \fcecUser'_j)$ such that $\fcecrequestid' \allowbreak = \allowbreak \fcecrequestid$ and $\fcecUser'_j \allowbreak = \allowbreak \fcecUser_j$ is not stored.

        \item $\fcecAuthority_i$ runs $\cecresponse \allowbreak \gets \allowbreak \cecWithdraw(\cecparams, \allowbreak \ssk_{\fcecAuthority_i}, \allowbreak \cecrequest)$.

        \item $\fcecAuthority_i$ deletes $(\sid, \allowbreak \fcecrequestid, \allowbreak \cecrequest, \allowbreak \fcecUser_j)$, sets $\sid_{\SMT} \allowbreak \gets \allowbreak (\fcecAuthority_i, \allowbreak \fcecUser_j, \allowbreak \sid')$ and sends $(\fsmtsendini, \allowbreak \sid_{\SMT}, \allowbreak \langle \fcecrequestid, \allowbreak \cecresponse \rangle)$ to $\Functionality_{\SMT}$.


        \item $\fcecUser_j$ receives $(\fsmtsendend, \allowbreak \sid_{\SMT}, \allowbreak \langle \fcecrequestid, \allowbreak \cecresponse \rangle)$ from $\Functionality_{\SMT}$.

        \item $\fcecUser_j$ parses $\sid_{\SMT}$ as $(\fcecAuthority_i, \allowbreak \fcecUser_j, \allowbreak \sid')$. $\fcecUser_j$ aborts if a tuple $(\sid, \allowbreak \fcecAuthority'_i, \allowbreak \fcecrequestid', \allowbreak \fcecwalletnum)$ such that $\fcecrequestid' \allowbreak = \allowbreak \fcecrequestid$ and $\fcecAuthority'_i \allowbreak = \allowbreak \fcecAuthority'_i$ is not stored. Else $\fcecUser_j$ takes the stored tuple $(\sid, \allowbreak \fcecwalletnum', \allowbreak \cecrequest, \allowbreak \cecrequestinfo)$ such that $\fcecwalletnum' = \fcecwalletnum$ and runs $\cecwallet_i \allowbreak \gets \allowbreak \cecWithdrawVf(\cecparams, \allowbreak \spk_{\fcecAuthority_i}, \allowbreak \ssk_{\fcecUser_j}, \allowbreak \cecresponse, \allowbreak \cecrequestinfo)$. If $\cecwallet_i \allowbreak = \allowbreak 0$, $\fcecUser_j$ aborts.

        \item If there is not a tuple $(\sid, \allowbreak \fcecwalletnum', \allowbreak \cecset, \allowbreak \fcecwalletset)$ such that $\fcecwalletnum' \allowbreak = \allowbreak \fcecwalletnum$, $\fcecUser_j$ stores $(\sid, \allowbreak \fcecwalletnum, \allowbreak \emptyset, \allowbreak \emptyset)$.

        \item $\fcecUser_j$ updates $(\sid, \allowbreak \fcecwalletnum, \allowbreak \cecset, \allowbreak \fcecwalletset)$ to $(\sid, \allowbreak \fcecwalletnum, \allowbreak \cecset \cup \{i\}, \allowbreak \fcecwalletset \cup \{\cecwallet_i\})$.

        \item If $(\sid, \allowbreak \fcecwalletnum', \allowbreak \cecwallet)$ such that $\fcecwalletnum' \allowbreak = \allowbreak \fcecwalletnum$ is not stored, and if $|\cecset| \allowbreak  \geq \allowbreak t$ in the tuple $(\sid, \allowbreak \fcecwalletnum', \allowbreak \cecset, \allowbreak \fcecwalletset)$ such that $\fcecwalletnum' \allowbreak = \allowbreak \fcecwalletnum$, then $\fcecUser_j$ runs $\cecwallet \allowbreak \gets \allowbreak \cecCreateWallet(\spk, \allowbreak \ssk_{\fcecUser_j}, \allowbreak \cecset, \allowbreak \fcecwalletset)$ and stores $(\sid, \allowbreak \fcecwalletnum, \allowbreak \cecwallet)$.

        \item $\fcecUser_j$ deletes $(\sid, \allowbreak \fcecAuthority_i, \allowbreak \fcecrequestid, \allowbreak \fcecwalletnum)$.

        \item Output $(\fcecissueend, \allowbreak \sid, \allowbreak \fcecrequestid, \allowbreak \fcecAuthority_i)$.

    \end{itemize}

    \item On input $(\fcecspendini, \allowbreak \sid, \allowbreak \fcecwalletnum, \allowbreak V, \allowbreak \fcecdsset, \allowbreak \cecpaymentinfo, \allowbreak \fcecnym, \fcecProvider_k)$, $\fcecUser_j$ and $\fcecAuthority_i$ do:

    \begin{itemize}[leftmargin=*]

        \item $\fcecUser_j$ aborts if $\fcecnym \allowbreak \notin \allowbreak \fcecunivnym$, or if $\cecpaymentinfo \allowbreak \notin \allowbreak \fcecunivpaymentinfo$, or if $\cecpaymentinfo$ does not contain $\fcecProvider_k$, or if $(\sid, \allowbreak \fcecwalletnum', \allowbreak \cecwallet)$ such that $\fcecwalletnum' \allowbreak = \allowbreak \fcecwalletnum$ is not stored, or if $V \allowbreak \notin \allowbreak [1,L]$.

        \item $\fcecUser_j$ runs $b \allowbreak \gets \allowbreak \cecSpend(\spk, \allowbreak \ssk_{\fcecUser_j}, \allowbreak \cecwallet, \allowbreak \cecpaymentinfo, \allowbreak V)$. If $b \allowbreak = \allowbreak 0$, $\fcecUser_j$ aborts. Else $\fcecUser_j$ parses $b$ as $(\cecwallet', \allowbreak \cecpayment)$ and updates the stored tuple $(\sid, \allowbreak \fcecwalletnum, \allowbreak \cecwallet)$ to $(\sid, \allowbreak \fcecwalletnum, \allowbreak \cecwallet')$.

        \item $\fcecUser_j$ sends $(\fnymsendini, \allowbreak \sid, \allowbreak \langle \cecpayment, \cecpaymentinfo \rangle, \allowbreak \fcecnym, \allowbreak \fcecProvider_k)$ to $\Functionality_{\NYM}$.

        \item $\fcecProvider_k$ receives $(\fnymsendend, \allowbreak \sid, \langle \cecpayment, \cecpaymentinfo \rangle, \allowbreak \fcecnym)$ from  $\Functionality_{\NYM}$.

        \item $\fcecProvider_k$ aborts if $(\sid, \allowbreak \cecparams, \allowbreak \spk, \allowbreak \langle  \spk_{\fcecAuthority_i} \rangle_{i\in[1,n]})$ is not stored, or if $\cecpaymentinfo$ does not contain $\fcecProvider_k$.

        \item $\fcecProvider_k$ runs $b \allowbreak \gets \allowbreak \cecSpendVf(\spk, \allowbreak \cecpayment, \allowbreak \cecpaymentinfo)$. If $b \allowbreak = \allowbreak 0$, $\fcecProvider_k$ aborts, else $\fcecProvider_k$ sets $V \gets b$, creates a random unique payment identifier $\fcecpaymentid$ and stores $(\sid, \allowbreak \fcecpaymentid, \allowbreak \cecpayment, \allowbreak \cecpaymentinfo, \allowbreak V, \allowbreak 0)$.

        \item $\fcecProvider_k$ outputs $(\fcecspendend, \allowbreak \fcecpaymentid, \allowbreak V, \allowbreak \cecpaymentinfo, \allowbreak \fcecnym)$.

    \end{itemize}

    \item On input $(\fcecdepositini, \allowbreak \sid, \allowbreak \fcecpaymentid)$, $\fcecProvider_k$ does the following:

    \begin{itemize}[leftmargin=*]

        \item Abort if a tuple $(\sid, \allowbreak \fcecpaymentid', \allowbreak \cecpayment, \allowbreak \cecpaymentinfo, \allowbreak V, \allowbreak b)$ such that $\fcecpaymentid' \allowbreak = \allowbreak \fcecpaymentid$ and $b \allowbreak = \allowbreak 0$ is not stored.

        \item Send $(\fbbwriteini, \allowbreak \sid, \allowbreak \langle \cecpayment, \allowbreak \cecpaymentinfo \rangle)$ to the functionality $\Functionality_{\BB}$. $\Functionality_{\BB}$ sends $(\fbbwriteend, \allowbreak \sid)$ to $\fcecProvider_k$.

        \item Update the tuple $(\sid, \allowbreak \fcecpaymentid, \allowbreak \cecpayment, \allowbreak \cecpaymentinfo, \allowbreak V, \allowbreak b)$ so that $b \allowbreak = \allowbreak 1$.

        \item Output $(\fcecdepositend, \allowbreak \sid, \allowbreak \fcecpaymentid)$.

    \end{itemize}

    \item On input $(\fcecdepvfini, \allowbreak \sid, \allowbreak \fcecusers, \allowbreak \cecpaymentinfo)$ from an authority $\fcecAuthority_i$:

    \begin{itemize}[leftmargin=*]

        \item Abort if $(\sid, \allowbreak \cecparams, \allowbreak \spk, \allowbreak \ssk_{\fcecAuthority_i}, \allowbreak \spk_{\fcecAuthority_i})$ is not stored.

        \item For all $\fcecUser_j \allowbreak \in \allowbreak \fcecusers$, if $(\sid, \allowbreak \fcecUser'_j, \spk_{\fcecUser_j})$ such that $\fcecUser'_j \allowbreak = \allowbreak \fcecUser_j$ is not stored, do the following:
        \begin{itemize}[leftmargin=*]

            \item Set $\sid_{\Freg} \allowbreak \gets \allowbreak (\fcecUser_j, \allowbreak \sid')$ and send $(\fregretrieveini, \allowbreak \sid_{\Freg})$ to $\Functionality_{\Freg}$. $\Functionality_{\Freg}$ sends back $(\fregretrieveend, \allowbreak \sid_{\Freg}, \allowbreak \spk_{\fcecUser_j})$.

            \item If $\spk_{\fcecUser_j} \allowbreak = \allowbreak \bot$, abort, else store $(\sid, \allowbreak \fcecUser_j, \spk_{\fcecUser_j})$.

        \end{itemize}

        \item Include in $\cecPK$ the public keys $\spk_{\fcecUser_j}$ in all the stored tuples $(\sid, \allowbreak \fcecUser_j, \allowbreak \spk_{\fcecUser_j})$ such that $\fcecUser_j \allowbreak \in \allowbreak \fcecusers$.

        \item If $(\sid, \allowbreak \fbbindex)$ is not stored, set $\fbbindex \allowbreak = \allowbreak 1$ and store $(\sid, \allowbreak \fbbindex)$.

        \item While $\fbbmessage' \neq \bot$, do the following:

        \begin{itemize}[leftmargin=*]

            \item Send $(\fbbgetbbini, \allowbreak \sid, \allowbreak \fbbindex)$ to $\Functionality_{\BB}$. $\Functionality_{\BB}$ sends back $(\fbbgetbbend, \allowbreak \sid, \allowbreak \fbbmessage')$. If $\fbbmessage' \allowbreak = \allowbreak (\fcecProvider_k, \cecpayment, \allowbreak \cecpaymentinfo)$, Store $(\sid, \allowbreak \fcecProvider_k, \cecpayment, \allowbreak \cecpaymentinfo)$.

            \item Increment $\fbbindex$.

        \end{itemize}

        \item Update $\fbbindex$ in the tuple $(\sid, \allowbreak \fbbindex)$.

        \item Find all the stored tuples $(\sid, \allowbreak \fcecProvider_k, \cecpayment, \allowbreak \cecpaymentinfo')$ such that $\cecpaymentinfo' \allowbreak = \allowbreak \cecpaymentinfo$ and do the following:

        \begin{itemize}[leftmargin=*]

            \item If there is no tuple such that $\cecpaymentinfo' \allowbreak = \allowbreak \cecpaymentinfo$, set $c \allowbreak \gets \allowbreak 0$.

            \item If there are $K \allowbreak > \allowbreak 1$ tuples such that $\cecpaymentinfo' \allowbreak = \allowbreak \cecpaymentinfo$, set $c \allowbreak \gets \allowbreak (\fcecProvider_1, \allowbreak \ldots, \allowbreak \fcecProvider_K)$, where $(\fcecProvider_1, \allowbreak \ldots, \allowbreak \fcecProvider_K)$ are providers that deposited payments with $\cecpaymentinfo$ more than once, or that deposited a payment such that $\fcecProvider_k$ is not included in $\cecpaymentinfo$.

            \item If there is one tuple such that $\cecpaymentinfo' \allowbreak = \allowbreak \cecpaymentinfo$, output $c \allowbreak \gets \allowbreak \fcecProvider_k$ if the identity of the provider is not in $\cecpaymentinfo$. Else, for all the remaining tuples $(\sid, \allowbreak \fcecProvider'_k, \cecpayment', \allowbreak \cecpaymentinfo')$, run the algorithm $c \allowbreak \gets \allowbreak \cecIdentify(\cecparams, \allowbreak \cecPK, \allowbreak \cecpayment, \allowbreak \cecpayment', \allowbreak \cecpaymentinfo, \allowbreak \cecpaymentinfo')$ until $c \allowbreak \neq \allowbreak 1$. If $c \allowbreak = \allowbreak 1$ for all tuples, set $c \allowbreak \gets \allowbreak 1$.


        \end{itemize}

    \item Output $(\fcecdepvfend, \allowbreak \sid, \allowbreak \cecpaymentinfo, \allowbreak c)$.

    \end{itemize}
    \end{enumerate}
    \end{multicols}
        }
   \end{framed}
\vspace*{-3mm}
\caption{Construction $\Pi_{\CEC}$}
\label{fig:constructionEC}
\end{figure*}

In Figure~\ref{fig:constructionEC}, we describe our construction $\mathrm{\Pi}_{\CEC}$ for $\Functionality_{\CEC}$. 
$\mathrm{\Pi}_{\CEC}$ uses
the ideal functionalities $\Functionality_{\SMT}$ for secure message transmission, $\Functionality_{\NYM}$ for a pseudonymous channel, $\Functionality_{\KG}$ for key generation, $\Functionality_{\Freg}$ for registration and $\Functionality_{\BB}$ for an authenticated bulletin board, which are described in~\ref{sec:functionalitiesFULL}. $\Functionality_{\SMT}$ is used for the communication channel between users and authorities in the $\fcecrequest$ and $\fcecissue$ interfaces, while $\Functionality_{\NYM}$ is used for the communication channel between users and providers in the $\fcecspend$ interface. $\Functionality_{\KG}$ runs algorithms $\cecSetup$ and $\cecKeyGenA$, and is used in the $\fcecsetup$ and $\fcecregister$ interfaces to generate and distribute both the parameters of the scheme and the keys of the authorities. $\Functionality_{\Freg}$ is used in the $\fcecregister$ interface to register the user public keys, and in the $\fcecissue$ and $\fcecdepvf$ interfaces to give those keys to authorities. $\Functionality_{\BB}$ is used in the $\fcecdeposit$ interface to deposit payments, and in the $\fcecdepvf$ interface to let authorities retrieve the deposited payments. We remark that $\mathrm{\Pi}_{\CEC}$ also uses a functionality for random oracle $\Functionality_{\RO}$ as in~\cite{cryptoeprint:2022:011} to model the random oracle queries done in the algorithms used as a building block. However, this is omitted in the description of $\mathrm{\Pi}_{\CEC}$.

Additionally, $\mathrm{\Pi}_{\CEC}$ uses the algorithms defined below. We define these algorithms to simplify the description of $\mathrm{\Pi}_{\CEC}$. In \S\ref{sec:instantiation}, we instantiate them for both our compact and divisible e-cash schemes.
\begin{description}[leftmargin=10pt]

\item[$\cecSetup(1^{\securityparameter}, L)$.] It computes the system parameters $\cecparams$ on input the security parameter $1^{\securityparameter}$ and 
the number of coins $L$ in a full wallet. These parameters are publicly available.

\item[$\cecKeyGenA(\cecparams, t, n)$.] Given $\cecparams$, the threshold $t$, and the number of authorities $n$, output the public verification key $\spk$ and the key pairs $(\ssk_{\fcecAuthority_i}, \allowbreak \spk_{\fcecAuthority_i})_{i\in[1,n]}$ for each of the authorities.

\item[$\cecKeyGenU(\cecparams)$.] It is run by each user $\fcecUser_j$ to generate 
a secret key and a public key $(\ssk_{\fcecUser_j}, \allowbreak \spk_{\fcecUser_j})$.

\item[$\cecRequest(\cecparams, \ssk_{\fcecUser_j})$.] On input $\cecparams$ and the secret key $\ssk_{\fcecUser_j}$ of the user $\fcecUser_j$, output a request $\cecrequest$ and request information $\cecrequestinfo$.

\item[$\cecRequestVf(\cecparams, \cecrequest, \spk_{\fcecUser_j})$.] Given $\cecparams$, a request $\cecrequest$ and the public key $\spk_{\fcecUser_j}$ of the user $\fcecUser_j$, output $1$ if the request is valid and $0$ otherwise.

\item[$\cecWithdraw(\cecparams, \ssk_{\fcecAuthority_i}, \cecrequest)$.] On input $\cecparams$, the secret key of authority $\fcecAuthority_i$ and a request $\cecrequest$, output a response $\cecresponse$.

\item[$\cecWithdrawVf(\cecparams, \spk_{\fcecAuthority_i}, \ssk_{\fcecUser_j}, \cecresponse, \cecrequestinfo)$.] Given $\cecparams$, the public key $\spk_{\fcecAuthority_i}$ of authority $\fcecAuthority_i$, the secret key $\ssk_{\fcecUser_j}$ of user $\fcecUser_j$, a response $\cecresponse$ and request information $\cecrequestinfo$, output a partial wallet $\cecwallet_i$ if the response $\cecresponse$ is correct and is associated to a request with request information $\cecrequestinfo$, else output $0$.

\item[$\cecCreateWallet(\spk, \ssk_{\fcecUser_j}, \cecset, \langle \cecwallet_i \rangle_{i \in \cecset})$.] 
Given the public key $\spk$, the secret key $\ssk_{\fcecUser_j}$, a set of indices $\cecset \in [1,n]$ and partial wallets $\langle \cecwallet_i \rangle_{i \in \cecset}$, output a wallet $\cecwallet$ if $|\cecset| \geq t$, else output $0$.

\item[$\cecSpend(\spk, \ssk_{\fcecUser_j}, \cecwallet, \cecpaymentinfo,  V)$.] Given the public key $\spk$, the secret key $\ssk_{\fcecUser_j}$, a wallet $\cecwallet$, payment information $\cecpaymentinfo$, and a number of coins $V$, outputs an updated wallet $\cecwallet'$ and a payment $\cecpayment$ if there are $V$ non-spent coins in $\cecwallet$ or $0$ otherwise.

\item[$\cecSpendVf(\spk, \cecpayment, \cecpaymentinfo)$.] Given the public key $\spk$, a payment $\cecpayment$, and payment information $\cecpaymentinfo$, 
output the number $V$ of coins received 
if the payment is correct, or $0$ otherwise.

\item[$\cecIdentify(\cecparams, \cecPK, \cecpayment_1, \cecpayment_2, \cecpaymentinfo_1, \cecpaymentinfo_2)$.] Given $\cecparams$, a list of user public keys $\cecPK$, and two payments $\cecpayment_1$ and $\cecpayment_2$ with respective payment information $\cecpaymentinfo_1$ and $\cecpaymentinfo_2$:
\begin{itemize}
    \item Output $1$ if $\cecpayment_1$ and $\cecpayment_2$ are payments where different coins were used.
    \item Else, output $\cecpaymentinfo_1$ if $\cecpaymentinfo_1 \allowbreak = \allowbreak \cecpaymentinfo_2$, which indicates that the payment has been double deposited.
    \item Else, output the public key $\spk_{\fcecUser_j} \allowbreak \in \allowbreak \cecPK$ of the user $\fcecUser_j$ that double spent a coin in payments $\cecpayment_1$ and $\cecpayment_2$.
    \item Else, output $\bot$.
\end{itemize}
\end{description}

\paragraph{Remark about $\Pi_{\CEC}$.} In the $\fcecregister$ interface, providers generate their own key pair, although it is not used later in the protocol. This is done to simplify the description of the protocol by making users and providers call the same $\fcecregister$ interface. Nevertheless, when the protocol is instantiated by replacing the ideal functionalities used as building blocks with concrete protocols that realize them, providers will need to generate their own keys.

%% file: 5InstantiationOfThresholdOfflineEcash.tex
\section{Instantiation of $\mathrm{\Pi}_{\CEC}$}\label{sec:instantiation} 

\subsection{Threshold Issuance Compact Ecash}\label{sec:compactecash}

Our compact $\CEC$ scheme is based on the scheme proposed in~\cite{DBLP:conf/eurocrypt/CamenischHL05}.
In order to provide threshold issuance, we use the Coconut protocol~\cite{DBLP:conf/ndss/SonninoABMD19} with the modifications in~\cite{cryptoeprint:2022:011}. We also make some changes in the scheme in~\cite{DBLP:conf/eurocrypt/CamenischHL05} to improve efficiency (see~\ref{sec:exts}).

\subsubsection{High-level Overview}


In~\cite{DBLP:conf/eurocrypt/CamenischHL05}, a central bank plays the role of the authority. In the setup phase, the bank generates a key pair for a signature scheme, and each of the users generates a key pair.

\begin{description}[leftmargin=0pt]

\item[Withdrawal Phase.] A wallet of $L$ coins is a signature under the bank's public key on a user secret key $\ssk_{\fcecUser_j}$ and two random values $\cecsn$ and $\cect$. The user $\fcecUser_j$ obtains the signature from the bank on $(\ssk_{\fcecUser_j}, \allowbreak \cecsn, \allowbreak \cect)$ through a blind signature protocol. The bank does not learn any of the signed values, but learns the user public key $\spk_{\fcecUser_j}$ associated with $\ssk_{\fcecUser_j}$.

\item[Spending Phase.] In order to spend coin $l \in [0,L-1]$, $\fcecUser_j$ proves in zero-knowledge (ZK) possession of a signature on $(\ssk_{\fcecUser_j}, \allowbreak \cecsn, \allowbreak \cect)$.  Additionally, $\fcecUser_j$ generates a serial number $S$ and a double-spending tag $T$, which are used to detect and identify double-spenders. $S$ and $T$ are computed by evaluating the pseudorandom function (PF) in \S\ref{subsec:pseudorandomfunction} on input $l$. Concretely, $S$ is the output of the PF $f_{\ga,\cecsn}$ on input $l$.
$T$ is computed on input $\ssk_{\fcecUser_j}$, the output of the PF $f_{\ga,\cect}$ on input $l$, and $R \allowbreak \gets \allowbreak H(\cecpaymentinfo)$. $\cecpaymentinfo$ is given by the provider and should be unique for each payment.
$\fcecUser_j$ also proves in ZK that $S$ and $T$ are correctly computed. 

\item[Deposit phase.] The provider sends to the bank the payment received from the user. To check whether the coin has been double-spent, the bank compares the serial number $S$ with the serial numbers of previously received coins. If there is a match, but the payment information $\cecpaymentinfo$ is the same in both coins, then the bank finds that the provider has deposited the coin twice. Else, the bank identifies the user that double-spent the coin by using the double-spending tags of both coins.

\end{description}


\subsubsection{Our extensions}\label{sec:exts} 
In our compact $\CEC$ scheme, the signature scheme is instantiated with PS signatures, which are described in \S\ref{subsec:signatureSchemes}. In the setup phase, the secret keys $\ssk_{\fcecAuthority_i}$ for each of the authorities $(\fcecAuthority_1, \allowbreak \ldots, \allowbreak \fcecAuthority_n)$ are generated by evaluating random polynomials of degree $t-1$ on input $[1,n]$, while the verification key $\spk$ used to verify wallets corresponds to a secret key that would be given by the evaluation of those polynomials on input $0$. In the withdrawal phase, the user runs a blind signature protocol with at least $t$ authorities. After obtaining at least $t$ valid signatures, the user uses Lagrange interpolation to obtain a signature verifiable with $\spk$.

A wallet is a signature on $(\ssk_{\fcecUser_j}, \allowbreak \cecsn)$. In comparison to~\cite{DBLP:conf/eurocrypt/CamenischHL05}, we remove the secret $\cect$. Thanks to this change, the size of the wallet is smaller, and the ZK proofs used in both the withdrawal and spending phase are more efficient in comparison to~\cite{DBLP:conf/eurocrypt/CamenischHL05}. To make this change possible, we modify the way the serial number $S$ and double spending tag $T$ are computed. Concretely, $S$ is the output of the PF $f_{\delta,\cecsn}$ on input $l$, and the computation of $T$ uses the evaluation of the PF $f_{\ga,\cecsn}$ on input $l$, i.e., we use a new generator $\delta$ for the computation of the serial numbers. This change allows us to use the same secret $\cecsn$ as the index of both PFs without compromising the security of our scheme. In \S\ref{sec:compactECashRangeProof}, we quantify the cost reduction attained by removing $\cect$.
We further improve the efficiency of the withdrawal phase, by removing the need for the bank to contribute randomness to create $\cecsn$ in the blind signature protocol in~\cite{DBLP:conf/eurocrypt/CamenischHL05}. In our protocol, the user picks $\cecsn$ and as discussed in \S\ref{sec:securityCompact}, this change does not compromise the security of our scheme. We also improve efficiency by using one ZK proof $\pi_v$ to spend $V$ coins, instead of repeating $V$ times the spending protocol for one coin. 


\subsubsection{Construction}\label{sec:cec:construction} 
 The algorithms of our compact $\CEC$ scheme are defined below. In~\S\ref{sec:securitydefinitionsbuildingblocks}, we describe the cryptographic primitives used by the algorithms.
\begin{description}[leftmargin=10pt,topsep=0pt]

\item[$\cecSetup(1^{\securityparameter}, L)$.] Execute the following steps:

\begin{itemize}[leftmargin=*]

    \item Run $(\p,\Ga,\Gb,\allowbreak \Gt,\e,\ga,\gb) \gets \BilinearSetup(1^\securityparameter)$.

    \item Pick $3$ random generators $(\gamma_1, \gamma_2, \delta) \gets \Ga$.

    \item Output $\cecparams \allowbreak \gets \allowbreak (\p, \allowbreak \Ga, \allowbreak \Gb, \allowbreak \Gt, \allowbreak \e, \allowbreak \ga, \allowbreak \gb, \allowbreak \gamma_1, \allowbreak \gamma_2, \allowbreak \delta, \allowbreak L)$.

\end{itemize}

\item[$\cecKeyGenA(\cecparams, t, n)$.] Execute the following steps:
\begin{itemize}[leftmargin=*]

    \item Choose $(1 + 2)$ polynomials $(v, w_1, w_2)$ of degree $(t - 1)$ with random coefficients in $\Zp$.

    \item Set $(x, y_1, y_2) \gets (v(0), w_1(0), w_2(0))$.

    \item For $i = 1$ to $n$, set the secret key $\ssk_{\fcecAuthority_i}$ of each authority $\fcecAuthority_i$ as $\ssk_{\fcecAuthority_i} \allowbreak = \allowbreak (x_i, \allowbreak y_{i,1}, \allowbreak y_{i,2}) \allowbreak \gets \allowbreak (v(i),\allowbreak w_1(i), \allowbreak w_2(i))$.

    \item For $i = 1$ to $n$, set the verification key $\spk_{\fcecAuthority_i}$ of each authority $\fcecAuthority_i$ as $\spk_{\fcecAuthority_i} \allowbreak = \allowbreak (\tilde{\alpha}_i, \allowbreak \beta_{i,1}, \allowbreak \tilde{\beta}_{i,1},  \allowbreak \beta_{i,2}, \allowbreak \tilde{\beta}_{i,2}) \allowbreak \gets \allowbreak (\gb^{x_i}, \allowbreak \ga^{y_{i,1}}, \allowbreak \gb^{y_{i,1}},  \allowbreak \ga^{y_{i,2}}, \allowbreak \gb^{y_{i,2}})$.

    \item Set the verification key $\spk = (\cecparams, \allowbreak \tilde{\alpha}, \allowbreak \beta_{1}, \allowbreak \tilde{\beta}_{1}, \allowbreak \beta_{2}, \tilde{\beta}_{2}) \gets \allowbreak (\cecparams, \allowbreak \gb^{x}, \allowbreak \ga^{y_{1}}, \allowbreak \gb^{y_{1}}, \allowbreak \ga^{y_{2}}, \allowbreak \gb^{y_{2}})$.

    \item Output $(\spk, \langle \spk_{\fcecAuthority_i}, \ssk_{\fcecAuthority_i} \rangle_{i =1}^{n})$.

\end{itemize}

\item[$\cecKeyGenU(\cecparams)$.] Execute the following steps:
\begin{itemize}[leftmargin=*]

    \item Pick random $\ssk_{\fcecUser_j} \gets \Zp$ and compute $\spk_{\fcecUser_j} \gets \ga^{\ssk_{\fcecUser_j}}$.

    \item Output $(\ssk_{\fcecUser_j}, \allowbreak \spk_{\fcecUser_j})$.

\end{itemize}

\item[$\cecRequest(\cecparams, \ssk_{\fcecUser_j})$.] Execute the following steps:

\begin{itemize}[leftmargin=*]

    \item Pick random $\cecsn \gets \Zp$ and set $(\cecmes_{1}, \allowbreak \cecmes_{2}) \allowbreak = (\ssk_{\fcecUser_j}, \allowbreak \cecsn)$.

    \item Pick random $o \gets \Zp$ and compute  $\com = \ga^{o} \prod_{j=1}^{2} \gamma_j^{\cecmes_{j}}$.

    \item Compute $\h \gets H(\com)$, where $H$ is a hash function modeled as a random oracle.

    \item Compute commitments to each of the messages. For $j = 1$ to $2$, pick random $o_{j} \gets \Zp$ and set $\com_{j} = \ga^{o_{j}} \h^{\cecmes_{j}}$.

    \item Compute a ZK argument of knowledge $\pi_s$ via the Fiat-Shamir heuristic for the following relation:
    \vspace{-3mm}
    {\small
    \begin{align*}
        \pi_s =  & \NIZK\{(\cecmes_{1}, \cecmes_{2}, o, o_{1}, o_{2}):
        \com = \ga^{o} \prod_{j=1}^{2} \gamma_j^{\cecmes_{j}}\ \land\ \\ & \spk_{\fcecUser_j} \gets \ga^{\cecmes_{1}} \land\
        \{\com_{j}= \ga^{o_{j}} \h^{\cecmes_{j}} \}_{\forall j \in [1,2]}  \}
    \end{align*}
    }
    \item Set $\cecrequestinfo \allowbreak \gets \allowbreak (\h, \allowbreak o_{1}, \allowbreak o_{2}, \allowbreak \cecsn)$ and $\cecrequest \allowbreak \gets \allowbreak (\h, \allowbreak \com, \allowbreak \com_{1}, \allowbreak \com_{2}, \allowbreak \pi_s)$.

    \item Output $\cecrequest$ and $\cecrequestinfo$.

\end{itemize}

\item[$\cecRequestVf(\cecparams, \cecrequest, \spk_{\fcecUser_j})$.] Execute the following steps:
\begin{itemize}[leftmargin=*]

    \item Parse $\cecrequest$ as $(\h, \allowbreak \com, \allowbreak \com_{1}, \allowbreak \com_{2}, \allowbreak \pi_s)$.

    \item Compute $\h' \gets H(\com)$ and output $0$ if $\h \neq \h'$.

    \item Verify the ZK argument $\pi_s$ by using the tuple $(\cecparams, \allowbreak \h, \allowbreak \com, \allowbreak \com_{1}, \allowbreak \com_{2}, \allowbreak \spk_{\fcecUser_j})$. Output $0$ if the proof $\pi_s$ is not correct, else output $1$.

\end{itemize}

\item[$\cecWithdraw(\cecparams, \ssk_{\fcecAuthority_i}, \cecrequest)$.] Execute the following steps:
\begin{itemize}[leftmargin=*]

    \item Parse $\cecrequest$ as $(\h, \allowbreak \com, \allowbreak \com_{1}, \allowbreak \com_{2}, \allowbreak \pi_s)$ and $\ssk_{\fcecAuthority_i}$ as $(x_i, \allowbreak y_{i,1}, \allowbreak y_{i,2})$.

    \item Compute $c = \h^{x_i} \prod_{j=1}^{2} \com_j^{y_{i,j}}$.

    \item Set the blinded signature share $\hat{\sigma}_{i} \allowbreak \gets \allowbreak (\h, \allowbreak c)$.

    \item Output $\cecresponse \gets \hat{\sigma}_{i}$.

\end{itemize}

\item[$\cecWithdrawVf(\cecparams, \spk_{\fcecAuthority_i}, \ssk_{\fcecUser_j}, \cecresponse, \cecrequestinfo)$.] Do the following:

\begin{itemize}[leftmargin=*]

    \item Parse $\cecrequestinfo$ as $(\h', \allowbreak o_{1},  \allowbreak o_{2}, \allowbreak \cecsn)$, $\cecresponse$ as $\hat{\sigma}_{i} \allowbreak = \allowbreak (\h, \allowbreak c)$, and $\spk_{\fcecAuthority_i}$ as $(\tilde{\alpha}_i, \allowbreak \beta_{i,1}, \allowbreak \tilde{\beta}_{i,1}, \allowbreak \beta_{i,2}, \allowbreak \tilde{\beta}_{i,2})$. Output $0$ if $\h \allowbreak \neq \allowbreak \h'$.

    \item Compute $\sigma_{i} \allowbreak = \allowbreak (\h, \allowbreak s) \allowbreak \gets \allowbreak (\h, \allowbreak c \prod_{j=1}^{2} \beta_{i,j}^{-o_{j}})$.

    \item Set $(\cecmes_{1}, \cecmes_{2}) \gets (\ssk_{\fcecUser_j}, \cecsn)$. Output $0$ if $\e(\h, \tilde{\alpha}_i \prod_{j=1}^{2} \tilde{\beta}_{i,j}^{\cecmes_{j}}) \allowbreak = \allowbreak \e(s, \gb)$ does not hold.

    \item Output $\cecwallet_i \gets (i,\sigma_{i},\cecsn)$.

\end{itemize}

\item[$\cecCreateWallet(\spk, \ssk_{\fcecUser_j}, \cecset, \langle \cecwallet_i \rangle_{i \in \cecset})$.] Execute the following steps:
\begin{itemize}[leftmargin=*]

    \item If $|\cecset| \neq t$, output $0$.

    \item For all $i \in \cecset$, evaluate at 0 the Lagrange basis polynomials $l_i = [\prod_{j \in \cecset,j\neq i} (0-j)] [\prod_{j\in\cecset,j\neq i} (i-j)]^{-1}\ \mathrm{mod}\ \p $

    \item For all $i \in \cecset$, parse $\cecwallet_{i}$ as  $(i,\sigma_{i},\cecsn)$ and $\sigma_{i}$ as $(\h, \allowbreak s_i)$.

    \item Compute the signature $\sigma = (\h, s) \gets (\h, \allowbreak \prod_{i \in \cecset} s_{i}^{l_i})$.

    \item Parse $\spk$ as $(\cecparams, \allowbreak \tilde{\alpha}, \allowbreak \beta_{1}, \allowbreak \tilde{\beta}_{1}, \beta_{2}, \tilde{\beta}_{2})$.

    \item Set $(\cecmes_{1}, \allowbreak \cecmes_{2}) \allowbreak = \allowbreak (\ssk_{\fcecUser_j}, \allowbreak \cecsn)$ and output $0$ if $\e(\h, \tilde{\alpha} \prod_{j=1}^{2} \tilde{\beta}_j^{\cecmes_{j}}) = \e(s, \gb)$ does not hold, else output $\cecwallet \allowbreak \gets \allowbreak (\sigma, \allowbreak \cecsn, l)$, where $l$ is a counter from $0$ to $L-1$ initialized to $0$.

\end{itemize}

\item[$\cecSpend(\spk, \ssk_{\fcecUser_j}, \cecwallet, \cecpaymentinfo, V)$.] Execute the following steps:

\begin{itemize}[leftmargin=*]

    \item Parse $\cecwallet$ as $(\sigma, \allowbreak \cecsn, \allowbreak l)$. If $l + V -1 \geq L$, output $0$.

    \item Parse $\sigma$ as $(\h, \allowbreak s)$ and $\spk$ as $(\cecparams, \allowbreak \tilde{\alpha}, \allowbreak \beta_{1}, \allowbreak \tilde{\beta}_{1}, \beta_{2}, \tilde{\beta}_{2})$.

    \item Pick random $r \gets \Zp$ and $r' \gets \Zp$.

    \item Compute $\sigma' = (\h',s') \gets (\h^{r'}, s^{r'}(\h')^{r})$ and $\kappa \gets \tilde{\alpha} \tilde{\beta}_1^{\ssk_{\fcecUser_j}} \tilde{\beta}_2^{\cecsn} \gb^{r}$.

    \item Pick random $o_c \gets \Zp$ and compute the commitment $C \gets \ga^{o_c} \gamma_1^{\cecsn}$.

    \item For $k \in [0,V-1]$, compute $R_k \gets H'(\cecpaymentinfo,k)$, where $\cecpaymentinfo$ must contain the identifier of the merchant, and $H'$ is a collision-resistant hash function.

    \item For $k \in [0,V-1]$, set $l_k \gets l+k$, pick random $o_{a_k}$ and compute $A_k = \ga^{o_{a_k}} \gamma_1^{l_k}$.

    \item For $k \in [0,V-1]$, compute the serial numbers $S_{k} \gets f_{\delta,\cecsn}(l_k) = \delta^{1/(\cecsn+l_k+1)}$ and also compute the double spending tags $T_k \gets \ga^{\ssk_{\fcecUser_j}} (f_{\ga,\cecsn}(l_k))^{R_{k}} = \ga^{\ssk_{\fcecUser_j} + R_{k}/(\cecsn+l_k+1)}$.

    \item For $k \in [0,V-1]$, compute the values $\mu_k \gets 1/(\cecsn+l_k+1)$ and $o_{\mu_k} \gets -(o_{a_k} + o_c) \mu_k$.

    \item Compute a ZK argument of knowledge $\pi_v$ via the Fiat-Shamir heuristic for the following relation:
    \vspace{-3mm}
    {\small
    \begin{align*}
        \pi_v =  & \NIZK\{(\ssk_{\fcecUser_j}, \cecsn, r, o_c, \langle l_k, o_{a_k}, \mu_k, o_{\mu_k} \rangle_{k=0}^{V-1}): \\
        &\kappa = \tilde{\alpha} \tilde{\beta}_1^{\ssk_{\fcecUser_j}} \tilde{\beta}_2^{\cecsn} \gb^{r}\ \land\ C = \ga^{o_c} \gamma_1^{\cecsn}\ \land\  \\&
        \langle A_k = \ga^{o_{a_k}} \gamma_1^{l_k}\ \land\ l_k \in [0,L-1]\ \land\ \\&
        S_k=\delta^{\mu_k}\ \land\ \gamma_1 = (A_k C\gamma_1)^{\mu_k} \ga^{o_{\mu_k}}\ \land\ \\&
        T_k=\ga^{\ssk_{\fcecUser_j}} (\ga^{R_{k}})^{\mu_k}\  \rangle_{k\in[0,V-1]}
        \}
    \end{align*}
    }
    In \S\ref{sec:compactECashRangeProof}, we explain this ZK proof and show how to prove the statement $l_k \in [0,L-1]$.

    \item Output a payment $\cecpayment \gets (\kappa, \sigma', \langle S_k, T_k, A_k \rangle_{k\in[0,V-1]}, V, C, \allowbreak \pi_v)$ and an updated wallet $\cecwallet' \gets (\sigma, \allowbreak \cecsn, \allowbreak l+V)$.

\end{itemize}

\item[$\cecSpendVf(\spk, \cecpayment, \cecpaymentinfo)$.] Execute the following steps:

\begin{itemize}[leftmargin=*]

    \item Parse $\spk$ as $(\cecparams, \allowbreak \tilde{\alpha}, \allowbreak \beta_{1}, \allowbreak \tilde{\beta}_{1}, \beta_{2}, \tilde{\beta}_{2})$.

    \item Parse $\cecpayment$ as $(\kappa, \sigma', \langle S_k, T_k, A_k \rangle_{k\in[0,V-1]}, V, C, \pi_v)$.

    \item Parse $\sigma'$ as $(\h', \allowbreak s')$ and output $0$ if $\h' = 1$ or if $\e(\h', \kappa) \allowbreak = \allowbreak \e(s',\gb)$ does not hold.

    \item Output $0$ if not all the serial numbers $\langle S_k \rangle_{k\in[0,V-1]}$ are different from each other.

    \item For $k \in [0,V-1]$, compute $R_k \gets H'(\cecpaymentinfo,k)$.

    \item Verify $\pi_v$ by using $\cecpaymentinfo$, $\spk$, $\langle S_k, T_k, A_k, R_{k} \rangle_{k\in[0,V-1]}$, $C$ and $\kappa$. Output $0$ if the proof is not correct, else output $V$.

\end{itemize}

\item[$\cecIdentify(\cecparams, \cecPK, \cecpayment_1, \cecpayment_2, \cecpaymentinfo_1, \cecpaymentinfo_2)$.] Do:
\begin{itemize}[leftmargin=*]

    \item Parse $\cecpayment_1$ as $(\kappa_1, \sigma'_1, \langle S_{k,1}, T_{k,1}, A_{k,1} \rangle_{k\in[0,V_1-1]}, V_1, C_1, \pi_{v,1})$ and $\cecpayment_2$ as $(\kappa_2, \sigma'_2, \langle S_{k,2}, T_{k,2}, A_{k,2} \rangle_{k\in[0,V_2-1]}, V_2, C_2, \pi_{v,2})$.

    \item For $k\in[0,V_1-1]$, for $j\in[0,V_2-1]$, check whether $S_{k,1} \allowbreak = \allowbreak S_{j,2}$. If the equality never holds, output $1$.

    \item Else, output $\cecpaymentinfo_1$ if $\cecpaymentinfo_1 = \cecpaymentinfo_2$.

    \item Else, for $k\in[0,V_1-1]$ and $j\in[0,V_2-1]$ such that $S_{k,1} = S_{j,2}$, compute
    $
    \spk_{\fcecUser_j} \gets (T_{j,2}^{R_{k,1}}/T_{k,1}^{R_{j,2}})^{(R_{k,1}-R_{j,2})^{-1}}
    $
    If $\spk_{\fcecUser_j} \in \cecPK$ output $\spk_{\fcecUser_j}$, else output $\bot$.

\end{itemize}

\end{description}

\subsubsection{Security Analysis of Compact E-Cash}
\label{sec:securityCompact}
In \S\ref{sec:securityProofCompact}, we prove formally that $\mathrm{\Pi}_{\CEC}$, when instantiated with the algorithms of our compact $\CEC$ scheme, realizes $\Functionality_{\CEC}$. In this section, we give intuition on why our scheme is secure. 
\begin{description}[leftmargin=0pt, topsep=2pt]
\item[Unlinkability.] In the withdrawal phase, a corrupt authority does not learn the user secrets $(\ssk_{\fcecUser_j}, \allowbreak \cecsn)$ thanks to the hiding property of the Pedersen commitment scheme and to the ZK property of the argument $\pi_s$. In the spend phase, a corrupt provider does not learn anything from a payment beyond the number of coins spent. To prove that, several properties are used. First, we use the ZK property of the argument $\pi_v$. Second, to prove that $C$ and $A_k$ do not reveal any information about $\cecsn$ or $l_k$, we use the hiding property of the Pedersen commitment scheme. Third, to prove that $S_k$ and $T_k$ do not reveal any information about $\spk_{\fcecUser_j}$, $\cecsn$ or $l_k$, we use the pseudorandomness property of the PF, along with the XDH assumption.\footnote{In contrast to~\cite{DBLP:conf/eurocrypt/CamenischHL05}, the XDH assumption is needed in our scheme because we use the same coin secret $\cecsn$ to compute $S_k$ and $T_k$.} Finally, as in the modified version of Coconut in~\cite{cryptoeprint:2022:011}, the PS signature is ``randomized'' in a way that enables us to prove signature possession without revealing any information about the original signature or the signed messages.

\item[Traceability, Exculpability and Clearance.] To prove that a user cannot spend coins that she has not withdrawn before, we use several properties of our building blocks. The weak simulation extractability property allows us to extract the witnesses from the ZK arguments $\pi_s$. Thanks to that extraction, we can use the binding property of the commitment scheme $\com$ included in a request message to ensure that different commitments $\com$ and $\com'$ commit to different tuples $(\ssk_{\fcecUser_j}, \allowbreak \cecsn)$. This is required for the unforgeability of PS signatures in the RO model. We remark that $\com$ is the input to the random oracle and that it is necessary to ensure that a different generator $\h$ is created to sign different message tuples. The binding property of $\com$ guarantees that. Second, we use the weak simulation extractability property of arguments $\pi_v$ to extract the witnesses. Thanks to that, in the spending phase, we can extract a signature on $(\ssk_{\fcecUser_j}, \allowbreak \cecsn)$ from a payment message and show that, if a user did not withdraw at least $t$ signatures from $t$ different authorities, then the user can be used to break the existential unforgeability property of PS signatures in the RO model.

Therefore, we know that, if more coins are deposited than those being withdrawn, it is the case that a user has double-spent coins or that a provider has double-deposited coins. In the deposit phase, an authority checks that payments that are deposited have different serial numbers. We show that the extractability of $\pi_v$, along with the discrete logarithm assumption, guarantees that the serial numbers and double spending tags are correctly computed. Therefore, if two payments have at least one common serial number, there are three possibilities:
\begin{enumerate*}
    \item Double depositing, 
    \item double spending, 
    \item none of the former.
\end{enumerate*}
Double depositing can be punished by checking whether two payments are associated with the same payment information $\cecpaymentinfo$, which contains the identifier of the provider. We recall that $\cecpaymentinfo$ is signed in $\pi_v$. Moreover, our construction $\mathrm{\Pi}_{\CEC}$ uses an authenticated bulletin board. Thus, an authority can check that the provider that deposits a payment is the same whose identity is in $\cecpaymentinfo$. The latter guarantees the clearance property. If double depositing did not happen because the payment information $\cecpaymentinfo$ and $\cecpaymentinfo'$ is different in those payments, the authority can retrieve the public key of the user who double spent a coin through the computation described in the algorithm $\cecIdentify$. This computation requires that  $R_k \gets H'(\cecpaymentinfo,k)$ is different from  $R'_{k'} \gets H'(\cecpaymentinfo',k')$. We show that, if the hash function $H'$ is collision-resistant, the double spender can always be identified. In our scheme, a corrupt user is able to compute two payments where there is no double spending, yet two serial numbers are equal. The reason is that, unlike in the scheme in~\cite{DBLP:conf/eurocrypt/CamenischHL05}, the user picks the coin secret $\cecsn$ on its own. When a corrupt user does that, our security analysis guarantees that, under the hardness of the discrete logarithm assumption, algorithm $\cecIdentify$ will never identify an honest user as the double spender. This guarantees the exculpability property. Moreover, we also show that, under the discrete logarithm assumption, a corrupt user cannot compute a payment with a serial number that is equal to a serial number in a payment computed by an honest user. Consequently, when $\cecIdentify$ detects that two serial numbers are equal, but is unable to find the public key of the user who double spent (i.e. $\cecIdentify$ outputs $\bot$), we are in a case in which in fact there is no double-spending.

\end{description}

\subsection{Threshold Issuance Divisible E-Cash}\label{sec:divisibleecash}

In our compact $\CEC$ scheme in~\S\ref{sec:compactecash}, the cost of the spending phase grows linearly with the number of coins spent. In divisible e-cash, the cost of the spending phase is independent of the number of coins spent. To make that possible, the main change in comparison to compact e-cash is that the serial numbers of coins are generated during the deposit phase, rather than the spending phase.

Our divisible $\CEC$ scheme is based on the work by Pointcheval et al.~\cite{DBLP:conf/pkc/PointchevalST17}, which 
proposes a scheme in the standard model with  Groth-Sahai proofs~\cite{DBLP:conf/eurocrypt/GrothS08}. We modify that scheme as follows. In~\cite{DBLP:conf/pkc/PointchevalST17}, a wallet is a signature on two group elements $(U_1,U_2) = (u_1^{\ssk_{\fcecUser_j}}, \allowbreak u_2^\cecsn)$ We replace the signature scheme used in~\cite{DBLP:conf/pkc/PointchevalST17} by the PS signature scheme, and we sign $(\ssk_{\fcecUser_j}, \allowbreak \cecsn)$. Thus, the wallet in our divisible $\CEC$ scheme has the same form as in our compact $\CEC$ scheme. Thanks to that, in the withdrawal phase we use the same algorithms $\cecRequest$, $\cecRequestVf$, $\cecWithdraw$, $\cecWithdrawVf$ and $\cecCreateWallet$ (see ~\S\ref{sec:cec:construction}) to provide a threshold issuance protocol. At setup, algorithms $\cecKeyGenA$ and $\cecKeyGenU$  also work as in~\S\ref{sec:cec:construction}.

In the spending phase in~\cite{DBLP:conf/pkc/PointchevalST17}, a Groth-Sahai non-interactive ZK proof and a Groth-Sahai non-interactive witness-indistinguishable proof are computed. The latter involves proof of possession of the signature on $(U_1, \allowbreak U_2)$. In our scheme, we use a NIZK argument computed via the Fiat-Shamir heuristic, which involves all the statements proven in both the Groth-Sahai ZK proof and witness-indistinguishable proof. To prove possession of a PS signature on $(\ssk_{\fcecUser_j}, \allowbreak \cecsn)$, we use the method depicted in algorithm $\cecSpend$ in~\S\ref{sec:cec:construction}.

Because Groth-Sahai proofs are randomizable, in~\cite{DBLP:conf/pkc/PointchevalST17}, the user needs to compute a one-time signature on the payment. A statement is added to the Groth-Sahai proof to certify the public key used for the one-time signature. In our scheme, this is not needed, because non-interactive ZK arguments computed via the Fiat-Shamir heuristic are signatures of knowledge.

The remaining values computed in the spending phase, and the statements proven about them, are the same in~\cite{DBLP:conf/pkc/PointchevalST17} and in our scheme. We note that our non-interactive ZK argument involves proving knowledge of group elements in addition to discrete logarithm representations. We show how this is done in~\S\ref{subsec:zkpk}.

\subsubsection{High-level Overview}

A wallet of $L$ coins is a signature on $(\ssk_{\fcecUser_j}, \allowbreak \cecsn)$, where $\ssk_{\fcecUser_j}$ is the secret key of user $\fcecUser_j$ and $\cecsn$ is a coin secret. The $L$ serial numbers of the coins in a wallet are given by
$
 \mathrm{SN}_l \allowbreak = \allowbreak \e(\varsigma,\gb)^{\cecsn y^l},\ l \in [1,L]
$
where values $y$ and $\varsigma$ are part of the parameters of the scheme.

\begin{description}[leftmargin=0pt]
    \item[Withdrawal Phase.] The withdrawal phase is the same as in our compact $\CEC$ scheme in~\S\ref{sec:cec:construction}.

    \item[Spending Phase.] In the spending phase, to spend $V$ coins, the user needs to give information that allows the authorities to compute the $V$ serial numbers of the spent coins, but no more than that. To this end, to spend $V$ coins with indices $[l,l+V-1]$, the user computes an ElGamal encryption $\phi_{V,l}$ of $\varsigma_l^\cecsn$ under the public key $\eta_V$.
    The values $\varsigma_l$ (for $l \allowbreak \in \allowbreak [1,L]$) and $\eta_V$ (for $V \allowbreak \in \allowbreak [1,L]$) are part of the public parameters of the scheme.
    $\phi_{V,l}$ is used in the deposit phase by the authorities to generate the serial numbers $(\mathrm{SN}_l,\ldots,\mathrm{SN}_{l+V-1})$. This ELGamal encryption with public key $\eta_V$ restricts the authorities to generate only those $V$ serial numbers.

    To enable identification of double spenders, the user computes the double spending tag $\varphi_{V,l}$ as an ElGamal encryption of  $(\ga^R)^{\ssk_{\fcecUser_j}} \theta_l^{\cecsn}$ under public key $\eta_{V}$, where $R$ is a hash of the payment information $\cecpaymentinfo$ given by the provider. 
    The values $\theta_l$, for $l \allowbreak \in \allowbreak [1,L]$,
    are part of the parameters of the scheme.

    The user also needs to prove in zero-knowledge that $\phi_{V,l}$ and $\varphi_{V,l}$ are correctly computed. To this end, the user proves possession of a signature on $(\ssk_{\fcecUser_j}, \allowbreak \cecsn)$ and proves that those values were used to compute $\phi_{V,l}$ and $\varphi_{V,l}$. Additionally, the user needs to prove that the correct values $\varsigma_l$ and $\theta_l$ in the public parameters have been used, and that $l \allowbreak \leq \allowbreak L-V+1$. To allow the user to prove those statements, the public parameters of the scheme contain signatures on the pairs $(\varsigma_l, \allowbreak \theta_l)$ (for $l \allowbreak \in \allowbreak [1,L]$). The user proves possession of the signature on the pair $(\varsigma_{l+V-1}, \allowbreak \theta_{l+V-1})$ and proves that $(\varsigma_l, \allowbreak \theta_l)$ are correctly chosen through the equations $\e(\varsigma_l, \tilde{\delta}_{V-1}) = \e(\varsigma_{l+V-1},\tilde{g})$ and $\e(\theta_l, \tilde{\delta}_{V-1}) = \e(\theta_{l+V-1},\tilde{g})$. The values  $\tilde{\delta}_k \gets \gb^{y^k}$ (for $k \in [0,L-1]$) are part of the parameters of the scheme. Although this proof does not prove that $l \geq 1$, in the security analysis it is shown that the user is unable to generate $(\varsigma_l, \allowbreak \theta_l)$ such that $l \allowbreak \notin \allowbreak [1,L]$.

    \item[Deposit Phase.] In the deposit phase, an authority checks whether a coin has been double spent. For this purpose, the authority computes the serial numbers of the spent coins by doing $\mathrm{SN}_{k} \allowbreak \gets \allowbreak \e(\phi_{V,l}[2], \tilde{\delta}_k) \allowbreak \e(\phi_{V,l}[1], \tilde{\eta}_{V,k})$ for $k \in [0,V-1]$.
    Here, the values $\tilde{\eta}_{l,k}$
    for $l \allowbreak \in \allowbreak [1,L]$ and $k \allowbreak \in \allowbreak [0, \allowbreak l-1]$
    are part of the parameters of the scheme. Because of those values, the size of the parameters is quadratic in the number $L$ of coins in a wallet, but this is only the case for the parameters that the authorities need, i.e., the size of parameters for users is linear in $L$.

    When a collision between serial numbers of two payments is detected, the authority uses the security tags of both payments for identification of the user who double spent. The mechanism used is similar to the one of our compact $\CEC$ scheme in~\S\ref{sec:compactecash}. However, in the divisible $\CEC$ scheme, the authority, rather than computing the user key, checks whether an equality holds for each of the public keys of the users one by one, which is a disadvantage.
\end{description}

\subsubsection{Construction}
Our divisible e-cash scheme works as follows:
\begin{description}[leftmargin=10pt]
\item[$\cecSetup(1^{\securityparameter}, L)$.] Execute the following steps:

\begin{itemize}[leftmargin=*]

    \item Run $\grp = (\p,\Ga,\Gb,\allowbreak \Gt,\e,\ga,\gb) \gets \BilinearSetup(1^\securityparameter)$.

    \item Pick random generators $\eta, \gamma_1, \gamma_2 \allowbreak \gets \allowbreak \Ga$.


    \item Generate random scalars $(z,y) \allowbreak \gets \allowbreak \Zp$ and compute $(\varsigma, \theta) \gets (\ga^z, \eta^z)$. Generate for $l \allowbreak \in \allowbreak [1,L]$, $a_l \gets \Zp$.

    \item For $l \allowbreak \in \allowbreak [1,L]$, compute $(\varsigma_l, \theta_l) \allowbreak \gets \allowbreak (\varsigma^{y^l}, \allowbreak \theta^{y^l})$.

    \item For $k \in [0,L-1]$, compute $\tilde{\delta}_k \gets \gb^{y^k}$.

    \item For $l \allowbreak \in \allowbreak [1,L]$, compute $\eta_l \gets \ga^{a_l}$.

    \item For $l \allowbreak \in \allowbreak [1,L]$, for $k \allowbreak \in \allowbreak [0, \allowbreak l-1]$, compute $\tilde{\eta}_{l,k} \allowbreak \gets \allowbreak \gb^{-a_l \cdot y^k}$.

    \item Run algorithm $(\spk_{sps}, \allowbreak \ssk_{sps}) \gets \SKeygen(\grp,2,0)$ of the SPS scheme in~\S\ref{subsec:signatureSchemes}.

    \item For $l \allowbreak \in \allowbreak [1,L]$, compute $\tau_l \gets \SSign(\ssk_{sps},\langle \varsigma_l, \theta_l \rangle)$.

    \item Set the parameters for users $\cecparams_u \gets (\p, \allowbreak \Ga, \allowbreak \Gb,\allowbreak \Gt, \allowbreak \e,\allowbreak \ga,\allowbreak \gb, \allowbreak \eta, \allowbreak \gamma_1, \allowbreak \gamma_2, \allowbreak \{\eta_l, \allowbreak \varsigma_l, \allowbreak \theta_l, \allowbreak \tau_l\}_{l=1}^L, \allowbreak \{\tilde{\delta}_k\}_{k=0}^{L-1}, \allowbreak \spk_{sps})$. Set the additional parameters for authorities $\cecparams_a \allowbreak \gets \allowbreak (\{\langle \tilde{\eta}_{l,k} \rangle_{k=0}^{l-1}\}_{l=1}^{L-1})$.

    \item Output $\cecparams \gets (\cecparams_u, \cecparams_a)$.

\end{itemize}

\item[$\cecSpend(\spk, \ssk_{\fcecUser_j}, \cecwallet, \cecpaymentinfo, V)$.] Execute the following steps:

\begin{itemize}[leftmargin=*]

    \item Parse $\cecwallet$ as $(\sigma, \allowbreak \cecsn, \allowbreak l)$ and $\sigma$ as $(\h, \allowbreak s)$. If $l+V-1 > L$, output $0$.

    \item Parse $\spk$ as $(\cecparams_u, \allowbreak \tilde{\alpha}, \allowbreak \beta_{1}, \allowbreak \tilde{\beta}_{1}, \beta_{2}, \tilde{\beta}_{2})$.

    \item Pick random scalars $r \gets \Zp$ and $r' \gets \Zp$ and compute $\sigma' = (\h',s') \gets (\h^{r'}, s^{r'}(\h')^{r})$.

    \item Compute $\kappa \gets \tilde{\alpha} \tilde{\beta}_1^{\ssk_{\fcecUser_j}} \tilde{\beta}_2^{\cecsn}  \gb^{r}$.

    \item Pick random $r_1,r_2 \gets \Zp$ and set $\phi_{V,l} = (\phi_{V,l}[1],\phi_{V,l}[2]) \gets (\ga^{r_1}, \varsigma_l^\cecsn \eta_{V}^{r_1})$.

    \item Set $R \gets H'(\cecpaymentinfo)$, where $H'$ is a collision-resistant hash function, and compute $$\varphi_{V,l} = (\varphi_{V,l}[1],\varphi_{V,l}[2]) \gets (\ga^{r_2}, (\ga^R)^{\ssk_{\fcecUser_j}} \theta_l^{\cecsn} \eta_{V}^{r_2})$$

    \item Take $\cecparams_u$ from $\spk$. Take the public key $\spk_{sps} \allowbreak = \allowbreak (Y, \allowbreak W_1, \allowbreak W_2, \allowbreak Z)$ and the signature $\tau_{l+V-1} \allowbreak = \allowbreak (R_{l+V-1}, \allowbreak S_{l+V-1}, \allowbreak T_{l+V-1})$.

    \item Compute a ZK argument of knowledge $\pi_v$ via the Fiat-Shamir heuristic for the following relation:
    {\small
    \begin{align*}
        \pi_v =  & \NIZK\{(\ssk_{\fcecUser_j}, \cecsn, r, r_1, r_2, \varsigma_l, \theta_l, \varsigma_{l+V-1}, \theta_{l+V-1}, R_{l+V-1}, S_{l+V-1}, T_{l+V-1}): \nonumber \\ &
        \kappa = \tilde{\alpha} \tilde{\beta}_1^{\ssk_{\fcecUser_j}} \tilde{\beta}_2^{\cecsn} \gb^{r}\ \land\  
        \phi_{V,l}[1] = \ga^{r_1}\ \land\ \phi_{V,l}[2] = \varsigma_l^{\cecsn} \eta_{V}^{r_1}\ \land\  \\&
        \varphi_{V,l}[1] = \ga^{r_2}\ \land\ \varphi_{V,l}[2] = (\ga^R)^{\ssk_{\fcecUser_j}} \theta_l^{\cecsn} \eta_{V}^{r_2}\ \land\  \\&
        \e(\varsigma_l, \tilde{\delta}_{V-1}) = \e(\varsigma_{l+V-1},\tilde{g})\ \land\  \\ & \e(\theta_l, \tilde{\delta}_{V-1}) = \e(\theta_{l+V-1},\tilde{g})\ \land\ \e(R_{l+V-1},T_{l+V-1})  \e(\ga, \gb)^{-1} =1  \\&
        \e(R_{l+V-1},Y) \e(S_{l+V-1},\gb)  e(\varsigma_{l+V-1}, W_1) e(\theta_{l+V-1}, W_2) \cdot\  \e(g, Z)^{-1} =1\
     \end{align*}
     }

     Non-interactive ZK arguments computed via the Fiat-Shamir heuristic are signatures of knowledge, i.e.\ they can be used to sign messages. This non-interactive argument signs the payment information $\cecpaymentinfo$. Since there are group elements in the witness, the transformation described in~\S\ref{subsec:zkpk} is needed. We depict the argument after applying the transformation in Appendix~\ref{sec:divisibleEcashSpend}.

    \item Output a payment $\cecpayment \gets (\kappa, \sigma', \phi_{V,l}, \varphi_{V,l}, R, \pi_v, V)$ and an updated wallet $\cecwallet' \gets (\sigma, \allowbreak \cecsn, \allowbreak l+V)$.

\end{itemize}

\item[$\cecSpendVf(\spk, \cecpayment, \cecpaymentinfo)$.] Execute the following steps:

\begin{itemize}[leftmargin=*]

    \item Parse $\cecpayment$ as $(\kappa, \sigma', \phi_{V,l}, \varphi_{V,l}, R, \pi_v, V)$ and $\sigma'$ as $(\h', \allowbreak s')$. Output $0$ if $\h' = 1$ or if $\e(\h', \kappa) \allowbreak = \allowbreak \e(s',\gb)$ does not hold.

    \item Output $0$ if $R \neq H'(\cecpaymentinfo)$.

    \item Verify $\pi_v$ by using $\cecpaymentinfo$, $\spk$,  $\phi_{V,l}$, $\varphi_{V,l}$, $V$, $R$ and $\kappa$. Output $0$ if the proof is not correct, else output $V$.

\end{itemize}

\item[$\cecIdentify(\cecparams, \cecPK, \cecpayment_1, \cecpayment_2, \cecpaymentinfo_1, \cecpaymentinfo_2)$.] Compute:
\begin{itemize}[leftmargin=*]

    \item Parse $\cecpayment_1$ as $(\kappa_1, \allowbreak \sigma'_1, \allowbreak \phi_{V_1,l_1,1}, \allowbreak \varphi_{V_1,l_1,1}, \allowbreak R_1, \allowbreak \pi_{v,1}, \allowbreak V_1)$ and $\cecpayment_2$ as $(\kappa_2, \allowbreak \sigma'_2, \allowbreak \phi_{V_2,l_2,2}, \allowbreak \varphi_{V_2,l_2,2}, \allowbreak R_2, \allowbreak \pi_{v,2}, \allowbreak V_2)$.

    \item For $k \in [0,V_1-1]$, compute the serial numbers
    \begin{align*}
        \mathrm{SN}_{k,1} \allowbreak \gets \allowbreak \e(\phi_{V_1,l_1,1}[2], \tilde{\delta}_k) \allowbreak \e(\phi_{V_1,l_1,1}[1], \tilde{\eta}_{V_1,k}).
    \end{align*}
    For $k \in [0,V_2-1]$, compute the serial numbers
    \begin{align*}
        \mathrm{SN}_{k,2} \allowbreak \gets \allowbreak \e(\phi_{V_2,l_2,2}[2], \tilde{\delta}_k) \allowbreak \e(\phi_{V_2,l_2,2}[1], \tilde{\eta}_{V_2,k}).
    \end{align*}

    \item Output $1$ if none of the serial numbers $\mathrm{SN}_{k_1,1}$, for $k_1 \in [0,V_1-1]$, is equal to $\mathrm{SN}_{k_2,2}$, for $k_2 \in [0,V_2-1]$.

    \item Else, output $\cecpaymentinfo_1$ if $\cecpaymentinfo_1 = \cecpaymentinfo_2$.

    \item Else, let $k_1 \in [0,V_1-1]$ and $k_2 \in [0,V_2-1]$ be two indices such that $\mathrm{SN}_{k_1,1}=\mathrm{SN}_{k_2,2}$. Compute
    \begin{align*}
        T_1 \gets \e(\varphi_{V_1,l_1,1}[2],\tilde{\delta}_{k_1}) \e(\varphi_{V_1,l_1,1}[1],\tilde{\eta}_{V_1,k_1})
    \end{align*}
    and
    \begin{align*}
        T_2 \gets \e(\varphi_{V_2,l_2,2}[2],\tilde{\delta}_{k_2}) \e(\varphi_{V_2,l_2,2}[1],\tilde{\eta}_{V_2,k_2}).
    \end{align*}
    For each $\spk_{\fcecUser_j} \in \mathrm{PK}$, check whether $T_1T_2^{-1} = \e(\spk_{\fcecUser_j},\tilde{\delta}_{k_1}^{R_1}\tilde{\delta}_{k_2}^{-R_2})$ and output $\spk_{\fcecUser_j}$ if the equality holds. Output $\bot$ if the equality does not hold for any $\spk_{\fcecUser_j} \in \mathrm{PK}$.
\end{itemize}

\end{description}

\subsubsection{Security Analysis of Divisible E-Cash}
\label{sec:securityDivisible}
In \S\ref{sec:securityProofDivisible}, we prove formally that $\mathrm{\Pi}_{\CEC}$, when instantiated with the algorithms of our divisible $\CEC$ scheme, realizes $\Functionality_{\CEC}$. In this section, we give intuition on why our scheme is secure. The security analysis of our divisible $\CEC$ scheme is based on the security analysis given for our compact $\CEC$ scheme regarding the withdrawal phase and the non-interactive ZK argument of possession of PS signatures used in the spending phase. As in our compact $\CEC$ scheme, the anonymity property also relies on the hiding property of Pedersen commitments and the ZK property of the proof system, as well as on the method to ``randomize'' signatures in the spend phase. The traceability property relies on the weak simulation extractability property of the non-interactive ZK arguments of knowledge, the binding property of the commitment scheme and the existential unforgeability property of PS signatures in the RO model.

The remaining part of our analysis follows the security proof given in~\cite{DBLP:conf/pkc/PointchevalST17} for the divisible e-cash scheme. The anonymity property holds under the $N$-MXDH' assumption (see~\S\ref{subsec:bilinearMaps}). In~\cite{DBLP:conf/pkc/PointchevalST17}, it is shown that this assumption holds in the generic bilinear group model. 


The traceability property relies on the existential unforgeability of the SPS scheme in~\cite{DBLP:conf/crypto/AbeGHO11}, which guarantees that the values $(\varsigma_{l+V-1}, \allowbreak \theta_{l+V-1})$ used as a witness in the ZK argument $\pi_v$ are correct. It also relies on the $BDHI$ assumption, which guarantees that the adversary cannot generate values $(\varsigma_{l}, \allowbreak \theta_{l})$ such that $l < 1$. Those two properties, along with the above-mentioned binding property of the commitment scheme and the existential unforgeability property of PS signatures in the RO model, guarantee that the ElGamal encryptions $\phi_{V,l}$ and $\varphi_{V,l}$ in payment are computed correctly. This ensures that, if there is double-spending, an authority can identify the user that double-spent a coin.

As noted in~\cite{DBLP:conf/asiacrypt/BoursePS19}, in an RO model version of the scheme in~\cite{DBLP:conf/pkc/PointchevalST17}, like our scheme, we can extract the user's secret key from the ZK argument $\pi_v$. Thus, we can show that an honest user cannot be found guilty of double spending under the discrete logarithm assumption. In our scheme, unlike in~\cite{DBLP:conf/pkc/PointchevalST17}, the authority does not contribute randomness to the generation of the coin secret $\cecsn$. This means that, as in our compact $\CEC$ scheme, the adversary is able to generate two payments where there is no double spending, and yet there is a match between serial numbers. In that case, it is guaranteed that an honest user will not be found guilty.

As for clearance, like in our compact $\CEC$ scheme, our construction in~\S\ref{sec:constructionCEC} guarantees that only the provider that receives a payment can deposit it. This is done by using an authenticated bulletin board and checking that the provider's identity is contained in the payment information $\cecpaymentinfo$.

%% file: 6EfficiencyComparison.tex
\section{Efficiency Analysis and Comparison}
\label{sec:efficiencycomparison}

For years research has focused almost exclusively on divisible e-cash because 
of the constant cost of the spending phase. However, this is achieved at the expense of much more expensive deposit and identification phases. In this section, we compare the efficiency of our compact and divisible $\CEC$ schemes. To this end, in \S\ref{sec:compactECashRangeProof}, we describe an instantiation of our compact $\CEC$ scheme with a concrete set membership proof, and in \S\ref{sec:divisibleEcashSpend}, we describe how the NIZK arguments of our divisible $\CEC$ scheme are instantiated. 
Our comparison shows that our compact EC with multiple denominations, an idea mentioned but never explored, keeps efficient deposit and identification phases, while dramatically reducing the spending phase cost as opposed to using one denomination. In fact, when the price range is not large, the concrete (as opposed to asymptotic) cost of compact EC with multiple denominations is smaller than that of divisible EC also in the spending phase. Such a comparison was not done before and can guide the choice of an EC scheme for a practical payment system. 


In~\S\ref{sec:multipledenominations}, we analyze the average number of coins that need to be spent depending on the choice of denominations. In~\S\ref{sec:performancemeasurements}, we describe the implementation of our schemes and compare their performance. 

\begin{table}
    \centering
    \caption{Average number of spent coins given $D$ and $P_{max}$}
    {\small
    \begin{tabular}{|r|r|r|} \hline
        \multicolumn{1}{|c|}{$P_{max}$} & \multicolumn{1}{|c|}{$D$} & \multicolumn{1}{|c|}{Avg.\ Num.\ Coins} \\ \hline
        10 & $[1,2,5]$ &  1.9 \\
        100 & $[1,2,5,10,20,50]$ &  3.4 \\
        1000 & $[1,2,5,\ldots,100,200,500]$ &  5.1 \\
        10000 & $[1,2,5,\ldots,1000,2000,5000]$ &  6.8 \\
        100000 & $[1,2,5,\ldots,10000,20000,50000]$ & 8.5 \\
        1000000 & $[1,2,5,\ldots,10000,20000,50000]$ & 17.5 \\ \hline
    \end{tabular}
    \label{tab:denominations}
}
\vspace{-5mm}
\end{table}

\subsection{Choice of Multiple Denominations}
\label{sec:multipledenominations}

The spending phase is arguably the phase in which time constraints are more demanding. Spending one coin is more efficient in our compact e-cash scheme than in our divisible e-cash scheme. However, for any real-world payment, multiple coins need to be spent, and consequently, the compact e-cash scheme is not practical. 
To counter this problem, a solution that has often been suggested in the e-cash literature, but not studied in depth, is to use multiple denominations. 
A question that arises when using multiple denominations is how to choose those denominations optimally, i.e., to minimize the average number of coins that need to be spent. This \emph{optimal denomination problem} has been studied in the context of Fiat currencies~\cite{shallit2003country}. Assuming that all the prices within a range $[1, \allowbreak P_{max}]$ are equally likely, and fixing the number $D$ of denominations, the problem is to find the set of denominations that minimizes the average number of spent coins. 
As calculated in~\cite{shallit2003country}, for a price range of $[1, \allowbreak 100]$ cents, and the denominations $[1,\allowbreak 5,\allowbreak 10,\allowbreak 25]$, the average number of coins spent is $4.7$. However, using the optimal sets, which are $[1, \allowbreak 5, \allowbreak 18, \allowbreak 25]$ and $[1, \allowbreak 5, \allowbreak 18, \allowbreak 29]$, the average is $3.89$.

In general, the optimal denomination problem is NP-hard. Given a naive approach of choosing $1$ as the smallest denomination (to be able to pay for the lowest price), we must calculate the number of all possible sets of $D-1$ denominations in the range $[1, \allowbreak P_{max}]$, which is given by the number of combinations without repetition $C_{D-1}(P_{max})$, and 
the average number of coins needed to pay for prices in $[1, \allowbreak P_{max}]$
for each set of those $D$ denominations. This computation is too expensive for practical values of $D$ and $P_{max}$ (e.g.\ $D=15$ and $P_{max}=1000000$ cents), although it can be optimized by restricting the set of denominations to those that fulfil certain properties.

A problem with using multiple denominations in e-cash is that a user may not be able to pay a price even given enough funds. For example, a user left with two coins of 10 cents is not able to pay for a price of 11 cents. The obvious solution to this problem would be to allow the user to exchange one coin of 10 cents for 10 coins of 1 cent, but this would require interaction with authorities, turning the scheme into an online one. For the scheme to remain offline, the only solution would be for the user to withdraw coins of 1 cent denomination.
Therefore, for practical use of the compact e-cash scheme, we need an easy-to-use set of denominations. 
Taking the denominations of the euro (i.e.\ $[1, \allowbreak 2,\allowbreak 5, \allowbreak 10, \allowbreak 20, \allowbreak 50, \allowbreak 100, \allowbreak 200, \allowbreak 500, \allowbreak 1000, \allowbreak 2000, \allowbreak 5000, \allowbreak 10000, \allowbreak 20000, \allowbreak 50000]$ cents) as an example, in Table~\ref{tab:denominations} we calculate the average number of coins that need to be spent for different values of $P_{max}$ and $D$. Although these denominations are not optimal, 
the results in~\cite{shallit2003country} suggest that the improvement derived from using an optimal denomination set for the same values of $D$ is not dramatic.
Furthermore, these denominations allow us to compute the optimal representation for a given price (i.e.\ the optimal number of coins of each denomination that are needed to pay), by running the simple \emph{greedy algorithm}~\cite{shallit2003country}, whereas the optimal denomination set may not allow that. Therefore, we use those denominations to compare the efficiency of our compact e-cash with that of the divisible e-cash scheme. 

\subsection{Implementation and Comparison}
\label{sec:performancemeasurements}

\paragraph{Implementation} We implement the protocols presented in~\S\ref{sec:compactecash} and~\S\ref{sec:divisibleecash} in $\mathsf{Rust 1.56.0}$. Our implementation is open source.\footnote{\url{https://github.com/nymtech/nym/tree/research/ecash}}
For the implementation of the elliptic curve we use a fork of \texttt{bls12\_381} library and further extend it to facilitate operations in $\Gt$.\footnote{\url{https://github.com/jstuczyn/bls12_381}} All benchmarks were run on a dedicated 16 GB Linode machine, with a $2.2$ GHz AMD CPU. To minimize accuracy errors, we execute each measurement 300 times and compute its average. 
With this specification, a single exponentiation in $\Ga$ takes approximately $531.24$ us, in $\Gb$ approximately $2.13$ ms, while in $\Gt$ approximately $3.85$ ms. A single pairing operation takes $2.59$ ms.

\paragraph{Evaluation} 
For our benchmarks, we set the number of authorities to $n = 100$ with a threshold of $t = 70$. The maximum number of coins in a wallet is set to $L=100$. We also set the number of users to $100$, which determines the size of the set of public keys used as input to algorithm $\cecIdentify$. We always place the target public key as the last one $PK$ list.

The withdrawal protocol is the same in both our compact and divisible $\CEC$ schemes. Algorithm $\cecRequest$ takes on average $9.16$ ms. The algorithms run by the authorities, i.e. $\cecRequestVf$ and $\cecWithdraw$, take in total $8.64$ ms on average. Algorithm $\cecWithdrawVf$ takes $8.47$ ms. We remark that to withdraw a wallet, a user needs to run $\cecRequest$ only once because the same request is sent to at least $t$ authorities. However, algorithm $\cecWithdrawVf$ needs to be run at least $t$ times to verify each of the responses. Finally, algorithm $\cecCreateWallet$, which is run just once after $t$ valid responses have been received from different authorities, takes $65.28$ ms. We consider that those timings are practical, and we stress that they do not depend on the number of coins in a full wallet.



\setlength{\tabcolsep}{10pt}
\begin{table}
\centering
    \caption{Performance benchmarks of Compact and Divisible $\CEC$. For Compact $\CEC$ we measure for $V=1$.}
    \begin{tabular}{@{}lll@{}}
        \hline
         & \thead{Compact $\CEC$} & 
            \thead{Divisible $\CEC$}\\\hline
         $ \cecSpend$ & 34.75 ms & 124.49 ms\\
         $ \cecSpendVf$ & 34.15 ms & 129.81 ms\\ \hline
         $ \cecIdentify$ & 1.61 ms & 133.81 ms\\
         \hline 
    \end{tabular}
    \label{tab:benchmarks}
\end{table}

Table~\ref{tab:benchmarks} presents a comparison between spending a single coin in our compact $\CEC$ scheme vs spending $V$ coins in our divisible $\CEC$ scheme. Algorithms $\cecSpend$ and $\cecSpendVf$ are almost four times faster in the compact than in the divisible $\CEC$ scheme. However, as we noted earlier, the cost of the spending phase in the divisible $\CEC$ scheme is independent of the number of coins spent, while in the compact $\CEC$ scheme, it grows linearly. As an example, let $1267$ be the price to be paid. The total execution time to settle a payment using the compact $\CEC$ scheme would take $44.03$ seconds, thus significantly more than in the divisible $\CEC$ scheme. 

However, the spending phase of our compact e-cash scheme can be optimized as follows. First, algorithm $\cecSpend$ allows us to spend $V$ coins with cost smaller than  the cost of running $V$ times $\cecSpend$ to spend one coin. Our benchmarks show that spending $V = 2$ takes on average $53.43$ ms, which is almost $25\%$ faster than executing the spend protocol twice sequentially to spend one coin. Second, as discussed in~\S\ref{sec:multipledenominations}, we can run several instances of the scheme in parallel and assign to each of them a different denomination. For example, let's consider a set of denominations $[1000, 500, 100, 50, 20, 10, 5, 2, 1]$. Given our price of $1267$, the spender now executes the pay function five times with value $V = 1$ for coins $[1000, 50, 10, 5, 2]$ and once with value $V = 2$ for coins with denomination $100$. 
Thus, the total execution time is $261.93$ ms, which is significantly faster. However, in case of large payments and a limited number of denominations, the compact e-cash is still significantly inefficient. For example, given denominations $[100, 50, 20, 10, 5, 2, 1]$ we need $425.34$ ms to complete a  payment of $1267$. A similar dependency can be observed in the case of payment verification. 

We use our analysis from~\S\ref{sec:multipledenominations} to estimate the average time required to complete a payment using our compact $\CEC$ scheme given different price ranges $[1, \allowbreak P_{max}]$ and sets of denominations. The results are summarised in Table~\ref{tab:avg_spend_time}. We observe that our compact $\CEC$ scheme is more efficient than our divisible $\CEC$ scheme for small price ranges and small sets of denominations. 

The major advantage of the compact $\CEC$ scheme over the divisible $\CEC$ scheme is a fast \emph{identification} phase. In the compact $\CEC$ scheme, to detect double-spending, an authority simply needs to compare serial numbers. In a practical setting, it is likely that double spending happens infrequently, and so the computation cost for the authority is negligible. However, in the divisible $\CEC$ scheme, the authority needs to compute the serial numbers of a payment by running $\cecIdentify$ before the authority can compare them with the serial numbers of other payments. The computation of a serial number involves two pairings. 

Once double spending is detected, the compact $\CEC$ scheme can identify the user guilty of double spending in $1.61$ ms, independently of the number of users in the system. However, in the divisible $\CEC$ scheme, the computation is more expensive and it grows linearly with the number of users in the system. We remark that the timings given in Table~\ref{tab:benchmarks} are calculated for 100 users, but in practice this number could be orders of magnitude bigger, which makes identification in compact e-cash much more efficient than in divisible e-cash.



\begin{table}
    \centering
    \caption{Average time required to complete a payment in Compact $\CEC$ for different values of $D$ and $P_{max}$}
    \begin{tabular}{|r|r|c|} \hline
        \multicolumn{1}{|c|}{$P_{max}$} & \multicolumn{1}{|c|}{$D$} & 
        \thead{Spend [ms]} \\ \hline
        10 & $[1,2,5]$ &  50.84 \\
        100 & $[1,2,5,10,20,50]$ &  90.98 \\
        1000 & $[1,2,5,\ldots,100,200,500]$ &  136.46 \\
        10000 & $[1,2,5,\ldots,1000,2000,5000]$ &  181.95 \\
        100000 & $[1,2,5,\ldots,10000,20000,50000]$ & 227.43 \\
        1000000 & $[1,2,5,\ldots,10000,20000,50000]$ & 468.26 \\ \hline
    \end{tabular}
    \label{tab:avg_spend_time}
\vspace{-5mm}    
\end{table}

%% file: 11Conclusion.tex
\section{Integration}\label{sec:integration}
In our schemes, deposited payments are verified by the authorities. Double-spending is detected by storing previous payments on a bulletin board and checking that serial
numbers in a deposited payment are not equal to any serial number
of previously submitted payments.
A blockchain can be used to implement the bulletin board. When a provider wishes to deposit a payment, the payment is broadcast to authorities, which could also be $n$ validators in a proof-of-stake blockchain. The consensus mechanism of the blockchain is then used to agree on whether the payment is valid and, in that case, recorded in the blockchain. This allows decentralizing also the deposit phase of our protocols. 

Also, note that authorities in this setting may leave and enter the system dynamically. However, we must take into account that an authority that leaves the system still possesses a valid share of the secret key, and we assume that, after leaving the system, the authority becomes corrupt.  Similarly to decentralized e-cash schemes, in our schemes we assume an honest majority, i.e. if $n = 2a + 1$, there are at most $a$ corrupt authorities. For our schemes to be secure, if $a$ is the number of corrupt authorities currently in the system, and $b$ is the number of authorities that have left, we need that $a + b < t$, where $t$ is the threshold. Therefore, algorithm $\cecKeyGenA$ needs to be run periodically to create a new public verification key $\spk$ and new key pairs $(\ssk_{\fcecAuthority_i}, \allowbreak \spk_{\fcecAuthority_i})_{i\in[1,n]}$ so as to ensure that $a + b < t$. This implies that e-cash expires whenever a new key is created. A time interval in which users can convert old wallets to use the new secret key can be given~\cite{chaum2021issue}. Once this interval ends, e-cash expiration makes it possible to delete the payments stored for double-spending detection and reset the bulletin board, which avoids an ever-growing blockchain~\cite{zcashattack1}. The blockchain can thus be used as a settlement layer for offline e-cash transactions. 

This scheme would get the best of both worlds, that of online blockchain and offline e-cash, but further research needs to be done to specify this model formally and parameterize a real-world implementation. Threshold-issuance offline e-cash may end up solving the pressing problems of privacy and scalability of payments in both blockchain and even CBDCs as e-cash moves from theory into practice.

\section{Conclusion}
\label{sec:conclusion} 

In this work, we proposed the first offline anonymous e-cash scheme with threshold issuance, motivated by the concrete scalability concerns of blockchains and concerns with centralization in CBDCs~\cite{digitaleuro}. We define the ideal functionality and propose two instantiations based on an improved compact and a divisible e-cash. We have shown that our schemes realize the ideal functionality and formally prove their security. We have also implemented both schemes and compared their efficiency, showing that compact e-cash is more efficient and feasible for smaller transactions, which would naturally compose the majority of offline e-cash transactions in application scenarios such as a user-facing blockchain or CBDC where practical deployment concerns would necessitate distributed authorities. 




%% file: ASecurityDefinitionsBuildingBlocks.tex
\section{Building Blocks}
\label{sec:securitydefinitionsbuildingblocks}

\subsection{Bilinear Maps}

Let $\Ga$, $\Gb$ and $\Gt$ be groups of prime order $\p$. A map $\e: \Ga \times \Gb \rightarrow \Gt$ must satisfy bilinearity, i.e., $\e(\ga^x,\gb^y)=e(\ga,\gb)^{xy}$; non-degeneracy, i.e., for all generators $\ga \in \Ga$ and $\gb \in \Gb$, $\e(\ga,\gb)$ generates $\Gt$; and efficiency, i.e., there exists an efficient algorithm $\BilinearSetup(1^\securityparameter)$ that outputs the pairing group setup $(\p,\Ga,\Gb,\allowbreak \Gt,\e,\ga,\gb)$ and an efficient algorithm to compute $\e(a,b)$ for any $a \in \Ga$, $b \in \Gb$. In type 3 pairings, $\Ga \neq \Gb$ and there exists no efficiently computable homomorphism $f : \Gb \rightarrow \Ga$.

\subsection{Assumptions}
\label{subsec:bilinearMaps}

We recall the assumptions that are needed to prove the security of our schemes.

\begin{definition}\label{xdh}[XDH and SXDH Assumptions ~\cite{DBLP:journals/iacr/BallardGMM05}]
Given $(\p, \allowbreak \Ga, \allowbreak \Gb,\allowbreak \Gt, \allowbreak \e, \allowbreak \ga, \allowbreak \gb)$, the external Diffie-Hellman assumption states that the decisional Diffie-Hellman problem is intractable in $\Ga$ or in $\Gb$. The symmetric external Diffie-Hellman assumption states that it is intractable in both $\Ga$ and $\Gb$.
\end{definition}

\begin{definition}\label{sdh}[$q$-SDH Assumption~\cite{DBLP:journals/joc/BonehB08}]
Given $(\ga,\allowbreak\ga^{x},\allowbreak\ga^{x^2},\allowbreak\ldots,\allowbreak\ga^{x^q})\allowbreak \in \allowbreak \Ga^{q+1}$, the strong Diffie-Hellman assumption states that it is hard to output a pair $(m,\allowbreak \ga^{1/(x+m)})\in \Zp \times \Ga$.
\end{definition}

\begin{definition}\label{ddhi}[$y$-DDHI Assumption~\cite{DBLP:conf/eurocrypt/CamenischHL05}]
Let $\Ga$ be a group of prime order $q$ and let $\ga$ be a generator of $\Ga$. Given $(\ga, \allowbreak \ga^{x}, \ldots, \allowbreak \ga^{(x^y)}, \allowbreak R)$ for a random $x \gets \Zp$, the decisional Diffie-Hellman inversion assumption states that it is hard to decide if $R= \ga^{1/x}$ or not.
\end{definition}

\begin{definition}\label{bdhi}[$N$-BDHI Assumption~\cite{DBLP:journals/iacr/BonehB04a}]
Given $(\p,\allowbreak\Ga,\allowbreak\Gb,\allowbreak \Gt,\allowbreak\e,\allowbreak\ga,\allowbreak\gb)$ and the tuple $(\{\ga^{y^i}\}_{i=0}^{N},\{\gb^{y^i}\}_{i=0}^{N}) \in \Ga^{N+1} \times \Gb^{N+1}$, the bilinear Diffie-Hellman inversion assumption states that it is hard to compute $\e(\ga,\gb)^{1/y} \in \Gt$.
\end{definition}

\begin{definition}\label{mxdh}[$N$-MXDH' Assumption~\cite{DBLP:conf/pkc/PointchevalST17}]
$\forall N \in \mathbb{N^\ast}$, we define $C = N^3 - N^2$, $S = C + 1$, $E = N^2 - N$, $D = S + E$, and $P = D + C$. Given $(\p,\Ga,\Gb,\allowbreak \Gt,\e,\ga,\gb)$ and $\{(\ga^{\gamma^k}, \allowbreak \h^{\gamma^k})_{k=0}^{P}, \allowbreak  (\ga^{\alpha \delta \gamma^{-k}}, \allowbreak  \h^{\alpha \delta \gamma^{-k}})_{k=0}^{E}, \allowbreak (\ga^{\chi \gamma^{k}}, \allowbreak \h^{\chi \gamma^{k}})_{k=D+1}^{P}, \allowbreak (\ga^{\alpha \gamma^{-k}}, \allowbreak \ga^{\chi \gamma^{k}/ \alpha}, \allowbreak \h^{\chi \gamma^{k}/ \alpha})_{k=0}^{C}\} \allowbreak \in \allowbreak \Ga^{2P+5S+2E+2}$, as well as $(\gb^{\gamma^{k}},\gb^{\alpha \gamma^{-k}})_{k=0}^{C} \in \Gb^{2S}$ and a pair $(\ga^{z_1},\h^{z_2}) \in \Ga^2$, it is hard to decide whether $z_1 = z_2 = \delta + \chi \gamma^{D}/\alpha$ or $(z_1, \allowbreak z_2)$ is random.

\end{definition}

\begin{definition}\label{dhe}[$n$-DHE~\cite{DBLP:conf/pkc/CamenischKS09}]
Let $(\p,\allowbreak\G,\allowbreak\Gb,\allowbreak \Gt,\allowbreak\e,\allowbreak\g,\allowbreak\gb) \allowbreak\leftarrow \allowbreak\BilinearSetup(1^k)$ and $\alpha \allowbreak\gets \allowbreak\Zp$. Given $(\p,\allowbreak\G,\allowbreak\Gb,\allowbreak \Gt,\allowbreak\e,\allowbreak\g,\allowbreak\gb)$ and a tuple $(\g_1,\allowbreak\gb_1,\allowbreak\ldots,\allowbreak\g_n,\allowbreak\gb_n,\allowbreak\g_{n+2},\allowbreak\ldots,\allowbreak\g_{2n})$ such that $\g_i \allowbreak = \allowbreak\g^{(\alpha^i)}$ and $\gb_i \allowbreak= \allowbreak \gb^{(\alpha^i)}$, for any p.p.t.\ adversary $\Adversary$, $\mathrm{Pr}[\g^{(\alpha^{n+1})} \allowbreak\leftarrow \allowbreak \Adversary(\p,\allowbreak\G,\allowbreak\Gb,\allowbreak \Gt,\allowbreak\e,\allowbreak \g,\allowbreak \gb,\allowbreak \g_1,\allowbreak\gb_1,\allowbreak\ldots,\allowbreak\g_n,\allowbreak\gb_n,\allowbreak\g_{n+2},\allowbreak\ldots,\allowbreak\g_{2n})] \allowbreak\leq \allowbreak \epsilon(k)$.
\end{definition}

\subsection{Zero-Knowledge Arguments of Knowledge}
\label{subsec:zkpk}

Informally speaking, a zero-knowledge argument of knowledge is a two-party protocol between a prover and a verifier with two properties. First, it should be a proof of knowledge, i.e., there should exist a knowledge extractor that extracts the secret input from a successful prover with all but negligible probability. Second, it should be zero-knowledge, i.e., for all possible verifiers there exists a simulator that, without knowledge of the secret input, yields a distribution that cannot be distinguished from the interaction with a real prover.

To express a zero-knowledge argument of knowledge, we follow the notation introduced by Camenisch and Stadler~\cite{camsta97a}, i.e., we denote as  $\ZKPK\{(w): y=f(w)\}$ a ``{\em zero-knowledge proof of knowledge of the secret input $w$ such that $y=f(w)$}'', where $w$ is a secret input, while $y$ and the function $f$ are publicly known.

Let $\mathcal{L}$ be a language in NP. We can associate to any NP-language $\mathcal{L}$ a polynomial time recognizable relation $\mathcal{R}_{\mathcal{L}}$ defining $\mathcal{L}$ as $\mathcal{L} = \{x: \exists w\ \mathrm{s.t.}\ (x,w) \in \mathcal{R}_{\mathcal{L}} \}$, where $|w| \leq \mathrm{poly}(|x|)$. The string $w$ is called a witness for membership of $x \in \mathcal{L}$.

A protocol $\Sigma = (\mathcal{P}, \mathcal{V})$ for an NP-language $\mathcal{L}$ is an interactive proof system. The prover $\mathcal{P}$ and the verifier $\mathcal{V}$ know an instance $x$ of the language $\mathcal{L}$. The prover $\mathcal{P}$ also knows a witness $w$ for membership of $x \in \mathcal{L}$. $\Sigma$-protocols have a 3-move shape where the first message $\alpha$, called \emph{commitment}, is sent by the prover. The second message $\beta$, called \emph{challenge}, is chosen randomly and sent by the verifier. The last message $\gamma$, called \emph{response}, is sent by the prover. A $\Sigma$-protocol fulfills the properties of completeness, honest-verifier zero-knowledge, and special soundness defined in Faust et al.~\cite{faust2012non}.

In our e-cash schemes, zero-knowledge arguments of knowledge based on the Fiat-Shamir transform~\cite{fiasha86} are used. The Fiat-Shamir transform removes the interaction between the prover $\mathcal{P}$ and the verifier $\mathcal{V}$ of a $\Sigma$ protocol by replacing the challenge with a hash value $H(\alpha,x)$  computed by the prover, where $H$ is modeled as a random oracle. (It is possible to include an additional message $m$ as input to $H$, i.e. $H(\alpha,x,m)$, turning the argument of knowledge into a \emph{signature of knowledge} of the message $m$.) An argument  $\pi$ consists of $(\alpha, H(\alpha,x), \gamma)$. The Fiat-Shamir system is denoted by $(\mathcal{P}^{H}, \mathcal{V}^{H})$ and fulfills the properties of zero-knowledge and weak simulation extractability defined in Faust et al.~\cite{faust2012non}, which we recall below.

\begin{definition}[Zero-Knowledge]\label{def:zero-knowledge}
Define the zero knowledge simulator $\mathcal{S}$ as follows. $\mathcal{S}$ is a stateful algorithm that can operate in two modes: $(h_i, st) \gets \mathcal{S}(1,st,q_i)$ answers random oracle queries $q_i$, while $(\pi, st) \gets \mathcal{S}(2,st,x)$ outputs a simulated proof $\pi$ for an instance $x$. $\mathcal{S}(1,\cdots)$ and $\mathcal{S}(2,\cdots)$ share the state $st$ that is updated after each operation.

Let $\mathcal{L}$ be a language in NP. Denote with $(\mathcal{S}_1, \mathcal{S}_2)$ the oracles such that $\mathcal{S}_1(q_i)$ returns the first output of $(h_i, st) \gets \mathcal{S}(1,st,q_i)$ and $\mathcal{S}_2(x,w)$ returns the first output of $(\pi, st) \gets \mathcal{S}(2,st,x)$ if $(x,w) \in \mathcal{R}_{\mathcal{L}}$. A protocol $(\mathcal{P}^{H}, \mathcal{V}^{H})$ is a non-interactive zero-knowledge proof for the language $\mathcal{L}$ in the random oracle model if there exists a ppt simulator $\mathcal{S}$ such that for all ppt distinguishers $\mathcal{D}$ we have
\begin{equation*}
\Prob[\mathcal{D}^{H(\cdot),\mathcal{P}^{H}(\cdot,\cdot)}(1^k) = 1] \approx \Prob[\mathcal{D}^{\mathcal{S}_1(\cdot),\mathcal{S}_2(\cdot,\cdot)}(1^k)=1],
\end{equation*}
where both $\mathcal{P}$ and $\mathcal{S}_2$ oracles output $\bot$ if $(x,w) \notin \mathcal{R}_{\mathcal{L}}$.

\end{definition}

\begin{definition}[Weak Simulation Extractability]\label{def:weakse}
Let $\mathcal{L}$ be a language in NP. Consider a non-interactive zero-knowledge proof system $(\mathcal{P}^{H}, \mathcal{V}^{H})$ for $\mathcal{L}$ with zero-knowledge simulator $\mathcal{S}$. Let $(\mathcal{S}_1,\mathcal{S}'_2)$ be oracles returning  the first output of $(h_i, st) \gets \mathcal{S}(1,st,q_i)$ and $(\pi, st) \gets \mathcal{S}(2,st,x)$ respectively. $(\mathcal{P}^{H}, \mathcal{V}^{H})$ is weakly simulation extractable  with extraction error $\nu$ and with respect to $\mathcal{S}$ in the random oracle model, if for all ppt adversaries $\Adversary$ there exists an efficient algorithm $\mathcal{E}_{\Adversary}$ with access to the answers $(\mathcal{T}_{H}, \mathcal{T})$ of $(\mathcal{S}_1,\mathcal{S}'_2)$ respectively such that the following holds. Let
\begin{align*}
& \mathrm{acc} = \Prob[(x^{\ast},\pi^{\ast}) \gets \Adversary^{\mathcal{S}_1(\cdot),\mathcal{S}'_2(\cdot)}(1^k;\rho): \\ & (x^{\ast},\pi^{\ast})\notin \mathcal{T}; \mathcal{V}^{\mathcal{S}_1}(x^{\ast},\pi^{\ast})=1] \\
& \mathrm{ext} = \Prob[(x^{\ast},\pi^{\ast}) \gets \Adversary^{\mathcal{S}_1(\cdot),\mathcal{S}'_2(\cdot)}(1^k;\rho); \\ & \indent w^{\ast} \gets \mathcal{E}_{\Adversary}(x^{\ast},\pi^{\ast};\rho,\mathcal{T}_H,\mathcal{T}): (x^{\ast},\pi^{\ast}) \notin \mathcal{T}; (x^{\ast},w^{\ast}) \in  \mathcal{R}_{\mathcal{L}}],
\end{align*}
where the probability space in both cases is over the random choices of $\mathcal{S}$ and the adversary's random tape $\rho$. Then, there exists a constant $d > 0$ and a polynomial $p$ such that whenever $\mathrm{acc} \geq \nu$, we have $\mathrm{ext} \geq (1/p)(\mathrm{acc}-\nu)^{d}$.
\end{definition}

\paragraph{Types of proofs.} We use known results for computing ZK proofs of discrete logarithms~\cite{camsta97a}. A protocol proving knowledge of exponents $(w_1, \allowbreak \ldots, \allowbreak w_n)$ that satisfy the formula  $\phi(w_1, \allowbreak \ldots, \allowbreak w_n)$ is described as
 \begin{equation}
\label{eq-knowsApp}
\ZKPK\{(w_1, \ldots, w_n): 1=\phi(w_1, \ldots, w_n)\} \end{equation}
The formula $\phi(w_1, \ldots,\allowbreak w_n)$ consists of conjunctions and disjunctions of ``atoms''. An atom expresses {\em  group relations}, such as
\begin{equation*}
    \prod_{j=1}^k g_j^{f_j}=1
\end{equation*}
where the $g_j$'s are elements of prime order groups and the $f_j$'s are polynomials in the variables $(w_1, \allowbreak \ldots, \allowbreak w_n)$.

A proof system for (\ref{eq-knowsApp}) can be transformed into a proof system for more expressive statements about secret exponents $\myvar{sexps}$ and secret bases $\myvar{sbases}$:
\begin{equation}
\label{eq-knows-secretApp}
\ZKPK\{(\myvar{sexps}, \myvar{sbases}): 1=\phi(\myvar{sexps}, \myvar{bases}\cup\myvar{sbases})\}
\end{equation}
The transformation adds an additional base $h$ to the public bases. For each $g_j \in \myvar{sbases}$, the transformation picks a random exponent $\rho_j$ and computes a blinded base $g_j'=g_j \h^{\rho_j}$. The transformation adds $g'_j$ to the public bases $\myvar{bases}$, $\rho_j$ to the secret exponents $\myvar{sexps}$, and rewrites $g_j^{f_j}$ into ${g_j'}^{f_j} h^{-f_j \rho_j}$.

The proof system supports pairing product equations
\begin{equation}
\prod_{j=1}^k e(g_j,\tilde{g}_j)^{f_j}=1
\end{equation}
in groups of prime order with a bilinear map $\e$, by treating the target group $\Gt$ as the group of the proof system. The embedding for secret bases is unchanged, except for the case in which both bases in a pairing are secret. In this case, $e(g_j,\tilde{g}_j)^{f_j}$ must be transformed into $e(g'_j,\tilde{g}'_j)^{f_j} e(g_j',\tilde{h})^{-f_j \tilde\rho_j} e(h,\tilde{g}'_j)^{-f_j \rho_j} \allowbreak e(h,\tilde{h})^{f_j \rho_j \tilde\rho_j}$.

\subsection{Commitment Schemes}
\label{subsec:commitmentSchemes}

A commitment scheme consists of algorithms $\ComSetup$, $\ComCommit$ and $\ComVerify$. The algorithm $\ComSetup(1^{\securityparameter})$ generates the parameters of the commitment scheme $\paramscom$, which include a description of the message space $\commessagespace$. $\ComCommit(\paramscom, \commessage)$ outputs a commitment $\com$ to $\commessage$ and auxiliary information $\open$. A commitment is opened by revealing $(\commessage, \allowbreak\open)$ and checking whether $\ComVerify(\paramscom,\com,\commessage,\open)$ outputs $1$ or $0$. 

A commitment scheme should fulfill the \emph{correctness}, \emph{hiding} and \emph{binding} properties. We recall the definitions of those properties below.

\begin{definition}[Correctness]
Correctness requires that $\ComVerify$ accepts all commitments created by algorithm $\ComCommit$, i.e., for all $\commessage \in \commessagespace$
\begin{equation*}
\Prob \left[
\begin{array}{c}
\paramscom \leftarrow \ComSetup(1^{\securityparameter});\
(\com, \open) \leftarrow \ComCommit(\paramscom, \commessage):\ \\
1 = \ComVerify(\paramscom, \com, \commessage, \open)
\end{array}
\right] = 1~.
\end{equation*}
\end{definition}

\begin{definition}[Hiding Property]\label{def:hiding}
The hiding property ensures that a commitment $\com$ to $\commessage$ does not reveal any information about $\commessage$. For any ppt adversary $\Adversary$, the hiding property is defined as follows:
\begin{equation*}
\Prob \left[
\begin{array}{l}
\paramscom \leftarrow \ComSetup(1^{\securityparameter});\ \\
(\commessage_0, \commessage_1, \cstate) \leftarrow \Adversary(\paramscom);\ \\
\bit \leftarrow \{0,1\};\ (\com, \open) \leftarrow \ComCommit(\paramscom, \commessage_{\bit});\ \\
\bit' \leftarrow \Adversary(\cstate,\com):\ \\
\commessage_0 \in \commessagespace\ \land\ \commessage_1 \in \commessagespace\ \land\ \bit = \bit'
\end{array}
\right] \leq \frac{1}{2} + \negligible(\securityparameter)~.
\end{equation*}
\end{definition}

\begin{definition}[Binding Property]
The binding property ensures that $\com$ cannot be opened to another value $\commessage'$. For any ppt adversary $\Adversary$, the binding property is defined as follows:
\begin{equation*}
\Prob \left[
\begin{array}{l}
\paramscom \leftarrow \ComSetup(1^{\securityparameter});\\
(\com, \commessage, \open, \commessage', \open') \leftarrow \Adversary(\paramscom):\ \\
\commessage \in \commessagespace\ \land\ \commessage' \in \commessagespace\ \land \commessage \neq \commessage' \\ \land\ 1 = \ComVerify(\paramscom, \com, \commessage, \open)\  \\
\land\ 1 = \ComVerify(\paramscom, \com, \commessage', \open')
\end{array}
\right] \leq \negligible(\securityparameter)~.
\end{equation*}
\end{definition}

Our e-cash schemes use the commitment scheme by Pedersen~\cite{DBLP:conf/crypto/Pedersen91} to commit to elements $x \in \Zp$, where $p$ is a prime. This commitment scheme is perfectly hiding and computationally binding under the discrete logarithm assumption. The Pedersen commitment scheme consists of the following algorithms.
\begin{description}[leftmargin=0pt]

\item[$\bullet$ $\ComSetup(1^{k})$.] On input the security parameter $1^k$, pick random generators $\g,\h$ of a group $\Gp$ of prime order $\p$. Output $\paramscom = (g,h, \commessagespace)$, where $\commessagespace = \mathbb{Z}_{p}$.

\item[$\bullet$ $\ComCommit(\paramscom,x)$.] Check that $x \in \commessagespace$. Pick random value $\open \in \Zp$, compute $\com = \g^{\open} \h^{x}$, and output $\com$.

\item[$\bullet$ $\ComVerify(\paramscom,\com,x',\open')$.] Recompute $\com' = \g^{\open'} \h^{x'}$. If $\com \allowbreak = \allowbreak \com'$ then output $\accept$ else $\reject$.

\end{description}



When committing to a tuple of messages, the Pedersen commitment scheme works as follows.
\begin{description}[leftmargin=0pt]

\item[$\bullet$ $\ComSetup(1^{\securityparameter},l)$.] On input the security parameter $1^\securityparameter$ and an upper bound $l$ on the number of elements to be committed, pick $l+1$ random generators $\h_1,\ldots,\h_l,\g$ of a group $\Gp$ of prime order $\p$. Output $\paramscom = (h_1,\ldots,h_l,g, \commessagespace)$, where $\commessagespace = \mathbb{Z}_{p}^{l}$.

\item[$\bullet$ $\ComCommit(\paramscom,\langle x_1,\ldots,x_l \rangle)$.] Pick random value $\open \gets \mathbb{Z}_{p}$, compute $\com = \g^{\open} \prod_{i = 1}^{l} \h_{i}^{x_i}$ and output $\com$.

\item[$\bullet$ $\ComVerify(\paramscom,\com,\langle x'_1,\ldots,x'_l \rangle,\open')$.] Recompute commitment $\com' = \g^{\open'} \prod_{i = 1}^{l} \h_{i}^{x'_i}$. If it is the case that $\com = \com'$ then output $\accept$ else $\reject$.

\end{description}

\subsection{Signature Schemes}
\label{subsec:signatureSchemes}

A signature scheme consists of the algorithms $\SKeygen$, $\SSign$, and $\SVerifySig$. $\SKeygen(1^\securityparameter)$ outputs a secret key $\ssk$ and a public key $\spk$, which include a description of the message space $\smessagespace$. $\SSign(\ssk, \smsg)$ outputs a signature $\ssig$ on message $\smsg \in \smessagespace$. $\SVerifySig(\spk,\ssig,\smsg)$ outputs $1$ if $\ssig$ is a valid signature on $\smsg$ and $0$ otherwise. This definition can be extended to blocks of messages $(\smsg_1, \ldots, \smsg_q)$. In this case, $\SKeygen(1^\securityparameter,q)$ receives the maximum number of messages as input. 
A signature scheme must fulfill the correctness and existential unforgeability properties~\cite{DBLP:journals/siamcomp/GoldwasserMR88}, which we recall below.
\begin{definition}[Correctness]
Correctness ensures that the algorithm $\SVerifySig$ accepts the signatures created by the algorithm $\SSign$ on input a secret key computed by algorithm  $\SKeygen$. More formally,  correctness is defined as follows.
\begin{equation*}
\Prob \left[
\begin{array}{l}
(\ssk, \spk)  \gets \SKeygen(1^\securityparameter);\
\smsg  \gets \smessagespace;\ \\
\ssig  \gets \SSign(\ssk, \smsg):\
1 = \SVerifySig(\spk, \ssig, \smsg)\
\end{array}
\right] = 1
\end{equation*}
\end{definition}

\begin{definition}[Existential Unforgeability]\label{def:unf}
The property of existential unforgeability  ensures that it is not feasible to output a signature on a message without knowledge of the secret key or of another signature on that message. Let $\SOracleSign$ be an oracle that, on input $\ssk$ and a message $\smsg \in \smessagespace$, outputs $\SSign(\ssk, \allowbreak \smsg)$, and let $\ssetsign$ be a set that contains the messages sent to $\SOracleSign$. More formally, for any ppt adversary $\Adversary$, existential unforgeability is defined as follows.
\begin{equation*}
\Prob \left[
\begin{array}{l}
(\ssk, \spk)  \gets \SKeygen(1^\securityparameter);\
(\smsg, \ssig)  \gets \Adversary(\spk)^{\SOracleSign(\ssk, \cdot)}:\ \\
1 = \SVerifySig(\spk, \ssig, \smsg)\ \land\ \smsg \in \smessagespace\ \land\ \smsg \notin \ssetsign\ \\
\end{array}
\right] \leq \epsilon(\securityparameter)
\end{equation*}
\end{definition}

\noindent \textbf{Pointcheval-Sanders (PS) signatures.}
The PS signature scheme is defined as follows~\cite{DBLP:conf/ctrsa/PointchevalS16}.
\begin{description}

\item[$\SKeygen(1^\securityparameter, q)$.] Run $\BilinearSetup(1^\securityparameter)$ to obtain a pairing group setup $\theta = (\p,\Ga,\Gb,\allowbreak \Gt,\e,\allowbreak \ga, \allowbreak \gb)$. Pick random secret key $(x, \allowbreak y_1, \ldots, \allowbreak y_q) \gets \mathbb{Z}_p^{q+1}$. Output the secret key $\ssk = (\theta, \allowbreak x, \allowbreak y_1, \allowbreak \ldots \allowbreak y_q)$ and the public key $\spk = (\theta, \tilde{\alpha}, \beta_1, \tilde{\beta}_1, \ldots, \beta_q, \tilde{\beta}_q) \gets (\theta, \allowbreak \gb^x, \allowbreak \ga^{y_1}, \allowbreak \gb^{y_1}, \allowbreak \ldots, \allowbreak \ga^{y_q}, \allowbreak \gb^{y_q})$.

\item[$\SSign(\ssk, \smsg_1, \ldots, \smsg_q)$.] Parse $\ssk$ as $(\theta, x, y_1, \ldots, y_q)$. Pick up random $r \gets \Zp$ and set $h \gets \ga^r$. Output the signature $\ssig = (h, s) \gets (h, h^{x+y_1\smsg_1+ \ldots +y_q\smsg_q})$.

\item[$\SVerifySig(\spk,\ssig, \smsg_1, \ldots, \smsg_q)$.] Output $1$ if $\e(h, \tilde{\alpha} \prod_{j=1}^{q} \tilde{\beta}_j^{\smsg_j}) \allowbreak = \allowbreak \e(s, \gb)$ and $h \neq 1$. Otherwise output $0$.

\end{description}
This signature scheme is randomizable. To randomize a signature $\ssig = (h, \allowbreak s)$, pick random $r' \gets \Zp$ and compute $\ssig' = (h^{r'}, s^{r'})$. The elements $(\beta_1, \allowbreak \ldots, \allowbreak \beta_q)$ in the public key are needed for the blind signature issuance protocol in~\cite{DBLP:conf/ctrsa/PointchevalS16}, as well as for the issuance protocols of Coconut and of our e-cash schemes.

\noindent \textbf{Pointcheval-Sanders signatures in the random oracle model.} Coconut and our e-cash schemes use a variant of PS signatures in which, in algorithm $\SSign$, the random generator $h$ is computed via a hash function, which is modeled as a random oracle (RO). This variant has been formalized in~\cite{cryptoeprint:2022:011} as PS signatures in the RO model.


In~\cite{cryptoeprint:2022:011}, the syntax of algorithm $\SSign$ is as follows. $\SSign$ uses a random oracle $H:\fromesspace \allowbreak \rightarrow \allowbreak \frolength$. $\SSign(\ssk, \allowbreak \smsg_1, \allowbreak \ldots \allowbreak \smsg_q, \allowbreak \sr, \allowbreak \sstate)$ receives as input a secret key $\ssk$, a tuple of messages $(\smsg_1, \allowbreak \ldots, \allowbreak \smsg_q)$, a value $\sr \allowbreak \in \allowbreak \fromesspace$ and state information $\sstate$, which stores tuples of the form $(\smsg_1, \allowbreak \ldots, \allowbreak \smsg_q, \allowbreak \sr)$. $\SSign$ outputs a signature $\ssig$ on $(\smsg_1, \allowbreak \ldots, \allowbreak \smsg_q)$ if $\sstate$ does not contain a tuple $(\smsg'_1, \allowbreak \ldots, \allowbreak \smsg'_q, \allowbreak \sr')$ such that $(\smsg_1, \allowbreak \ldots, \allowbreak \smsg_q) \neq (\smsg'_1, \allowbreak \ldots, \allowbreak \smsg'_q)$ and $\sr = \sr'$. $\SSign$ also outputs updated state information $\sstate'$.

The PS signature scheme in the RO model scheme works as follows. The algorithms $\SKeygen$ and $\SVerifySig$ remain unmodified.
\begin{description}

\item[$\SSign(\ssk, \smsg_1, \ldots, \smsg_q, \sr, \sstate)$.] Parse $\ssk$ as $(\theta, x, y_1, \ldots, y_q)$. If $\sstate$ contains a tuple $(\smsg'_1, \allowbreak \ldots, \allowbreak \smsg'_q, \allowbreak \sr')$ such that $(\smsg_1, \allowbreak \ldots, \allowbreak \smsg_q) \neq (\smsg'_1, \allowbreak \ldots, \allowbreak \smsg'_q)$ and $\sr = \sr'$, output $\ssig = \bot$ and $\sstate' =\sstate$. Otherwise compute $\h \gets H(\sr)$ and output the signature $\ssig = (h, s) \gets (h, h^{x+y_1\smsg_1+ \ldots +y_q\smsg_q})$ and the updated state information $\sstate' = \sstate \cup \{(\smsg_1, \allowbreak \ldots, \allowbreak \smsg_q,\sr)\}$.

\end{description}
We recall the definition of the existential unforgeability property in the RO model below.
\begin{definition}[Existential Unforgeability in the RO~\cite{cryptoeprint:2022:011}]\label{def:unfro}
For any ppt adversary $\Adversary$, existential unforgeability in the RO model is defined as follows.
\begin{equation*}
\Prob \left[
\begin{array}{l}
(\ssk, \spk)  \gets \SKeygen(1^\securityparameter);\ \\
(\smsg, \ssig)  \gets \Adversary(\spk)^{\SOracleSign(\ssk, \cdot, \cdot), H(\cdot)}:\ \\
1 = \SVerifySig(\spk, \ssig, \smsg)\ \land\ \smsg \in \smessagespace\ \land\ \smsg \notin \ssetsign\ \\
\end{array}
\right] \leq \epsilon(\securityparameter)
\end{equation*}
$\SOracleSign(\ssk, \cdot, \cdot)$ works as follows. On input $\ssk$, a message $\smsg = (\smsg_1, \allowbreak \ldots, \allowbreak \smsg_q)$ and the value $\sr$, $\SOracleSign$ runs $(\ssig, \sstate') \gets \SSign(\ssk, \smsg_1, \ldots, \smsg_q, \sr, \sstate)$. $\SOracleSign$ replaces $\sstate$ by $\sstate'$ and returns $\ssig$ to $\Adversary$. ($\sstate$ is empty in the first invocation of $\SOracleSign$.) $\ssetsign$ is a set that contains the messages sent to $\SOracleSign$.
\end{definition}
In comparison to the definition of existential unforgeability (see Definition~\ref{def:unf}), $\Adversary$ has access to the random oracle $H$, and the signing oracle is modified to follow the new syntax. The PS scheme in the RO model is existentially unforgeable under the generalized PS assumption proposed in~\cite{DBLP:journals/istr/KimLAP21,DBLP:journals/iacr/KimSAP21}.

\noindent \textbf{Structure-Preserving Signature (SPS) scheme.} In a SPS scheme, the public key, the messages, and the signatures are group elements in $\Ga$ and $\Gb$, and verification must consist purely in the checking of pairing product equations. Our divisible e-cash scheme uses the SPS scheme in~\cite{DBLP:conf/crypto/AbeGHO11}. In this SPS scheme, $a$ elements in $\Ga$ and $b$ elements in $\Gb$ are signed.
\begin{description}[leftmargin=10pt]

\item[$\SKeygen(\grp,a,b)$.] Let $\grp \allowbreak \gets \allowbreak (\p, \allowbreak \Ga, \allowbreak \Gb, \allowbreak \Gt, \allowbreak \e, \allowbreak \g, \allowbreak \gb)$ be the bilinear map parameters. Pick at random $u_1, \allowbreak \dots, \allowbreak u_b, \allowbreak y, \allowbreak w_1, \dots\allowbreak w_a, \allowbreak z \gets \Zp^*$ and compute $U_i  \allowbreak = \allowbreak g^{u_i}$, $i \allowbreak \in \allowbreak [1..b]$, $Y \allowbreak = \allowbreak \gb^y$, $W_i \allowbreak = \allowbreak \gb^{w_i}$, $i \allowbreak \in \allowbreak [1..a]$ and $Z \allowbreak = \allowbreak \gb^z$. Return the verification key $\spk \allowbreak \gets \allowbreak (\grp, \allowbreak U_1, \allowbreak \dots, \allowbreak U_b, \allowbreak Y, \allowbreak W_1, \allowbreak \dots, \allowbreak W_a,\allowbreak Z)$ and the signing key $\ssk \allowbreak \gets \allowbreak (\spk, \allowbreak u_1, \allowbreak \dots, \allowbreak u_b, \allowbreak y, \allowbreak w_1,\allowbreak \dots, \allowbreak w_a, \allowbreak z)$.

\item[$\SSign(\ssk,\langle m_1,\dots,m_{a+b} \rangle)$.] Pick $r \allowbreak \gets \allowbreak \Zp^*$, and set
\begin{align*}
    R \allowbreak \gets \allowbreak \g^r, \quad S \allowbreak \gets \allowbreak \g^{z-ry} \prod_{i=1}^a m_i^{-w_i}, \quad T \allowbreak \gets \allowbreak (\gb \prod_{i=1}^b  \allowbreak m_{a+i}^{-u_i})^{1/r},
\end{align*}
and output the signature $\ssig \gets (R, \allowbreak S, \allowbreak T)$.

\item[$\SVerifySig(\spk, \ssig,\langle m_1,\dots,m_{a+b} \rangle)$.] Output $\accept$ if it is satisfied that
\begin{equation*}
 \e(R,Y)\allowbreak \e(S,\gb)\prod_{i=1}^a \e(m_i,\allowbreak W_i) \allowbreak = \allowbreak \e(g, \allowbreak Z)
\end{equation*}
\noindent and
\begin{equation*}
\e(R,T) \prod_{i=1}^b \e(U_i,m_{a+i}) \allowbreak = \allowbreak \e(\ga, \allowbreak \gb)
\end{equation*}
\end{description}




\subsection{Pseudorandom Functions}
\label{subsec:pseudorandomfunction}

Pseudorandom functions (PF)~\cite{DBLP:journals/jacm/GoldreichGM86,DBLP:conf/pkc/BoyleGI14} are a family of indexed
functions $F = \{F_s\}$ such that: (1) given the index $s$, $F_s$ can be efficiently evaluated on all inputs; (2) no
probabilistic polynomial-time algorithm without $s$ can distinguish evaluations $F_s(x_i)$ for inputs $x_i$ of its choice from random values. 
We recall the definition of pseudorandom functions in~\cite{DBLP:conf/pkc/BoyleGI14}.
\begin{definition}[Pseudorandom Function Family]\label{def:pseudorandom}
A family of functions $\mathcal{F} = \{F_s\}_{s \in S}$, indexed by a set $S$, and where $F_s: D \rightarrow R$ for all $s$, is a pseudorandom function (PRF) family if for a randomly chosen $s$, and all PPT $\mathcal{A}$, the distinguishing advantage $\Pr_{s \leftarrow S}[\mathcal{A}^{f_s(\cdot)} \allowbreak = \allowbreak 1] \allowbreak - \Pr_{f\leftarrow(D\rightarrow R)}[\mathcal{A}^{f(\cdot)}=1]$ is negligible, where $(D \rightarrow R)$ denotes the set of all functions from $D$ to $R$.
\end{definition}

Our compact e-cash scheme uses the PF in~\cite{DBLP:conf/eurocrypt/CamenischHL05}, which works as follows. For every $n$, a function $f$ is defined by the tuple $(\Ga, \p, \ga, s)$, where $\Ga$ is a group of order $\p$, $\p$ is an $n$-bit prime, $\ga$ is a generator of $\Ga$, and $s$ is a seed in $\Zp$. For any input $x \in \Zp$ (except for $x = -1\ \mathrm{mod}\ \p$), the function $f_{\G,\p,\ga,s}(\cdot)$, which we denote as $f_{\ga,s}$ for fixed values of $(\Ga,\p,\ga)$, is defined as $f_{\ga,s}(x)=\ga^{1/(s+x+1)}$. This PF is secure under the $y$-DDHI assumption in $\Ga$, which we recall in \S\ref{subsec:bilinearMaps}. This PF is based on the verifiable random function in~\cite{DBLP:conf/pkc/DodisY05}, which is secure under the $y$-DBDHI assumption.

\subsection{Notation}
\label{subsec:notation}

In Table~\ref{tab:notation}, we summarize the notation used in our paper.
\begin{table}
    \centering
    \caption{Table of symbols}
    {\small
    \begin{tabular}{|l|l|} \hline
        \multicolumn{1}{|c|}{Symbol} & \multicolumn{1}{|c|}{Meaning}  \\ \hline
        \multicolumn{2}{|c|}{Bilinear Maps} \\ \hline
        $\e$ & Bilinear map \\
        $\BilinearSetup$ & Bilinear setup \\
        $\p$ & Prime number  \\ 
        $\Ga$ & Group of order $\p$   \\ 
        $\Gb$ & Group of order $\p$ \\ 
        $\Gt$ & Group of order $\p$ \\ 
        $\ga$ & Generator of $\Ga$  \\ 
        $\gb$ & Generator of $\Gb$  \\
        $\Zp$ & Integers modulo $\p$ \\ \hline
        \multicolumn{2}{|c|}{Security Definitions} \\ \hline
        $\Adversary$ & Adversary \\
        $\Environment$ & Environment \\
        $\Simulator$ & Simulator \\
        $\Functionality$ & Ideal Functionality \\
        $\pid$ & Party identifier \\
        $\sid$ & Session identifier \\
        $\qid$ & Query identifier \\ 
        $\Prob$ & Probability \\
        $\mathcal{O}$ & Oracle \\
        $\securityparameter$ & Security parameter \\
        $\negligible$ & Negligible function \\ \hline
        \multicolumn{2}{|c|}{E-cash} \\ \hline
        $\CEC$ & Threshold issuance offline anonymous e-cash \\ 
        $\fcecUser$ & User \\
        $\fcecProvider$ & Provider \\
        $\fcecAuthority$ & Authority \\
        $n$ & Number of authorities \\
        $t$ & Threshold \\
        $L$ & Number of coins in a full wallet \\
        $V$ & Number of coins spent in a payment \\
        $\cecparams$ & Parameters \\
        $\ssk_{\fcecAuthority}$ & Authority secret key \\
        $\spk_{\fcecAuthority}$ & Authority public key  \\
        $\ssk_{\fcecUser}$ & User secret key \\
        $\spk_{\fcecUser}$ & User public key \\
        $\cecrequest$ & Withdrawal request \\
        $\cecresponse$ & Withdrawal response \\
        $\cecwallet$ & Wallet \\
        $\cecpayment$ & Payment \\
        $\cecpaymentinfo$ & Payment information \\
        $\BB$ & Bulletin Board \\ \hline
    \end{tabular}
    \label{tab:notation}
}
\end{table}

%% file: 7UCSecurity.tex
\section{Ideal-World/Real-World Paradigm}
\label{sec:securitymodel}

The security of a protocol $\varphi$ is analyzed by comparing the view of an environment $\Environment$ in a real execution of $\varphi$ against that of $\Environment$ in the ideal protocol defined in $\Functionality_{\varphi}$. $\Environment$ chooses the inputs of the parties and collects their outputs. In the real world, $\Environment$ can communicate freely with an adversary $\Adversary$ who controls both the network and any corrupt parties.
In the ideal world, $\Environment$ interacts with dummy parties, who simply relay inputs and outputs between $\Environment$ and $\Functionality_{\varphi}$, and a simulator $\Simulator$.
We say that a protocol $\varphi$ securely realizes $\Functionality_{\varphi}$ if $\Environment$ cannot distinguish the real world from the ideal world, i.e., $\Environment$ cannot distinguish whether it is interacting with $\Adversary$ and parties running protocol $\varphi$ or with $\Simulator$ and dummy parties relaying to $\Functionality_{\varphi}$

A protocol $\varphi^{\FunctionalityG}$ securely realizes $\Functionality$ in the $\FunctionalityG$-hybrid model when $\varphi$ is allowed to invoke the ideal functionality $\FunctionalityG$. Therefore, for any protocol $\psi$ that securely realizes $\FunctionalityG$, the composed protocol $\varphi^{\psi}$, which is obtained by replacing each invocation of an instance of $\FunctionalityG$ with an invocation of an instance of $\psi$, securely realizes $\Functionality$.

In the ideal functionalities described in this paper, we consider static corruptions. When describing ideal functionalities, we use the conventions introduced in~\cite{DBLP:conf/crypto/CamenischDR16}, which are summarised
in~\ref{sec:securitymodel}.

\begin{description}[leftmargin=0pt]

\item[Interface Naming Convention.]
An ideal functionality can be invoked by using one or more interfaces. The name of a message in an interface consists of three fields separated by dots, e.g., $\fcecsetupini$ in $\Functionality_{\CEC}$ in~\S\ref{sec:idealfun}.
The first field indicates the name of the functionality and is the same in all interfaces of the functionality.
This field is useful for distinguishing between invocations of different functionalities in a hybrid protocol that uses two or more different functionalities.
The second field indicates the kind of action performed by the functionality and is the same in all messages that the functionality exchanges within the same interface.
The third field distinguishes between the messages that belong to the same interface, and can take the following different values.
 A message $\fcecsetupini$ is the incoming message received by the functionality, i.e., the message through which the interface is invoked.
 A message $\fcecsetupend$ is the outgoing message sent by the functionality, i.e., the message that ends the execution of the interface.
 The message $\fcecsetupsim$ is used by the functionality to send a message to $\Simulator$, and the message $\fcecsetuprep$ is used to receive a message from $\Simulator$.

\item[Network vs local communication.] The identity of an interactive Turing machine instance (ITI) consists of a party identifier $\pid$ and a session identifier $\sid$. A set of parties in an execution of a system of interactive Turing machines is a protocol instance if they have the same session identifier $\sid$.
    ITIs can pass direct inputs to and outputs from ``local'' ITIs that have the same $\pid$.
    An ideal functionality $\Functionality$ has $\pid=\bot$ and is considered local to all parties.     An instance of $\Functionality$ with the session identifier $\sid$ only accepts inputs from and passes outputs to machines with the same session identifier $\sid$. Some functionalities require the session identifier to have some structure. Those functionalities check whether the session identifier possesses the required structure in the first message that invokes the functionality. For the subsequent messages, the functionality implicitly checks that the session identifier equals the session identifier used in the first message.
    Communication between ITIs with different party identifiers must take place over the network.
    The network is controlled by $\Adversary$, meaning that he can arbitrarily delay, modify, drop, or insert messages.

\item[Query identifiers.] Some interfaces in a functionality can be invoked more than once. When the functionality sends a message $\fcecsetupsim$ to $\Simulator$ in such an interface, a query identifier $\qid$ is included in the message. The query identifier must also be included in the response $\fcecsetuprep$ sent by $\Simulator$. The query identifier is used to identify the message $\fcecsetupsim$ to which $\Simulator$ replies with a message $\fcecsetuprep$. We note that, typically, $\Simulator$ in the security proof may not be able to provide an immediate answer to the functionality after receiving a message $\fcecsetupsim$. The reason is that $\Simulator$ typically needs to interact with the copy of $\Adversary$ it runs in order to produce the message $\fcecsetuprep$, but $\Adversary$ may not provide the desired answer or may provide a delayed answer. In such cases, when the functionality sends more than one message $\fcecsetupsim$ to $\Simulator$, $\Simulator$ may provide delayed replies, and the order of those replies may not follow the order of the messages received.

\item[Aborts.]  When an ideal functionality $\Functionality$ aborts after being activated with a message sent by a party, we mean that $\Functionality$ halts the execution of its program and sends a special abortion message to the party that invoked the functionality. When an ideal functionality $\Functionality$ aborts after being activated with a message sent by $\Simulator$, we mean that $\Functionality$ halts the execution of its program and sends a special abortion message to the party that receives the outgoing message from $\Functionality$ after $\Functionality$ is activated by $\Simulator$.

\end{description}

%% file: 9DefinitionsIdealFunctionalities.tex
\section{Definitions of Ideal Functionalities}
\label{sec:functionalitiesFULL}

\subsection{Secure Message Transmission}
\label{sec:IdealFunctionalitySMT}

Our e-cash schemes use the functionality $\Functionality_{\SMT}$ for secure message transmission described in~\cite{DBLP:conf/focs/Canetti01}. $\Functionality_{\SMT}$ interacts with a sender $\Sender$ and a receiver $\Receiver$, and consists of one interface $\fsmtsend$. $\Sender$ uses the $\fsmtsend$ interface to send a message $\SMTmessage$ to $\Functionality_{\SMT}$. $\Functionality_{\SMT}$ leaks $\SMTfleakage(\SMTmessage)$, where $\SMTfleakage: \fsmtmessagespace \rightarrow \mathbb{N}$ is a function that leaks the message length, to the simulator $\Simulator$. After receiving a response from $\Simulator$, $\Functionality_{\SMT}$ sends $\SMTmessage$ to  $\Receiver$. $\Simulator$ cannot modify $\SMTmessage$. The session identifier $\sid$ contains the identities of $\Sender$ and $\Receiver$. 



\noindent \textbf{Ideal Functionality $\Functionality_{\SMT}$.} $\Functionality_{\SMT}$ is parameterized by a message space $\fsmtmessagespace$ and by a leakage function $\SMTfleakage: \fsmtmessagespace \rightarrow \mathbb{N}$, which leaks the message length.
\begin{enumerate}[leftmargin=*]

\item On input $(\fsmtsendini, \allowbreak \sid, \allowbreak \SMTmessage)$ from a party $\Sender$:

\begin{itemize}

\item Abort if $\sid \neq ( \Sender, \Receiver, \sid')$ or if $\SMTmessage \notin \fsmtmessagespace$.

\item Create a fresh $\ssid$ and store $(\ssid, \Receiver, \SMTmessage)$.

\item Send $(\fsmtsendsim, \allowbreak \sid, \allowbreak \ssid, \allowbreak \SMTfleakage(\SMTmessage))$ to $\Simulator$.

\end{itemize}

\item[S.] On input $(\fsmtsendrep, \allowbreak \sid, \allowbreak \ssid)$ from $\Simulator$:

\begin{itemize}

\item Abort if $(\ssid, \Receiver, \SMTmessage)$ is not stored.

\item Delete the record $(\ssid, \Receiver, \SMTmessage)$.

\item Send $(\fsmtsendend, \allowbreak \sid, \allowbreak \SMTmessage)$ to $\Receiver$.

\end{itemize}

\end{enumerate}

\subsection{Key Genearation}
\label{sec:funcKG}

In our e-cash constructions, the setup phase generates the parameters $\cecparams$ through algorithm $\cecSetup(1^{\securityparameter}, \allowbreak L)$. Moreover, algorithm  $\cecKeyGenA(\cecparams, \allowbreak t, \allowbreak n)$ generates a key pair for the Pointcheval-Sanders signature scheme in such a way that the shares of the secret key are given to $n$ authorities, so that $t \allowbreak \leq \allowbreak n$ authorities are needed to produce a signature.

To simplify our security analysis, in a manner similar to~\cite{cryptoeprint:2022:011}, we define an ideal functionality $\Functionality_{\KG}$ that runs both algorithms. In our construction in~\S\ref{sec:constructionCEC}, $\Functionality_{\KG}$ gives $\cecparams$ and the public keys $\spk$ and $(\spk_{\fcecAuthority_i})_{i\in[1,n]}$ to any party running the protocol, while each authority $\fcecAuthority_i$ also receives his secret key $\ssk_{\fcecAuthority_i}$.

$\Functionality_{\KG}$ could be replaced by an ideal functionality for distributed key generation (DKG)~\cite{DBLP:journals/iacr/KateHG12,cryptoeprint:2021:339}. DKG would avoid the need of a trusted party to generate the keys. With that replacement, our constructions would realize a modified version of our functionality in~\S\ref{sec:idealfunctionalityCEC}, where authorities cannot finalize the execution of the setup interface without involvement of other authorities. The security analysis of the remaining phases of our e-cash schemes is not affected by the fact that the authorities keys are generated by a trusted party or through a DKG protocol.

$\Functionality_{\KG}$ interacts with $n$ authorities $(\fkgAuthority_1, \ldots, \fkgAuthority_n)$. $\Functionality_{\KG}$ consists of two interfaces $\fkggetkey$ and $\fkgretrieve$. The interface $\fkggetkey$ is used by $\fkgAuthority_i$ to obtain its public key  $\spk_{\fcecAuthority_i}$ and secret key $\ssk_{\fcecAuthority_i}$, as well as the public key $\spk$ and the parameters $\cecparams$. The interface $\fkgretrieve$ is used by any party $\Party$ to obtain $\spk$, $\cecparams$ and the public keys $\langle \spk_{\fcecAuthority_i} \rangle_{i=1}^{n}$ of each of the authorities.


\noindent \textbf{Ideal Functionality $\Functionality_{\KG}$.}   $\Functionality_{\KG}$ is parameterized by probabilistic algorithms $\cecSetup$ and $\cecKeyGenA$, a security parameter $1^\securityparameter$, a threshold $t$ and a number $L$ of coins in a wallet.
\begin{enumerate}[leftmargin=*]

\item On input $(\fkggetkeyini, \sid)$ from an authority $\fkgAuthority_i$:

\begin{itemize}

\item Abort if $\sid \neq (\fkgAuthority_1, \ldots, \fkgAuthority_n, \sid')$, or if $n < t$.

\item If $(\sid, \cecparams, \allowbreak \spk, \allowbreak \langle \ssk_{\fcecAuthority_i}, \allowbreak \spk_{\fcecAuthority_i} \rangle_{i\in[1,n]})$ is not stored, run  $\cecparams \allowbreak \gets \allowbreak \cecSetup(1^{\securityparameter}, \allowbreak \allowbreak L)$ and $(\spk, \allowbreak \langle \ssk_{\fcecAuthority_i}, \allowbreak \spk_{\fcecAuthority_i} \rangle_{i\in[1,n]}) \allowbreak \gets \allowbreak \cecKeyGenA(\cecparams, \allowbreak t, \allowbreak n)$  and store $(\sid, \cecparams, \allowbreak \spk, \allowbreak \langle \ssk_{\fcecAuthority_i}, \allowbreak \spk_{\fcecAuthority_i} \rangle_{i\in[1,n]})$.

\item Create a fresh $\ssid$ and store $(\qid, \allowbreak \fkgAuthority_i)$.

\item Send $(\fkggetkeysim, \allowbreak \sid, \allowbreak \qid, \allowbreak \cecparams, \allowbreak \spk, \allowbreak \langle  \spk_{\fcecAuthority_i} \rangle_{i\in[1,n]})$ to $\Simulator$.

\end{itemize}

\item[S.] On input $(\fkggetkeyrep, \allowbreak \sid, \allowbreak \qid)$ from the simulator $\Simulator$: 

\begin{itemize}

\item Abort if $(\qid, \allowbreak \fkgAuthority_i)$ such that $\qid \allowbreak \neq \allowbreak \qid'$ is not stored.

\item Delete $(\qid, \allowbreak \fkgAuthority_i)$.

\item Send $(\fkggetkeyend, \allowbreak \sid, \allowbreak \cecparams, \allowbreak \spk, \allowbreak \ssk_{\fcecAuthority_i}, \allowbreak \spk_{\fcecAuthority_i})$ to $\fkgAuthority_i$.

\end{itemize}

\item On input $(\fkgretrieveini, \allowbreak \sid)$ from any party $\Party$:

\begin{itemize}

\item Abort if $\sid \neq (\fkgAuthority_1, \ldots, \fkgAuthority_n, \sid')$.

\item If $(\sid, \cecparams, \allowbreak \spk, \allowbreak \langle \ssk_{\fcecAuthority_i}, \allowbreak \spk_{\fcecAuthority_i} \rangle_{i\in[1,n]})$ is stored, set $\fkgvalue \gets  (\cecparams, \allowbreak \spk, \allowbreak \langle  \spk_{\fcecAuthority_i} \rangle_{i\in[1,n]})$, else set $\fkgvalue \gets \bot$.

\item Create a fresh $\ssid$ and store $(\ssid, \Party, \fkgvalue)$.

\item Send $(\fkgretrievesim, \allowbreak \sid, \allowbreak \ssid, \allowbreak \fkgvalue)$ to $\Simulator$.

\end{itemize}

\item[S.] On input $(\fkgretrieverep, \allowbreak \sid, \allowbreak \ssid)$ from  $\Simulator$:

\begin{itemize}

\item Abort if $(\ssid', \Party, \fkgvalue)$ such that $\ssid' \neq \ssid$ is not stored.

\item Delete the record $(\ssid, \Party, \fkgvalue)$.

\item Send $(\fkgretrieveend, \allowbreak \sid, \allowbreak \fkgvalue)$ to $\Party$.

\end{itemize}

\end{enumerate}


\subsection{Registration}
\label{sec:funcREG}

Our protocol uses the functionality $\Functionality_{\Freg}$ for key registration by Canetti~\cite{DBLP:conf/focs/Canetti01}. $\Functionality_{\Freg}$ interacts with any party $\FregT$ that registers a message $\fregvalue$ and with any party $\Party$ that retrieves the registered message. $\Functionality_{\Freg}$ consists of two interfaces $\fregregister$ and $\fregretrieve$. The interface $\fregregister$ is used by $\FregT$ to register a message $\fregvalue$ with $\Functionality_{\Freg}$. A party $\Party$ uses $\fregretrieve$ to retrieve $\fregvalue$ from $\Functionality_{\Freg}$.


\noindent \textbf{Ideal Functionality $\Functionality_{\Freg}$.} $\Functionality_{\Freg}$ is parameterized by a message space $\fregmessagespace$.
\begin{enumerate}[leftmargin=*]
\item On input $(\fregregisterini, \allowbreak \sid, \allowbreak \fregvalue)$ from a party $\FregT$: 

\begin{itemize}

\item Abort if $\sid \neq (\FregT, \sid')$, or if $\fregvalue \notin \fregmessagespace$ or if there is a tuple $(\sid,\fregvalue',0)$ stored.

\item Store $(\sid,\fregvalue,0)$.

\item Send $(\fregregistersim, \allowbreak \sid, \allowbreak \fregvalue)$ to $\Simulator$.

\end{itemize}

\item[S.] On input $(\fregregisterrep, \allowbreak \sid)$ from the simulator $\Simulator$: 

\begin{itemize}

\item Abort if $(\sid, \fregvalue, 0)$ is not stored or if $(\sid, \fregvalue, 1)$ is already stored.

\item Store $(\sid,\fregvalue,1)$ and parse $\sid$ as $(\FregT, \sid')$.

\item Send $(\fregregisterend, \allowbreak \sid)$ to $\FregT$.

\end{itemize}

\item On input $(\fregretrieveini, \allowbreak \sid)$ from any party $\Party$:

\begin{itemize}

\item If $(\sid,\fregvalue,1)$ is stored, set $\fregvalue' \gets \fregvalue$; else set $\fregvalue' \gets \bot$.

\item Create a fresh $\ssid$ and store $(\ssid, \Party, \fregvalue')$.

\item Send $(\fregretrievesim, \sid, \ssid, \fregvalue')$ to $\Simulator$.

\end{itemize}

\item[S.] On input $(\fregretrieverep, \sid, \ssid)$ from  $\Simulator$:

\begin{itemize}

\item Abort if $(\ssid, \Party, \fregvalue')$ is not stored.

\item Delete the record $(\ssid, \Party, \fregvalue')$.

\item Send $(\fregretrieveend, \sid, \fregvalue')$ to $\Party$.

\end{itemize}

\end{enumerate}

\subsection{Pseudonymous Channel}
\label{sec:funcNYM}

Our e-cash schemes use the functionality $\Functionality_{\NYM}$ for a secure idealized pseudonymous channel. We use $\Functionality_{\NYM}$ to describe our e-cash schemes for simplicity, in order to hide the details of real-world pseudonymous channels. $\Functionality_{\NYM}$ is similar to the functionality for anonymous secure message transmission in~\cite{DBLP:conf/esorics/CamenischLNR14}. $\Functionality_{\NYM}$ interacts with senders $\fnymSender$ and receivers $\fnymReceiver$.  $\Functionality_{\NYM}$ is parameterized by a message space $\fnymmessagespace$, a security parameter $\securityparameter$, a universe of pseudonyms $\fnymunivnym$, and a leakage function $\fnymleakage$, which leaks the message length. $\Functionality_{\NYM}$ consists of one interfaces $\fnymsend$. $\fnymSender$ uses the $\fnymsend$ interface to send a message $\fnymmessage \in \fnymmessagespace$, a pseudonym $\fnymnym \in \fnymunivnym$ and a receiver identifier $\fnymReceiver$ to $\Functionality_{\NYM}$. $\Functionality_{\NYM}$ sends $\fnymleakage(\fnymmessage)$ to the simulator $\Simulator$. After receiving a response from $\Simulator$, $\Functionality_{\NYM}$ sends $\fnymmessage$ and $\fnymnym$ to $\fnymReceiver$.


$\fnymReceiver$ does not learn the identifier $\fnymSender$. Instead $\fnymReceiver$ learns a pseudonym $\fnymnym$ chosen by $\fnymSender$. $\fnymSender$ can choose different pseudonyms to make the messages sent unlinkable towards $\fnymReceiver$.


\noindent \textbf{Ideal Functionality $\Functionality_{\NYM}$.} $\Functionality_{\NYM}$ is parameterized by a message space $\fnymmessagespace$, a security parameter $\securityparameter$, a universe of pseudonyms $\fnymunivnym$, and a leakage function $\fnymleakage$, which leaks the message length.
\begin{enumerate}[leftmargin=*]
\item On input $(\fnymsendini, \allowbreak \sid, \allowbreak \fnymmessage, \allowbreak \fnymnym, \allowbreak \fnymReceiver)$ from $\fnymSender$:

\begin{itemize}

\item Abort if $\fnymmessage \allowbreak \notin \allowbreak \fnymmessagespace$, or if $\fnymnym \allowbreak \notin \allowbreak \fnymunivnym$.

\item Create a fresh $\qid$ and store $(\qid, \allowbreak \fnymnym, \allowbreak \fnymSender, \allowbreak \fnymmessage, \allowbreak \fnymReceiver)$.

\item Send $(\fnymsendsim, \allowbreak \sid, \allowbreak \qid, \allowbreak \fnymleakage(\fnymmessage))$ to $\Simulator$.

\end{itemize}

\item[S.] On input $(\fnymsendrep, \allowbreak \sid, \allowbreak \qid)$ from $\Simulator$:

\begin{itemize}

\item Abort if $(\qid',  \allowbreak \fnymnym, \allowbreak \fnymSender, \allowbreak \fnymmessage, \allowbreak \fnymReceiver)$ such that $\qid = \qid'$ is not stored.


\item Delete the record $(\qid, \allowbreak \fnymnym, \allowbreak \fnymSender, \allowbreak \fnymmessage, \allowbreak \fnymReceiver)$.

\item Send $(\fnymsendend, \allowbreak \sid, \allowbreak \fnymmessage, \allowbreak \fnymnym)$ to $\fnymReceiver$.

\end{itemize}















\end{enumerate}


\subsection{Authenticated Bulletin Board}
\label{sec:funcBB}

Our e-cash schemes use the functionality $\Functionality_{\BB}$ for an authenticated bulletin board $\fbbbb$~\cite{DBLP:conf/tcc/Wikstrom04}. A $\fbbbb$ is used to store the payments deposited by providers, and to verify that there are not double spendings or double deposits. $\Functionality_{\BB}$ interacts with writers $\fbbWriter_j$ and readers $\fbbReader_k$. $\fbbWriter_j$ uses the $\fbbwrite$ interface to send a message $\fbbmessage$ to $\Functionality_{\BB}$. $\Functionality_{\BB}$ increments a counter $\fbbcounter$ of the number of messages stored in $\fbbbb$ and appends $[\fbbcounter, \allowbreak \fbbWriter_j, \allowbreak \fbbmessage]$ to $\fbbbb$. $\fbbReader_k$ uses the $\fbbgetbb$ interface on input an index $\fbbindex$. If $\fbbindex \allowbreak \in \allowbreak [1, \allowbreak \fbbcounter]$, $\Functionality_{\BB}$ takes the tuple $[\fbbindex, \allowbreak \fbbWriter_j, \allowbreak \fbbmessage]$ in $\fbbbb$ and sends $(\fbbWriter_j, \allowbreak \fbbmessage)$ to $\fbbReader_k$.


\noindent \textbf{Ideal Functionality $\Functionality_{\BB}$.} $\Functionality_{\BB}$ is parameterized by a universe of messages $\fbbunivmes$.  $\Functionality_{\BB}$ interacts with writers $\fbbWriter_j$ and readers $\fbbReader_k$.
\begin{enumerate}[leftmargin=*]

\item On input $(\fbbwriteini, \allowbreak \sid, \allowbreak \fbbmessage)$ from $\fbbWriter_j$:

\begin{itemize}


\item Abort if $\fbbmessage \allowbreak \notin \allowbreak \fbbunivmes$.

\item Create a fresh $\qid$ and store $(\qid, \allowbreak \fbbWriter_j, \allowbreak \fbbmessage)$.

\item Send $(\fbbwritesim, \allowbreak \sid, \allowbreak \qid)$ to $\Simulator$.

\end{itemize}

\item[S.] On input $(\fbbwriterep, \allowbreak \sid, \allowbreak \qid)$ from $\Simulator$:

\begin{itemize}

\item Abort if $(\qid', \allowbreak \fbbWriter_j, \allowbreak \fbbmessage)$ such that $\qid' \allowbreak = \allowbreak \qid$  is not stored.

\item If $(\sid, \allowbreak \fbbbb, \allowbreak \fbbcounter)$ is not stored, set $\fbbbb \gets \bot$ and $\fbbcounter \gets 0$.

\item Increment $\fbbcounter$, append $[\fbbcounter, \allowbreak \fbbWriter_j, \allowbreak \fbbmessage]$ to $\fbbbb$ and update $\fbbcounter$ and $\fbbbb$ in $(\sid, \allowbreak \fbbbb, \allowbreak \fbbcounter)$.

\item Delete $(\qid, \allowbreak \fbbWriter_j, \allowbreak \fbbmessage)$.

\item Send $(\fbbwriteend, \allowbreak \sid)$ to $\fbbWriter_j$.

\end{itemize}

\item On input $(\fbbgetbbini, \allowbreak \sid, \allowbreak \fbbindex)$ from $\fbbReader_k$:

\begin{itemize}

\item Create a fresh $\qid$ and store $(\qid, \allowbreak \fbbReader_k, \allowbreak \fbbindex)$.

\item Send $(\fbbgetbbsim, \allowbreak \sid, \allowbreak \qid)$ to $\Simulator$.

\end{itemize}

\item[S.] On input $(\fbbgetbbrep, \allowbreak \sid, \allowbreak \qid)$ from $\Simulator$:

\begin{itemize}

\item Abort if $(\qid', \fbbReader_k, \allowbreak \fbbindex)$ such that $\qid' = \qid$ is not stored.

\item If $(\sid, \allowbreak \fbbbb, \allowbreak \fbbcounter)$ is stored and $\fbbindex \in [1, \fbbcounter]$, take $[\fbbindex, \allowbreak \fbbWriter_j, \allowbreak \fbbmessage]$ from $\fbbbb$ and set $\fbbmessage' \gets (\fbbWriter_j, \allowbreak \fbbmessage)$, else set $\fbbmessage' \allowbreak \gets \allowbreak \bot$.

\item Send $(\fbbgetbbend, \allowbreak \sid, \allowbreak \fbbmessage')$ to $\fbbReader_k$.

\end{itemize}

\end{enumerate}


%% file: BSecurityAnalysisCompact.tex
\section{Security Proof for Our Compact E-Cash Scheme}
\label{sec:securityProofCompact}

To prove that construction $\mathrm{\Pi}_\CEC$, instantiated with the algorithms of the compact e-cash scheme in~\S\ref{sec:compactecash}, securely realizes the ideal functionality $\Functionality_{\CEC}$, we have to show that for any environment $\Environment$ and any adversary $\Adversary$ there exists a simulator $\Simulator$ such that $\Environment$ cannot distinguish between whether it is interacting with $\Adversary$ and the protocol in the real world or with $\Simulator$ and $\Functionality_{\CEC}$. The simulator thereby plays the role of all honest parties in the real world and interacts with $\Functionality_{\CEC}$ for all corrupt parties in the ideal world.

$\Simulator$ runs a copy of any adversary $\Adversary$, which is used to provide to $\Environment$ a view that is indistinguishable from the view given by $\Adversary$ in the real world. To achieve that, $\Simulator$ must simulate the real-world protocol towards the copy of $\Adversary$, in such a way that $\Adversary$ cannot distinguish an interaction with $\Simulator$ from an interaction with the real-world protocol. $\Simulator$ uses the information provided by $\Functionality_{\CEC}$ to provide a simulation of the real-world protocol.

Our simulator $\Simulator$ runs copies of the functionalities $\Functionality_{\SMT}$, $\Functionality_{\NYM}$, $\Functionality_{\KG}$, $\Functionality_{\Freg}$ and $\Functionality_{\BB}$. When any of the copies of these functionalities aborts, $\Simulator$ implicitly forwards the abortion message to the adversary if the functionality sends the abortion message to a corrupt party.

$\Simulator$ also runs copies of the extractors $\mathcal{E}_{s}$ and $\mathcal{E}_{v}$ and simulators  $\mathcal{S}_{s}$ and $\mathcal{S}_{v}$ for the non-interactive zero-knowledge arguments of knowledge $\pi_s$ and $\pi_v$, which are computed through the Fiat-Shamir transform. We remark that they involve calls to the random oracle and rewinding of the adversary. For simplicity, we omit those details.

\paragraph{Challenges in the proof.} Our compact e-cash scheme with threshold issuance is based on an existing compact e-cash scheme~\cite{DBLP:conf/eurocrypt/CamenischHL05}. We instantiate that scheme with the Pointcheval-Sanders signature scheme, and then we use an existing threshold issuance protocol for Pointcheval-Sanders signatures~\cite{cryptoeprint:2022:011} to provide threshold issuance. Nevertheless, the security proof does not follow straightforwardly from the proofs in~\cite{DBLP:conf/eurocrypt/CamenischHL05,cryptoeprint:2022:011}. 

First, in the blind issuance protocol of the compact e-cash scheme in~\cite{DBLP:conf/eurocrypt/CamenischHL05} one of the coin secrets is chosen jointly between the user and the bank, i.e. both the bank and the user add randomness to compute the secret. In a threshold issuance setting, it would be necessary that all the authorities add the same randomness without communicating between them. While adding a mechanism to do that is possible, in order to improve efficiency we decided to have the secret be chosen by the user only. This implies that, when a corrupt user picks up the secrets of two wallets, the user could pick up the secrets in such a way that two coins have the same serial number. Consequently, when those coins are spent, there is a situation in which double spending is detected (because serial numbers are the same), although no double spending has happened. In that case, we prove that the identification algorithm cannot output the public key of an honest user, and we are able to do that under the discrete logarithm assumption. 

Second, to improve efficiency, we use only one coin secret instead of the two coins secrets used in [14]. When using two coin secrets, it is straightforward to show that serial numbers and double spending tags do not reveal any information about the secrets by using the pseudorandomness property of the pseudorandom function. However, when using the same secret for both, but a different generator, we need to add an additional reduction to the XDH assumption. 


\paragraph{Simulator.} We describe the simulator $\Simulator$ for the case in which a subset of users $\fcecUser_j$, a subset of providers $\fcecProvider_k$ and up to $t-1$ authorities are corrupt. $\Simulator$ simulates the honest parties in the protocol $\mathrm{\Pi_{\CEC}}$ and runs copies of the ideal functionalities involved.

\begin{description}[leftmargin=*]

    \item[Honest authority $\fcecAuthority_i$ starts setup.] When the functionality $\Functionality_{\CEC}$ sends the message $(\fcecsetupsim, \allowbreak \sid, \allowbreak \fcecAuthority_i)$,  the simulator $\Simulator$ runs a copy of the functionality $\Functionality_{\KG}$ on input $(\fkggetkeyini, \allowbreak \sid)$. When the functionality $\Functionality_{\KG}$ sends the message $(\fkggetkeysim, \allowbreak \sid, \allowbreak \qid, \allowbreak \cecparams, \allowbreak \spk, \allowbreak \langle  \spk_{\fcecAuthority_i} \rangle_{i\in[1,n]})$, $\Simulator$ forwards that message to $\Adversary$.

    \item[Honest authority $\fcecAuthority_i$ ends setup.] When the adversary $\Adversary$ sends the message $(\fkggetkeyrep, \allowbreak \sid, \allowbreak \qid)$, $\Simulator$ runs a copy of $\Functionality_{\KG}$ on input that message. When $\Functionality_{\KG}$ sends $(\fkggetkeyend, \allowbreak \sid, \allowbreak \cecparams, \allowbreak \spk, \allowbreak \ssk_{\fcecAuthority_i}, \allowbreak \spk_{\fcecAuthority_i})$, $\Simulator$ sends $(\fcecsetuprep, \allowbreak \sid, \allowbreak \fcecAuthority_i)$ to $\Functionality_{\CEC}$.

    \item[Corrupt authority $\tilde{\fcecAuthority}_i$ starts setup.] When a corrupt authority $\tilde{\fcecAuthority}_i$ sends the message $(\fkggetkeyini, \allowbreak \sid)$, $\Simulator$ runs a copy of $\Functionality_{\KG}$ on input that message. When $\Functionality_{\KG}$ sends $(\fkggetkeysim, \allowbreak \sid, \allowbreak \qid, \allowbreak \cecparams, \allowbreak \spk, \allowbreak \langle  \spk_{\fcecAuthority_i} \rangle_{i\in[1,n]})$, $\Simulator$ forwards that message to $\tilde{\fcecAuthority}_i$.

    \item[Corrupt authority $\tilde{\fcecAuthority}_i$ ends setup.] When  $\tilde{\fcecAuthority}_i$ sends the message $(\fkggetkeyrep, \allowbreak \sid, \allowbreak \qid)$, $\Simulator$ runs a copy of $\Functionality_{\KG}$ on input that message. When $\Functionality_{\KG}$ sends $(\fkggetkeyend, \allowbreak \sid, \allowbreak \cecparams, \allowbreak \spk, \allowbreak \ssk_{\tilde{\fcecAuthority}_i}, \allowbreak \spk_{\tilde{\fcecAuthority}_i})$, $\Simulator$ sends $(\fcecsetupini, \allowbreak \sid)$ to the functionality $\Functionality_{\CEC}$. When $\Functionality_{\CEC}$ sends $(\fcecsetupsim, \allowbreak \sid, \allowbreak \tilde{\fcecAuthority}_i)$, the simulator $\Simulator$ sends the message $(\fcecsetuprep, \allowbreak \sid, \allowbreak \tilde{\fcecAuthority}_i)$ to $\Functionality_{\CEC}$. When $\Functionality_{\CEC}$ sends $(\fcecsetupend, \allowbreak \sid)$, $\Simulator$ sends the message $(\fkggetkeyend, \allowbreak \sid, \allowbreak \cecparams, \allowbreak \spk, \allowbreak \ssk_{\tilde{\fcecAuthority}_i}, \allowbreak \spk_{\tilde{\fcecAuthority}_i})$ to $\tilde{\fcecAuthority}_i$.

    \item[Honest user (or provider) requests keys.] When $\Functionality_{\CEC}$ sends the message $(\fcecregistersim, \allowbreak \sid, \allowbreak \fcecUser_j)$, $\Simulator$ runs a copy of $\Functionality_{\KG}$ on input $(\fkgretrieveini, \allowbreak \sid)$. When the functionality $\Functionality_{\KG}$ sends the message $(\fkgretrievesim, \allowbreak \sid, \allowbreak \ssid, \allowbreak \fkgvalue)$, $\Simulator$ forwards that message to $\Adversary$.

    \item[Honest user (or provider) receives and registers keys.] When $\Adversary$ sends $(\fkgretrieverep, \allowbreak \sid, \allowbreak \ssid)$, $\Simulator$ runs $\Functionality_{\KG}$ on input that message. When the functionality $\Functionality_{\KG}$ sends $(\fkgretrieveend, \allowbreak \sid, \allowbreak \fkgvalue)$, the simulator $\Simulator$ runs $(\ssk_{\fcecUser_j}, \allowbreak \spk_{\fcecUser_j}) \gets \cecKeyGenU(\cecparams)$, sets $\sid_{\Freg} \gets (\fcecUser_j, \sid')$ and runs a copy of $\Functionality_{\Freg}$ on input the message $(\fregregisterini, \allowbreak \sid_{\Freg}, \allowbreak \spk_{\fcecUser_j})$. When $\Functionality_{\Freg}$ sends the message $(\fregregistersim, \allowbreak \sid_{\Freg}, \allowbreak \spk_{\fcecUser_j})$, $\Simulator$ forwards that message to $\Adversary$.

    \item[Honest user (or provider) finalizes the registration of keys.] When $\Adversary$ sends the message $(\fregregisterrep, \allowbreak \sid_{\Freg})$, $\Simulator$ runs $\Functionality_{\Freg}$ on input that message. When $\Functionality_{\Freg}$ sends $(\fregregisterend, \allowbreak \sid_{\Freg})$, $\Simulator$ sends $(\fcecregisterrep, \allowbreak \sid, \allowbreak \fcecUser_j)$ to $\Functionality_{\CEC}$.

    \item[Corrupt user (or provider) requests keys.] When $\Adversary$ sends the message $(\fkgretrieveini, \allowbreak \sid)$, $\Simulator$ runs $\Functionality_{\KG}$ on input that message. When $\Functionality_{\KG}$ sends $(\fkgretrievesim, \allowbreak \sid, \allowbreak \ssid, \allowbreak \fkgvalue)$, $\Simulator$ forwards that message to $\Adversary$.

    \item[Corrupt user (or provider) receives keys.]  When $\Adversary$ sends the message $(\fkgretrieverep, \allowbreak \sid, \allowbreak \ssid)$, $\Simulator$ runs $\Functionality_{\KG}$ on input that message. When $\Functionality_{\KG}$ sends $(\fkgretrieveend, \allowbreak \sid, \allowbreak \fkgvalue)$, $\Simulator$ forwards that message to $\Adversary$.

    \item[Corrupt user (or provider) registers keys.] When $\Adversary$ sends the message $(\fregregisterini, \allowbreak \sid_{\Freg}, \allowbreak \spk_{\fcecUser_j})$, $\Simulator$ runs $\Functionality_{\Freg}$ on input that message. When the functionality $\Functionality_{\Freg}$ sends $(\fregregistersim, \allowbreak \sid_{\Freg}, \allowbreak \spk_{\fcecUser_j})$, $\Simulator$ forwards that message to $\Adversary$.

    \item[Corrupt user (or provider) finishes the registration of keys.]  When $\Adversary$ sends $(\fregregisterrep, \allowbreak \sid_{\Freg})$, $\Simulator$ runs $\Functionality_{\Freg}$ on input that message. When $\Functionality_{\Freg}$ sends $(\fregregisterend, \allowbreak \sid_{\Freg})$, the simulator $\Simulator$, acting as the corrupt user $\fcecUser_j$, sends $(\fcecregisterini, \allowbreak \sid)$ to  $\Functionality_{\CEC}$. When the functionality $\Functionality_{\CEC}$ sends $(\fcecregistersim, \allowbreak \sid, \allowbreak \fcecUser_j)$, the simulator $\Simulator$ sends $(\fcecregisterrep, \allowbreak \sid, \allowbreak \fcecUser_j)$ to $\Functionality_{\CEC}$. When $\Functionality_{\CEC}$ sends the message $(\fcecregisterend, \allowbreak \sid)$, $\Simulator$  sends $(\fregregisterend, \allowbreak \sid_{\Freg})$ to $\Adversary$.

    \item[Honest user sends a request to honest authority.]  When $\Functionality_{\CEC}$ sends $(\fcecrequestsim, \allowbreak \sid, \allowbreak \qid, \allowbreak \fcecUser_j, \allowbreak \fcecAuthority_i)$, the simulator $\Simulator$ parses the session identifier $\sid$ as $(\fcecAuthority_1, \allowbreak \ldots, \allowbreak \fcecAuthority_n, \allowbreak \sid')$, sets $\sid_{\SMT} \allowbreak \gets \allowbreak (\fcecUser_j, \allowbreak \fcecAuthority_i, \allowbreak \sid')$ and sends $(\fsmtsendsim, \allowbreak \sid_{\SMT}, \ssid, \allowbreak \SMTfleakage(\SMTmessage))$ to $\Adversary$, where $\SMTfleakage(\SMTmessage)$ is equal to the length of the message that the honest user sends to an authority in the $\fcecrequest$ interface.

    \item[Adversary forwards request.] When the adversary $\Adversary$ sends the message $(\fsmtsendrep, \allowbreak \sid_{\SMT}, \allowbreak \ssid)$ to reply to $(\fcecrequestsim, \allowbreak \sid, \allowbreak \qid, \allowbreak \fcecUser_j, \allowbreak \fcecAuthority_i)$, $\Simulator$ checks if a tuple $(\sid, \allowbreak \fcecUser_j, \allowbreak \fcecAuthority_i)$ is stored. If not, $\Simulator$ runs the procedure in Figure~\ref{fig:proc4}. After that, $\Simulator$ sends the message $(\fcecrequestrep, \allowbreak \sid, \allowbreak \qid)$ to $\Functionality_{\CEC}$.

    \item[Honest user sends a request to corrupt authority.]  When $\Functionality_{\CEC}$ sends $(\fcecrequestsim, \allowbreak \sid, \allowbreak \qid, \allowbreak \fcecUser_j, \allowbreak \fcecAuthority_i, \allowbreak \fcecwalletid)$, $\Simulator$ checks if a tuple $(\sid, \allowbreak \fcecUser'_j, \allowbreak \fcecwalletid', \allowbreak  \fcecwalletnum)$ such that $\fcecUser'_j \allowbreak = \allowbreak \fcecUser_j$ and $\fcecwalletid' \allowbreak = \allowbreak \fcecwalletid$ is stored. If not, $\Simulator$ does the following:
    \begin{itemize}

        \item Pick up the stored tuple $(\sid, \allowbreak \fcecUser'_j, \allowbreak \fcecwalletcount)$ such that $\fcecUser'_j \allowbreak = \allowbreak \fcecUser_j$. (This tuple is initialized to $\fcecwalletcount \allowbreak = \allowbreak 0$.)

        \item Set $\fcecwalletcount \allowbreak \gets \allowbreak \fcecwalletcount \allowbreak + \allowbreak 1$.

        \item Set $\fcecwalletnum \allowbreak \gets \allowbreak \fcecwalletcount$.

        \item Store $(\sid, \allowbreak \fcecUser_j, \allowbreak \fcecwalletid, \allowbreak  \fcecwalletnum)$.

        \item Update $\fcecwalletcount$ in the tuple $(\sid, \allowbreak \fcecUser_j, \allowbreak \fcecwalletcount)$.

    \end{itemize}
    $\Simulator$ parses the session identifier $\sid$ as $(\fcecAuthority_1, \allowbreak \ldots, \allowbreak \fcecAuthority_n, \allowbreak \sid')$, sets $\sid_{\SMT} \allowbreak \gets \allowbreak (\fcecUser_j,\fcecAuthority_i,\sid')$ and sends $(\fsmtsendsim, \allowbreak \sid_{\SMT}, \ssid, \allowbreak \SMTfleakage(\SMTmessage))$ to $\Adversary$, where $\SMTfleakage(\SMTmessage)$ is equal to the length of the message that the honest user sends to an authority in the $\fcecrequest$ interface.

    \item[Corrupt authority receives request from honest user.] When $\Functionality_{\CEC}$ sends $(\fcecrequestend, \allowbreak \sid, \allowbreak \fcecUser_j, \allowbreak \fcecrequestid)$ to a corrupt authority $\tilde{\fcecAuthority}_i$, $\Simulator$ retrieves the stored tuple $(\sid, \allowbreak \fcecUser_j, \allowbreak \fcecwalletid, \allowbreak  \fcecwalletnum)$ such that $\fcecUser'_j \allowbreak = \allowbreak \fcecUser_j$ and $\fcecwalletid$ was received before in the message $(\fcecrequestsim, \ldots)$ associated with the message $(\fcecrequestend, \ldots)$ received now. $\Simulator$ runs a copy of a user on input $(\fcecrequestini, \allowbreak \sid, \allowbreak \tilde{\fcecAuthority}_i, \allowbreak \fcecrequestid, \allowbreak \fcecwalletnum)$. (We remark that $\fcecwalletnum$ used a input to the copy of the user is not necessarily equal to the value of $\fcecwalletnum$ received by the honest user from the environment. However, it is enough to ensure that the same $\fcecwalletnum$ is used for each $\fcecwalletid$ received from $\Functionality_{\CEC}$, so that the copy of the user produces the same request for each $\fcecwalletid$.) When the copy of the user runs algorithm $\cecRequest$, $\Simulator$ uses the public key $\spk_{\fcecUser_j}$ that was previously computed for user $\fcecUser_j$. $\Simulator$ also uses the simulator $\mathcal{S}_s$ to compute a simulated proof $\pi_s$, and sets $\com$, $\com_{1}$ and $\com_{2}$ to be commitments to random messages. When the copy of the user sends the message $(\fsmtsendini, \allowbreak \sid_{\SMT}, \allowbreak \langle \fcecrequestid, \allowbreak \cecrequest \rangle)$ to $\Functionality_{\SMT}$, $\Simulator$ runs $\Functionality_{\SMT}$ on input that message. When $\Functionality_{\SMT}$ sends the message $(\fsmtsendsim, \allowbreak \sid, \allowbreak \ssid, \allowbreak \SMTfleakage(\SMTmessage))$, $\Simulator$ runs $\Functionality_{\SMT}$ on input $(\fsmtsendrep, \allowbreak \sid, \allowbreak \ssid)$. When $\Functionality_{\SMT}$ sends $(\fsmtsendend, \allowbreak \sid, \allowbreak \langle \fcecrequestid, \allowbreak \cecrequest \rangle)$, $\Simulator$ forwards that message to $\tilde{\fcecAuthority}_i$.

    \item[Corrupt authority requests user keys.] When the adversary $\Adversary$ sends $(\fregretrieveini, \allowbreak \sid_{\Freg})$, $\Simulator$ runs a copy of $\Functionality_{\Freg}$ on input that message. When $\Functionality_{\Freg}$ sends $(\fregregistersim, \allowbreak \sid, \allowbreak \fregvalue)$, $\Simulator$ forwards that message to $\Adversary$.

    \item[Corrupt authority receives user keys.] When the adversary $\Adversary$ sends $(\fregregisterrep, \allowbreak \sid)$, $\Simulator$ runs $\Functionality_{\Freg}$ on input that message. When $\Functionality_{\Freg}$ sends $(\fregretrieveend, \allowbreak \sid_{\Freg}, \allowbreak \fregvalue)$, $\Simulator$ forwards that message to $\Adversary$.

    \item[Corrupt user requests credential.] When the adversary $\Adversary$ sends the message $(\fsmtsendini, \allowbreak \sid_{\SMT}, \allowbreak \langle \fcecrequestid, \allowbreak \cecrequest \rangle)$, $\Simulator$ runs $\Functionality_{\SMT}$ on input that message. When $\Functionality_{\SMT}$ sends $(\fsmtsendsim, \allowbreak \sid_{\SMT}, \allowbreak \ssid, \allowbreak \SMTfleakage(\SMTmessage))$, $\Simulator$ sends that message to $\Adversary$.

      \begin{figure}
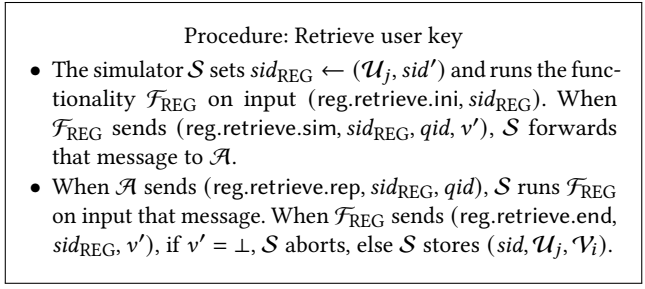

    \begin{framed}
    \begin{center}
    Procedure: Retrieve user key
    \end{center}

      \begin{itemize}[leftmargin=*]

            \item The simulator $\Simulator$ sets $\sid_{\Freg} \allowbreak \gets \allowbreak (\fcecUser_j, \allowbreak \sid')$ and runs the functionality $\Functionality_{\Freg}$ on input $(\fregretrieveini, \allowbreak \sid_{\Freg})$. When $\Functionality_{\Freg}$ sends
            $(\fregretrievesim, \allowbreak \sid_{\Freg}, \allowbreak \ssid, \allowbreak \fregvalue')$, $\Simulator$ forwards that message to $\Adversary$.

            \item When $\Adversary$ sends $(\fregretrieverep, \allowbreak \sid_{\Freg}, \allowbreak \ssid)$, $\Simulator$ runs $\Functionality_{\Freg}$ on input that message. When $\Functionality_{\Freg}$ sends $(\fregretrieveend, \allowbreak \sid_{\Freg}, \allowbreak \fregvalue')$, if $\fregvalue' = \bot$, $\Simulator$ aborts, else $\Simulator$ stores $(\sid, \allowbreak \fcecUser_j, \allowbreak \fcecAuthority_i)$.

        \end{itemize}

       \end{framed}
    \caption{Procedure for simulating user key retrieval.}
    \label{fig:proc4}
    \end{figure}

    \item[Honest authority receives request from corrupt user.] When \newline the adversary $\Adversary$ sends $(\fsmtsendrep, \allowbreak \sid_{\SMT}, \allowbreak \ssid)$, $\Simulator$ runs $\Functionality_{\SMT}$ on input that message. When $\Functionality_{\SMT}$ sends the message $(\fsmtsendend, \allowbreak \sid_{\SMT}, \allowbreak \SMTmessage)$, $\Simulator$ parses the message $\SMTmessage$ as $\langle \fcecrequestid, \allowbreak \cecrequest \rangle$ and $\sid_{\SMT}$ as $(\fcecUser_j, \allowbreak \fcecAuthority_i, \allowbreak \sid')$. $\Simulator$ does the following:
    \begin{itemize}

        \item Abort if the authority $\fcecAuthority_i$ did not end the setup.

        \item Check if a tuple $(\sid, \allowbreak \fcecUser_j, \allowbreak \fcecAuthority_i)$ is stored. If not, $\Simulator$ runs the procedure in Figure~\ref{fig:proc4}.

        \item Abort if there is a stored tuple $(\fcecUser'_j, \allowbreak \fcecwalletnum, \allowbreak \fcecsetrequestid, \allowbreak \cecrequest, \allowbreak \langle \cecmes_{1}, \allowbreak \cecmes_{2}, \allowbreak o, \allowbreak o_{1}, \allowbreak o_{2}  \rangle)$ such that $\fcecUser'_j \allowbreak = \allowbreak \fcecUser_j$ and $\{\fcecrequestid, \fcecAuthority_i\} \allowbreak \in \allowbreak \fcecsetrequestid$. Here $\Simulator$ aborts because there is a request pending with the same request identifier $\fcecrequestid$ from $\fcecUser_j$ for $\fcecAuthority_i$, like it is done in the real protocol.

        \item If there is a stored tuple $(\fcecUser'_j, \allowbreak \fcecwalletnum, \allowbreak \fcecsetrequestid, \allowbreak \cecrequest', \allowbreak \langle \cecmes_{1}, \allowbreak \cecmes_{2}, \allowbreak o, \allowbreak o_{1}, \allowbreak o_{2}  \rangle)$ such that $\fcecUser'_j \allowbreak = \allowbreak \fcecUser_j$ and $\cecrequest' \allowbreak = \allowbreak \cecrequest$, then update the set $\fcecsetrequestid \allowbreak \gets \allowbreak \fcecsetrequestid \cup \{\fcecrequestid, \fcecAuthority_i\}$ in that tuple and take $\fcecwalletnum$ from that tuple. Else do the following:

        \begin{itemize}

            \item Follow the steps of an honest authority in $\mathrm{\Pi}_{\CEC}$ to verify $\cecrequest$. This involves verifying the request by running the algorithm $b \allowbreak \gets \allowbreak \cecRequestVf(\cecparams, \allowbreak \cecrequest, \allowbreak \spk_{\fcecUser_j})$. ($\Simulator$ checked before that the corrupt user registered $\spk_{\fcecUser_j}$  through the procedure in Figure~\ref{fig:proc4}.) Abort if $b \allowbreak = \allowbreak 0$.

            \item Parse $\cecrequest$ as $(\h, \allowbreak \com, \allowbreak \com_{1}, \allowbreak \com_{2}, \allowbreak \pi_s)$. Run the extractor $\mathcal{E}_{s}$ to extract the witness $\langle \cecmes_{1}, \allowbreak \cecmes_{2}, \allowbreak o, \allowbreak o_{1}, \allowbreak o_{2} \rangle$ from $\pi_s$.

            \item If there is a tuple stored $(\fcecUser_j, \allowbreak \fcecwalletnum, \allowbreak \fcecsetrequestid, \allowbreak \langle \h, \allowbreak \com', \allowbreak \com_{1}, \allowbreak \com_{2}, \allowbreak \pi_s \rangle, \langle \cecmes'_{1}, \allowbreak \cecmes'_{2}, \allowbreak o, \allowbreak o_{1}, \allowbreak o_{2}  \rangle)$  such that $\com' \allowbreak = \allowbreak \com$ but $(\cecmes'_{1}, \allowbreak \cecmes'_{2}) \allowbreak \neq \allowbreak  (\cecmes_{1}, \allowbreak \cecmes_{2})$, $\Simulator$ outputs failure.

            \item If there is a stored tuple $(\fcecUser_j, \allowbreak \fcecwalletnum, \allowbreak \fcecsetrequestid, \allowbreak \cecrequest, \allowbreak \langle \cecmes_{1}, \allowbreak \cecmes_{2}, \allowbreak o, \allowbreak o_{1}, \allowbreak o_{2}  \rangle)$ such that $\cecrequest \allowbreak = \allowbreak \allowbreak \langle \h, \allowbreak \com', \allowbreak \com_{1}, \allowbreak \com_{2}, \allowbreak \pi_s \rangle$ and $\com' \allowbreak = \allowbreak \com$, then take $\fcecwalletnum$ from that tuple and store a tuple $(\fcecUser_j, \allowbreak \fcecwalletnum, \allowbreak \fcecsetrequestid, \allowbreak \cecrequest, \allowbreak \langle \cecmes_{1}, \allowbreak \cecmes_{2}, \allowbreak o, \allowbreak o_{1}, \allowbreak o_{2}  \rangle)$, where $\fcecsetrequestid \allowbreak \gets \allowbreak \{\fcecrequestid, \allowbreak \fcecAuthority_i\}$ and $\langle \cecmes_{1}, \allowbreak \cecmes_{2}, \allowbreak o, \allowbreak o_{1}, \allowbreak o_{2}  \rangle$ is the witness that was output by the extractor $\mathcal{E}_{s}$. (We remark that $\fcecwalletnum$ is the same because, if $\com$ is the same in two requests, the partial signatures obtained from two different authorities can be aggregated, even if the values $(\com_{1}, \allowbreak \com_{2}, \allowbreak  \allowbreak \pi_s)$ are different.) Else do the following:

            \begin{itemize}

                \item Pick up the stored tuple $(\sid, \allowbreak \fcecUser'_j, \allowbreak \fcecwalletcount)$ such that $\fcecUser'_j \allowbreak = \allowbreak \fcecUser_j$. (This tuple is initialized to $\fcecwalletcount \allowbreak = \allowbreak 0$.)

                \item Set $\fcecwalletcount \allowbreak \gets \allowbreak \fcecwalletcount \allowbreak + \allowbreak 1$.

                \item Set $\fcecwalletnum \allowbreak \gets \allowbreak \fcecwalletcount$.

                \item Store $(\fcecUser_j, \allowbreak \fcecwalletnum, \allowbreak \fcecsetrequestid, \allowbreak \cecrequest, \allowbreak \langle \cecmes_{1}, \allowbreak \cecmes_{2}, \allowbreak o, \allowbreak o_{1}, \allowbreak o_{2}  \rangle)$, where $\fcecsetrequestid \allowbreak \gets \allowbreak \{\fcecrequestid, \fcecAuthority_i\}$.

                \item Update $\fcecwalletcount$ in the tuple $(\sid, \allowbreak \fcecUser_j, \allowbreak \fcecwalletcount)$.

            \end{itemize}

        \end{itemize}

        $\Simulator$ sends $(\fcecrequestini, \allowbreak \sid, \allowbreak \fcecAuthority_i, \allowbreak \fcecrequestid, \allowbreak \fcecwalletnum)$ to $\Functionality_{\CEC}$. When $\Functionality_{\CEC}$ sends $(\fcecrequestsim, \allowbreak \sid, \allowbreak \qid, \allowbreak \fcecUser_j, \allowbreak \fcecAuthority_i)$, $\Simulator$ sends the message $(\fcecrequestrep, \allowbreak \sid, \allowbreak \qid)$ to $\Functionality_{\CEC}$.

    \end{itemize}

    \item[Honest authority issues attribute.] When the functionality $\Functionality_{\CEC}$ sends the message $(\fcecissuesim, \allowbreak \sid, \allowbreak \qid, \allowbreak \fcecAuthority_i, \allowbreak \fcecUser_j)$, $\Simulator$ parses $\sid$ as $(\fcecAuthority_1, \allowbreak \ldots, \allowbreak \fcecAuthority_n, \allowbreak \sid')$, sets $\sid_{\SMT} \allowbreak \gets \allowbreak (\fcecAuthority_i, \allowbreak \fcecUser_j, \allowbreak \sid')$ and sends the message $(\fsmtsendsim, \allowbreak \sid_{\SMT}, \allowbreak \ssid, \allowbreak \SMTfleakage(\SMTmessage))$, where $\SMTfleakage(\SMTmessage)$ is the length of the message that an honest authority sends in the $\fcecissue$ interface.

    \item[Adversary forwards issuance.] When the adversary $\Adversary$ sends the message $(\fsmtsendrep, \allowbreak \sid_{\SMT}, \allowbreak \ssid)$, $\Simulator$ sends $(\fcecissuerep, \allowbreak \sid, \allowbreak \qid)$ to $\Functionality_{\CEC}$.

    \item[Corrupt user receives issuance.] When $\Functionality_{\CEC}$ sends the message $(\fcecissueend, \allowbreak \sid, \allowbreak \fcecrequestid, \allowbreak \fcecAuthority_i)$ to a corrupt user $\fcecUser_j$, the simulator $\Simulator$ finds the stored tuple $(\fcecUser'_j, \allowbreak \fcecwalletnum, \allowbreak \fcecsetrequestid, \allowbreak \cecrequest, \allowbreak \langle \cecmes_{1}, \allowbreak \cecmes_{2}, \allowbreak o, \allowbreak o_{1}, \allowbreak o_{2}  \rangle)$ such that $\fcecUser'_j \allowbreak = \allowbreak \fcecUser_j$ and there is element $\{\fcecrequestid', \allowbreak \fcecAuthority'_i\} \allowbreak \in \allowbreak \fcecsetrequestid$ such that $\fcecrequestid' \allowbreak = \allowbreak \fcecrequestid$ and $\fcecAuthority'_i \allowbreak = \allowbreak \fcecAuthority_i$. $\Simulator$ parses $\cecrequest$ as $(\h, \allowbreak \com, \allowbreak \com_{1}, \allowbreak \com_{2}, \allowbreak \pi_s)$. $\Simulator$ parses the secret key $\ssk_{\fcecAuthority_i}$ as $(x_i, \allowbreak y_{i,1}, \allowbreak y_{i,2})$, computes $c = \h^{x_i} \prod_{j=1}^{2} \com_j^{y_{i,j}}$ and sets the blinded signature share $\hat{\sigma}_{i} \allowbreak \gets \allowbreak (\h, \allowbreak c)$ as in $\mathrm{\Pi}_{\CEC}$. If a tuple $(\sid, \allowbreak \fcecUser'_j, \allowbreak \fcecwalletnum', \allowbreak \cecset)$ such that $\fcecUser'_j \allowbreak = \allowbreak \fcecUser_j$ and $\fcecwalletnum' \allowbreak = \allowbreak \fcecwalletnum$ is not stored, $\Simulator$ stores a tuple $(\sid, \allowbreak \fcecUser_j, \allowbreak \fcecwalletnum, \allowbreak \{i\})$, else updates $\cecset \allowbreak \gets \allowbreak \cecset \cup \{i\}$ in the tuple $(\sid, \allowbreak \fcecUser_j, \allowbreak \fcecwalletnum, \allowbreak \cecset)$. After that, $\Simulator$ removes  $\{\fcecrequestid, \allowbreak \fcecAuthority_i\}$ from the set $\fcecsetrequestid$ in the tuple $(\fcecUser_j, \allowbreak \fcecwalletnum, \allowbreak \fcecsetrequestid, \allowbreak \cecrequest, \allowbreak \langle \cecmes_{1}, \allowbreak \cecmes_{2}, \allowbreak o, \allowbreak o_{1}, \allowbreak o_{2}  \rangle)$. Finally, $\Simulator$ sets $\sid_{\SMT} \allowbreak \gets \allowbreak (\fcecAuthority_i, \allowbreak \fcecUser_j, \allowbreak \sid')$, sets $\cecresponse \gets \hat{\sigma}_{i}$ and sends $(\fsmtsendend, \allowbreak \sid_{\SMT}, \allowbreak \langle \fcecrequestid, \allowbreak \cecresponse \rangle)$ to $\Adversary$.

    \item[Corrupt authority issues attribute.] When a corrupt authority $\tilde{\fcecAuthority}_i$ sends the message $(\fsmtsendini, \allowbreak \sid_{\SMT}, \allowbreak \langle \fcecrequestid, \allowbreak \cecresponse \rangle)$, $\Simulator$ runs $\Functionality_{\SMT}$ on input that message. When the functionality $\Functionality_{\SMT}$ sends $(\fsmtsendsim, \allowbreak \sid, \allowbreak \ssid, \allowbreak \SMTfleakage(\SMTmessage))$, $\Simulator$ sends that message to $\Adversary$.

    \item[Honest user receives issuance from corrupt authority.] After the adversary $\Adversary$ sends the message $(\fsmtsendrep, \allowbreak \sid, \allowbreak \ssid)$, the simulator $\Simulator$ runs $\Functionality_{\SMT}$ on input that message. When  $\Functionality_{\SMT}$ sends $(\fsmtsendend, \allowbreak \sid_{\SMT}, \allowbreak \langle \fcecrequestid, \allowbreak \cecresponse \rangle)$, the simulator $\Simulator$  parses $\sid_{\SMT}$ as $(\fcecAuthority_i, \allowbreak \fcecUser_j, \allowbreak \sid')$ and runs the copy of the user $\fcecUser_j$ on input that message. We remark that the copy of the user finds the request identifier $\fcecrequestid$ of the request associated with this issuance message, or aborts if it is not found. When the copy of the user outputs $(\fcecissueend, \allowbreak \sid, \allowbreak \fcecrequestid, \allowbreak \fcecAuthority_i)$, $\Simulator$ sends $(\fcecissueini, \allowbreak \sid, \allowbreak \fcecUser_j, \allowbreak \fcecrequestid)$  to $\Functionality_{\CEC}$. When $\Functionality_{\CEC}$ sends $(\fcecissuesim, \allowbreak \sid, \allowbreak \qid, \allowbreak \fcecAuthority_i, \allowbreak \fcecUser_j)$, $\Simulator$ sends $(\fcecissuerep, \allowbreak \sid, \allowbreak \qid)$ to $\Functionality_{\CEC}$.

    \item[Honest user begins spending.] When $\Functionality_{\CEC}$ sends $(\fcecspendsim, \allowbreak \sid, \allowbreak \qid)$, $\Simulator$ sends $(\fnymsendsim, \allowbreak \sid, \allowbreak \qid, \allowbreak \fnymleakage(\fnymmessage))$ to $\Adversary$, where $\fnymleakage(\fnymmessage)$ is the length of the message $\langle \cecpayment, \allowbreak \cecpaymentinfo \rangle$ sent by an honest user in the spend interface. Here we consider that the length does not change depending on the number of spent coins. This can be achieved by setting a maximum number of coins to be spent and sending always messages whose length is the length of the message when the maximum number of coins is spent.

    \item[Adversary forwards spending.] When $\Adversary$ sends $(\fnymsendrep, \allowbreak \sid, \allowbreak \qid)$, $\Simulator$ sends $(\fcecspendrep, \allowbreak \sid, \allowbreak \qid)$ to $\Functionality_{\CEC}$.

    \item[Corrupt provider receives spending.]  When $\Functionality_{\CEC}$ sends the message $(\fcecspendend, \allowbreak \fcecpaymentid, \allowbreak V, \allowbreak \cecpaymentinfo, \allowbreak \fcecnym)$, $\Simulator$ sets the payment $\cecpayment \allowbreak \gets \allowbreak (\kappa, \allowbreak \sigma', \allowbreak \langle S_k, \allowbreak T_k, \allowbreak A_k \rangle_{k\in[0,V-1]}, \allowbreak V, \allowbreak C, \allowbreak \pi_v)$ by running the procedure in Figure~\ref{fig:proc2}. $\Simulator$ stores $(\sid, \allowbreak \fcecpaymentid, \allowbreak \cecpayment, \allowbreak \cecpaymentinfo)$ and sends the message $(\fnymsendend, \allowbreak \sid, \langle \cecpayment, \cecpaymentinfo \rangle, \allowbreak \fcecnym)$  to $\Adversary$.

    \item[Corrupt user begins spending.] When the adversary $\Adversary$ sends  $(\fnymsendini, \allowbreak \sid, \allowbreak \langle \cecpayment, \allowbreak \cecpaymentinfo \rangle, \allowbreak \fcecnym, \allowbreak \fcecProvider_k)$,  $\Simulator$ runs $\Functionality_{\NYM}$ on input that message. When $\Functionality_{\NYM}$ sends $(\fnymreplysim, \allowbreak \sid, \allowbreak \qid, \allowbreak \fnymleakage(\fnymmessage))$, $\Simulator$ sends that message to $\Adversary$.

    \item[Honest provider receives spending from corrupt user.] After the adversary $\Adversary$ sends the message $(\fnymreplyrep, \allowbreak \sid, \allowbreak \qid)$, the simulator $\Simulator$ runs $\Functionality_{\NYM}$ on input that message. When the functionality $\Functionality_{\NYM}$ sends $(\fnymsendend, \allowbreak \sid, \langle \cecpayment, \cecpaymentinfo \rangle, \allowbreak \fcecnym)$, $\Simulator$ runs $b \allowbreak \gets \allowbreak \cecSpendVf(\spk, \allowbreak \cecpayment, \allowbreak \cecpaymentinfo)$. If $b \allowbreak = \allowbreak 0$, $\Simulator$ aborts. Otherwise $\Simulator$ runs the procedure in Figure~\ref{fig:proc1} to check the payment and store a tuple $(\cecpayment, \allowbreak \cecpaymentinfo, \langle \ssk_{\fcecUser_j}, \allowbreak \cecsn, \allowbreak r, \allowbreak o_c, \allowbreak \langle l_k, \allowbreak o_{a_k}, \allowbreak \mu_k, \allowbreak o_{\mu_k} \rangle_{k=0}^{V-1} \rangle)$. After that, $\Simulator$ checks that the algorithm for identification of double spenders identifies the double spender properly by running the procedure in Figure~\ref{fig:proc3}.

    $\Simulator$ sets $\fcecdsset \allowbreak \gets \allowbreak \{l_k | k \in [0,V-1]\}$ and sends $(\fcecspendini, \allowbreak \sid, \allowbreak \fcecwalletnum, \allowbreak V, \allowbreak \fcecdsset, \allowbreak \cecpaymentinfo, \allowbreak \fcecnym, \allowbreak \fcecProvider_k)$ to $\Functionality_{\CEC}$. When the functionality $\Functionality_{\CEC}$ sends the message $(\fcecspendsim, \allowbreak \sid, \allowbreak \qid)$, $\Simulator$ sends  $(\fcecspendrep, \allowbreak \sid, \allowbreak \qid)$ to $\Functionality_{\CEC}$.

    \item[Honest provider begins deposit of honest user $\fcecUser_j$ payment.] If there are not corrupt authorities, $\Functionality_{\CEC}$ sends $(\fcecdepositsim, \allowbreak \sid, \allowbreak \qid)$ and $\Simulator$ picks up random $\cecpaymentinfo \allowbreak \gets \allowbreak \fcecunivpaymentinfo$ and $V \allowbreak \gets \allowbreak [1,L]$. If there is at least one corrupt authority, $\Functionality_{\CEC}$ sends the message $(\fcecdepositsim, \allowbreak \sid, \allowbreak \qid, \allowbreak V, \allowbreak \cecpaymentinfo)$.

    After receiving the message from $\Functionality_{\CEC}$, $\Simulator$ runs the procedure in Figure~\ref{fig:proc2} to simulate an honest user payment $\cecpayment$. $\Simulator$ runs a copy of $\Functionality_{\BB}$ on input $(\fbbwriteini, \allowbreak \sid, \allowbreak \langle \cecpayment, \allowbreak \cecpaymentinfo \rangle)$. When $\Functionality_{\BB}$ sends $(\fbbwritesim, \allowbreak \sid, \allowbreak \qid)$, $\Simulator$ forwards that message to $\Adversary$.

    \item[Honest provider starts deposit of corrupt user $\fcecUser_j$ payment.] When $\Functionality_{\CEC}$ sends the message $(\fcecdepositsim, \allowbreak \sid, \allowbreak \qid, \allowbreak \fcecUser_j, \allowbreak \fcecwalletnum, \allowbreak V, \allowbreak \fcecdsset, \allowbreak \cecpaymentinfo, \allowbreak \fcecnym, \allowbreak \fcecProvider_k)$, $\Simulator$ finds the tuple $(\cecpayment, \allowbreak \cecpaymentinfo', \langle \ssk_{\fcecUser_j}, \allowbreak \cecsn, \allowbreak r, \allowbreak o_c, \allowbreak \langle l_k, \allowbreak o_{a_k}, \allowbreak \mu_k, \allowbreak o_{\mu_k} \rangle_{k=0}^{V-1} \rangle)$, where $\cecpayment$ is  $(\kappa, \allowbreak \sigma', \allowbreak \langle S_k, \allowbreak T_k, \allowbreak A_k \rangle_{k\in[0,V'-1]}, \allowbreak V', \allowbreak C, \allowbreak \pi_v)$, such that $\ssk_{\fcecUser_j}$ is the secret key corresponding to the public key registered by the corrupt user $\fcecUser_j$, $\cecsn$ is the serial number associated with the wallet $\fcecwalletnum$, $V' \allowbreak = \allowbreak V$, $\langle l_k \rangle_{k\in[0,V-1]} = \fcecdsset$, and $\cecpaymentinfo' \allowbreak = \allowbreak \cecpaymentinfo$. $\Simulator$ runs  $\Functionality_{\BB}$ on input $(\fbbwriteini, \allowbreak \sid, \allowbreak \langle \cecpayment, \allowbreak \cecpaymentinfo \rangle)$. When $\Functionality_{\BB}$ sends $(\fbbwritesim, \allowbreak \sid, \allowbreak \qid)$, $\Simulator$ forwards that message to $\Adversary$. We remark that $\Functionality_{\CEC}$ sends $(\fcecdepositsim, \allowbreak \sid, \allowbreak \qid, \allowbreak \fcecUser_j, \allowbreak \fcecwalletnum, \allowbreak V, \allowbreak \fcecdsset, \allowbreak \cecpaymentinfo, \allowbreak \fcecnym, \allowbreak \fcecProvider_k)$ instead $(\fcecdepositsim, \allowbreak \sid, \allowbreak \qid)$ when the deposit corresponds to a payment made by a corrupt user and when at least one authority is corrupt. When all the authorities are honest, the adversary in the real world does not have access to the information in $\Functionality_{\BB}$, and thus the simulator does not need to learn from $\Functionality_{\CEC}$ whether a payment made by a corrupt user has been deposited.

    \item[Honest provider ends deposit.] When the adversary $\Adversary$ sends the message $(\fbbwriterep, \allowbreak \sid, \allowbreak \qid)$, $\Simulator$ runs $\Functionality_{\BB}$ on input that message. When $\Functionality_{\BB}$ sends $(\fbbwriteend, \allowbreak \sid)$, $\Simulator$ sends $(\fcecdepositrep, \allowbreak \sid, \allowbreak \qid)$ to $\Functionality_{\CEC}$.

    \begin{figure}
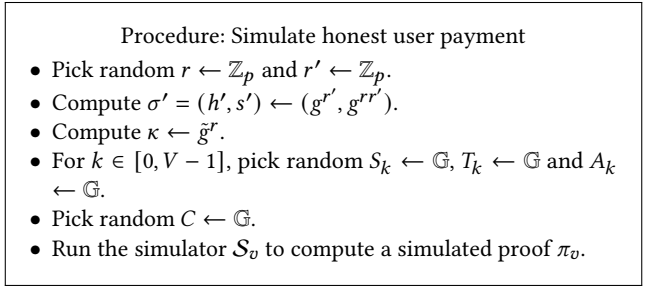

    \begin{framed}
    \begin{center}
    Procedure: Simulate honest user payment
    \end{center}

    \begin{itemize}[leftmargin=*]

        \item Pick random $r \gets \Zp$ and $r' \gets \Zp$.

        \item Compute $\sigma' = (\h',s') \gets (\ga^{r'}, \ga^{r r'})$.

        \item Compute $\kappa \gets \gb^{r}$.

        \item For $k \allowbreak \in \allowbreak [0, \allowbreak V-1]$, pick random $S_k \allowbreak \gets \allowbreak \Ga$, $T_k \allowbreak \gets \allowbreak \Ga$ and $A_k \allowbreak \gets \allowbreak \Ga$.

        \item Pick random $C \allowbreak \gets \allowbreak \Ga$.

        \item Run the simulator $\mathcal{S}_{v}$ to compute a simulated proof $\pi_v$.

    \end{itemize}

    \end{framed}
    \caption{Procedure for simulating an honest user payment.}
    \label{fig:proc2}
    \end{figure}

    \begin{figure}
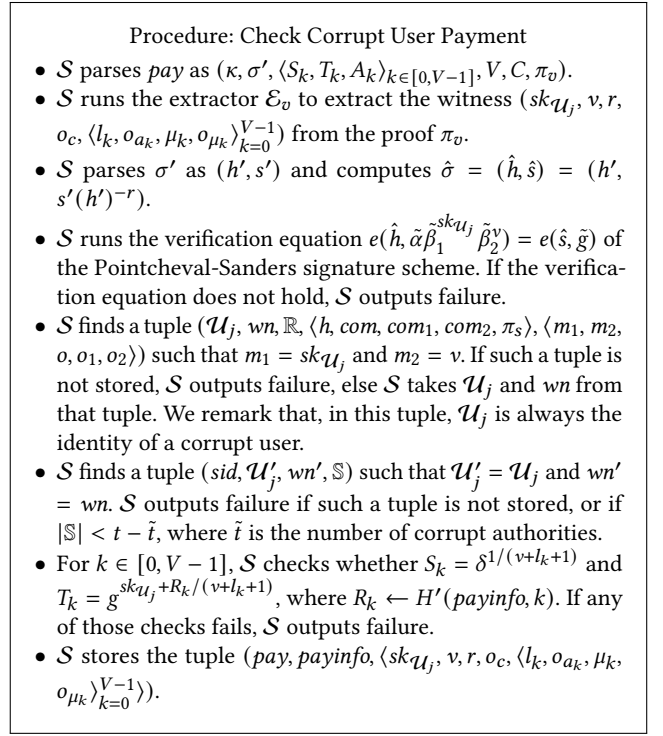

    \begin{framed}
    \begin{center}
    Procedure: Check Corrupt User Payment
    \end{center}

    \begin{itemize}[leftmargin=*]

        \item $\Simulator$ parses $\cecpayment$ as $(\kappa, \allowbreak \sigma', \allowbreak \langle S_k, \allowbreak T_k, \allowbreak A_k \rangle_{k\in[0,V-1]}, \allowbreak V, \allowbreak C, \allowbreak \pi_v)$.

        \item $\Simulator$ runs the extractor $\mathcal{E}_v$ to extract the witness $(\ssk_{\fcecUser_j}, \allowbreak \cecsn, \allowbreak r, \allowbreak o_c, \allowbreak \langle l_k, \allowbreak o_{a_k}, \allowbreak \mu_k, \allowbreak o_{\mu_k} \rangle_{k=0}^{V-1})$ from the proof $\pi_v$.

        \item $\Simulator$ parses $\sigma'$ as $(\h', \allowbreak s')$ and computes $\hat{\sigma} \allowbreak = \allowbreak (\hat{\h}, \allowbreak \hat{s}) \allowbreak = \allowbreak (\h', \allowbreak s' (\h')^{-r})$.

        \item $\Simulator$ runs the verification equation $\e(\hat{\h}, \allowbreak \tilde{\alpha} \allowbreak \tilde{\beta}_1^{\ssk_{\fcecUser_j}} \tilde{\beta}_2^{\cecsn}) \allowbreak =  \allowbreak \e(\hat{s}, \allowbreak \gb)$ of the Pointcheval-Sanders signature scheme. If the verification equation does not hold, $\Simulator$ outputs failure.

        \item $\Simulator$ finds a tuple $(\fcecUser_j, \allowbreak \fcecwalletnum, \allowbreak \fcecsetrequestid, \allowbreak \langle \h, \allowbreak \com, \allowbreak \com_{1}, \allowbreak \com_{2}, \allowbreak \pi_s \rangle, \allowbreak \langle \cecmes_{1}, \allowbreak \cecmes_{2}, \allowbreak o, \allowbreak o_{1}, \allowbreak o_{2}  \rangle)$ such that $\cecmes_{1} \allowbreak = \allowbreak \ssk_{\fcecUser_j}$ and $\cecmes_{2} \allowbreak = \allowbreak \cecsn$. If such a tuple is not stored, $\Simulator$ outputs failure, else $\Simulator$ takes $\fcecUser_j$ and $\fcecwalletnum$ from that tuple. We remark that, in this tuple, $\fcecUser_j$ is always the identity of a corrupt user.

        \item $\Simulator$ finds a tuple $(\sid, \allowbreak \fcecUser'_j, \allowbreak \fcecwalletnum', \allowbreak \cecset)$ such that $\fcecUser'_j \allowbreak = \allowbreak \fcecUser_j$ and $\fcecwalletnum' \allowbreak = \allowbreak \fcecwalletnum$.  $\Simulator$ outputs failure if such a tuple is not stored, or if $|\cecset| < t-\tilde{t}$, where $\tilde{t}$ is the number of corrupt authorities.

       \item For $k \in [0,V-1]$, $\Simulator$ checks whether $S_k = \delta^{1/(\cecsn+l_k+1)}$ and $T_k \allowbreak = \allowbreak \ga^{\ssk_{\fcecUser_j} + R_k/(\cecsn+l_k+1)}$, where $R_k \gets H'(\cecpaymentinfo,k)$. If any of those checks fails, $\Simulator$ outputs failure.

        \item $\Simulator$ stores the tuple $(\cecpayment, \allowbreak \cecpaymentinfo, \langle \ssk_{\fcecUser_j}, \allowbreak \cecsn, \allowbreak r, \allowbreak o_c, \allowbreak \langle l_k, \allowbreak o_{a_k}, \allowbreak \mu_k, \allowbreak o_{\mu_k} \rangle_{k=0}^{V-1} \rangle)$.

  \end{itemize}

  \end{framed}
  \caption{Procedure for checking a corrupt user payment.}
  \label{fig:proc1}
\end{figure}

    \begin{figure}
    \begin{framed}
    \begin{center}
        Procedure: Check Double Spending
    \end{center}

       \begin{itemize}[leftmargin=*]

        \item $\Simulator$ takes the tuple $(\cecpayment, \allowbreak \cecpaymentinfo, \langle \ssk_{\fcecUser_j}, \allowbreak \cecsn, \allowbreak r, \allowbreak o_c, \allowbreak \langle l_k, \allowbreak o_{a_k}, \allowbreak \mu_k, \allowbreak o_{\mu_k} \rangle_{k=0}^{V-1} \rangle)$ for the payment made by a corrupt user.

        \item For all the stored tuples $(\cecpayment', \allowbreak \cecpaymentinfo', \langle \ssk'_{\fcecUser_j}, \allowbreak \cecsn', \allowbreak r', \allowbreak o'_c, \allowbreak \langle l'_k, \allowbreak o'_{a_k}, \allowbreak \mu'_k, \allowbreak o'_{\mu_k} \rangle_{k=0}^{V'-1} \rangle)$, $\Simulator$ checks whether $\ssk'_{\fcecUser_j} \allowbreak = \allowbreak \ssk_{\fcecUser_j}$, $\cecsn' \allowbreak = \allowbreak \cecsn$ and $\langle l'_k \rangle_{k\in[0,V'-1]} \cap \langle l_k \rangle_{k\in[0,V-1]} \allowbreak \neq \allowbreak \emptyset$, which means that there is a double spending. $\Simulator$ sets $\cecPK$ to contain all the registered public keys and runs the algorithm $c \allowbreak \gets \allowbreak \cecIdentify(\cecparams, \allowbreak \cecPK, \allowbreak \cecpayment, \allowbreak \cecpayment', \allowbreak \cecpaymentinfo, \allowbreak \cecpaymentinfo')$. Then $\Simulator$ proceeds as follows:

        \begin{itemize}

            \item If there is double spending, $\Simulator$ does the following:
            \begin{itemize}

                \item If $\cecpaymentinfo \allowbreak = \allowbreak \cecpaymentinfo'$ and $c \allowbreak \neq \allowbreak \cecpaymentinfo$, $\Simulator$ outputs failure. We note that authorities verify that $\cecpaymentinfo$ contains an identifier of the provider that deposited the payment. This avoids that an honest provider can be found guilty when an adversarial provider e.g.\ double deposits a payment with payment information $\cecpaymentinfo$ that contains the identifier of an honest provider.

                \item If $\cecpaymentinfo \allowbreak \neq \allowbreak \cecpaymentinfo'$ and $c \allowbreak \neq \allowbreak \spk_{\fcecUser_j}$, where $\spk_{\fcecUser_j}$ is the public key associated with secret key $\ssk_{\fcecUser_j}$, $\Simulator$ outputs failure.

            \end{itemize}

            \item If there is not double spending, and if $c \allowbreak = \allowbreak \spk_{\fcecUser_j}$, where $\spk_{\fcecUser_j}$ is a public key associated with an honest user, $\Simulator$ outputs failure. We remark that it is possible for the adversary to produce two payments where there is not double-spending, and yet the algorithm $\cecIdentify$ detects double spending. When that happens, we must ensure that algorithm $\cecIdentify$ does not output the public key of an honest user. We recall that authorities verify that $\cecpaymentinfo$ contains an identifier of the provider that deposited the payment, which also avoids an honest provider from being framed in this case.

        \end{itemize}

    \end{itemize}

      \end{framed}
  \caption{Procedure for checking double spending.}
  \label{fig:proc3}
\end{figure}

    \item[Corrupt provider begins deposit.] When $\Adversary$ sends $(\fbbwriteini, \allowbreak \sid, \allowbreak \langle \cecpayment, \allowbreak \cecpaymentinfo \rangle)$, the simulator $\Simulator$ checks if a tuple $(\sid, \allowbreak \fcecpaymentid, \allowbreak \cecpayment', \allowbreak \cecpaymentinfo')$ such that $\cecpayment' \allowbreak = \allowbreak \cecpayment$ and  $\cecpaymentinfo' \allowbreak = \allowbreak \cecpaymentinfo$ is stored. If it is stored, which means that this is a payment that was sent to $\Adversary$ by the simulator acting as an honest user, $\Simulator$ takes $\fcecpaymentid$ from that tuple and proceeds with step 2 below. Otherwise we are in the case of a payment that a corrupt user sent to a corrupt provider, and thus $\Simulator$ did not receive it during the spending phase. $\Simulator$ proceeds with step 1.

    \noindent \emph{Step 1.} $\Simulator$ proceeds as follows:
    \begin{itemize}

        \item $\Simulator$ runs $b \allowbreak \gets \allowbreak \cecSpendVf(\spk, \allowbreak \cecpayment, \allowbreak \cecpaymentinfo)$. If $b \allowbreak = \allowbreak 0$, the simulator $\Simulator$ aborts.

        \item $\Simulator$ runs the procedure in Figure~\ref{fig:proc1} to store $(\cecpayment, \allowbreak \cecpaymentinfo, \langle \ssk_{\fcecUser_j}, \allowbreak \cecsn, \allowbreak r, \allowbreak o_c, \allowbreak \langle l_k, \allowbreak o_{a_k}, \allowbreak \mu_k, \allowbreak o_{\mu_k} \rangle_{k=0}^{V-1} \rangle)$.

        \item The simulator $\Simulator$ checks that the algorithm for identification of double spenders identifies the double spender properly by running the procedure in Figure~\ref{fig:proc3}.

    \end{itemize}
    $\Simulator$ sets $\fcecdsset \allowbreak \gets \allowbreak \{l_k | k \in [0,V-1]\}$, picks a random pseudonym $\fcecnym \allowbreak \gets \allowbreak \fcecunivnym$, takes the identity $\fcecProvider_k$ of the corrupt provider that sent the message $(\fbbwriteini, \ldots)$, and sends $(\fcecspendini, \allowbreak \sid, \allowbreak \fcecwalletnum, \allowbreak V, \allowbreak \fcecdsset, \allowbreak \cecpaymentinfo, \allowbreak \fcecnym, \allowbreak \fcecProvider_k)$ to $\Functionality_{\CEC}$. When $\Functionality_{\CEC}$ sends the message $(\fcecspendsim, \allowbreak \sid, \allowbreak \qid)$, $\Simulator$ sends  $(\fcecspendrep, \allowbreak \sid, \allowbreak \qid)$ to $\Functionality_{\CEC}$. When $\Functionality_{\CEC}$ sends the message $(\fcecspendend, \allowbreak \fcecpaymentid, \allowbreak V, \allowbreak \cecpaymentinfo, \allowbreak \fcecnym)$, $\Simulator$ proceeds with step 2.

    \noindent \emph{Step 2.} $\Simulator$ sends $(\fcecdepositini, \allowbreak \sid, \allowbreak \fcecpaymentid)$ to $\Functionality_{\CEC}$. When $\Functionality_{\CEC}$ sends $(\fcecdepositsim, \allowbreak \sid, \allowbreak \qid)$, $\Simulator$ runs a copy of $\Functionality_{\BB}$ on input the message $(\fbbwriteini, \allowbreak \sid, \allowbreak \langle \cecpayment, \allowbreak \cecpaymentinfo \rangle)$. When $\Functionality_{\BB}$ sends the message $(\fbbwritesim, \allowbreak \sid, \allowbreak \qid)$, $\Simulator$ forwards that message to $\Adversary$.

    \item[Corrupt provider ends deposit.] When the adversary $\Adversary$ sends the message $(\fbbwriterep, \allowbreak \sid, \allowbreak \qid)$, $\Simulator$ runs the copy of $\Functionality_{\BB}$ on input that message. When $\Functionality_{\BB}$ sends $(\fbbwriteend, \allowbreak \sid)$, $\Simulator$ sends $(\fcecdepositrep, \allowbreak \sid, \allowbreak \qid)$ to $\Functionality_{\CEC}$. When $\Functionality_{\CEC}$ sends $(\fcecdepositend, \allowbreak \sid, \allowbreak \fcecpaymentid)$, $\Simulator$ sends  $(\fbbwriteend, \allowbreak \sid)$ to $\Adversary$.

    \item[Honest authority runs deposit verification.] When $\Functionality_{\CEC}$ sends the message $(\fcecdepvfsim, \allowbreak \sid, \allowbreak \qid, \allowbreak \fcecusers', \allowbreak \fcecnumdeposits)$, for all $\fcecUser_j \allowbreak \in \allowbreak \fcecusers'$, $\Simulator$ runs the procedure in Figure~\ref{fig:proc4}. After that, for $i=1$ to $\fcecnumdeposits$, $\Simulator$ does the following:
    \begin{itemize}

        \item Send $(\fbbgetbbsim, \allowbreak \sid, \allowbreak \qid)$ to $\Adversary$.

        \item Receive $(\fbbgetbbrep, \allowbreak \sid, \allowbreak \qid)$ from $\Adversary$.

    \end{itemize}
    $\Simulator$ sends $(\fcecdepvfrep, \allowbreak \sid, \allowbreak \qid)$ to $\Functionality_{\CEC}$.

    \item[Corrupt authority starts reading of bulletin board.] When $\Adversary$ sends $(\fbbgetbbini, \allowbreak \sid, \allowbreak \fbbindex)$, $\Simulator$ runs $\Functionality_{\BB}$ on input that message. When $\Functionality_{\BB}$ sends $(\fbbgetbbsim, \allowbreak \sid, \allowbreak \qid)$, $\Simulator$ forwards that message to $\Adversary$.

    \item[Corrupt authority ends reading of bulletin board.] When the adversary $\Adversary$ sends $(\fbbgetbbrep, \allowbreak \sid, \allowbreak \qid)$, $\Simulator$ runs $\Functionality_{\BB}$ on input that message. When $\Functionality_{\BB}$ sends $(\fbbgetbbend, \allowbreak \sid, \allowbreak \fbbmessage')$, $\Simulator$ parses $\fbbmessage'$ as $(\fcecProvider_k, \allowbreak \cecpayment, \allowbreak \cecpaymentinfo)$. If $(\cecpayment, \allowbreak \cecpaymentinfo)$ is a payment that was computed by $\Simulator$, $\Simulator$ parses $\cecpayment$ as $(\kappa, \allowbreak \sigma', \allowbreak \langle S_k, \allowbreak T_k, \allowbreak A_k \rangle_{k\in[0,V-1]}, \allowbreak V, \allowbreak C, \allowbreak \pi_v)$ and checks whether any serial number $S_k$ is equal to any of the other serial numbers stored in the payments in $\Functionality_{\BB}$. If that is the case, $\Simulator$ outputs failure. Otherwise $\Simulator$ sends $(\fbbgetbbend, \allowbreak \sid, \allowbreak \fbbmessage')$ to $\Adversary$. We remark that payments computed by the adversary were already checked before by $\Simulator$ through the procedure in Figure~\ref{fig:proc3}.

\end{description}

\begin{theorem} \label{th:all}
	When a subset of users $\fcecUser_j$, a subset of providers $\fcecProvider_k$ and up to $t-1$ authorities $\fcecAuthority_i$ are corrupt, $\mathrm{\Pi}_{\CEC}$ securely realizes $\Functionality_{\CEC}$ in the random oracle model and in the $(\Functionality_{\SMT}, \allowbreak \Functionality_{\NYM}, \allowbreak \Functionality_{\KG}, \allowbreak \Functionality_{\Freg}, \allowbreak \Functionality_{\BB})$-hybrid model if the non-interactive proof of knowledge scheme is zero-knowledge and provides weak simulation extractability, the signature scheme by Pointcheval-Sanders in the RO model is unforgeable, the commmitment scheme is hiding and binding, the function in~\S\ref{subsec:pseudorandomfunction} is pseudorandom, and the hash function $H'$ is collision-resistant.
\end{theorem}

\paragraph{Proof of Theorem~\ref{th:all}.}
We show by means of a series of hybrid games that the environment $\Environment$ cannot distinguish between the ensemble $\RealEnsemble_{\mathrm{\Pi}_{\CEC},\Adversary,\Environment}$ and the ensemble $\IdealEnsemble_{\Functionality_{\CEC},\Simulator,\Environment}$ with non-negligible probability. We denote by $\Prob[\Gam i]$ the probability that the environment distinguishes $\Gam i$ from the real-world protocol.
\begin{description}[leftmargin=*]

	\item[$\Gam 0$:] This game corresponds to the execution of the real-world protocol. Therefore, $\Prob[\Gam 0] = 0$.

	\item[$\Gam 1$:] This game proceeds as $\Gam 0$, except that $\Gam 1$ runs the extractor $\mathcal{E}_s$ for the non-interactive ZK proofs of knowledge $\pi_s$ sent by the adversary. Under the weak simulation extractability property of the proof system (Definition~\ref{def:weakse}), we have that $|\Pr[\Gam 1]-\Pr[\Gam 0]| \allowbreak \leq \allowbreak \Adv_{\Adversary}^{\mathsf{ext}}$.

	\item[$\Gam 2$:] This game proceeds as $\Gam 1$, except that $\Gam 2$ outputs failure if two request messages were received from the adversary with commitments $\com'$ and $\com$ and proofs $\pi'_s$ and $\pi_s$ such that $\com' \allowbreak = \allowbreak \com$ but, after extraction of the witnesses $(\cecmes'_{1}, \allowbreak \cecmes'_{2}, \allowbreak o', \allowbreak o'_{1}, \allowbreak o'_{2})$ from $\pi'_s$ and $(\cecmes_{1}, \allowbreak \cecmes_{2}, \allowbreak o, \allowbreak o_{1}, \allowbreak o_{2})$ from $\pi_s$, $(\cecmes'_{1}, \allowbreak \cecmes'_{2}) \allowbreak \neq \allowbreak  (\cecmes_{1}, \allowbreak \cecmes_{2})$. Under the binding property of the commitment scheme, we have that $|\Pr[\Gam 2]-\Pr[\Gam 1]| \allowbreak \leq \allowbreak \Adv_{\Adversary}^{\mathsf{bin}}$.

	\begin{proof}
	 Given an adversary that makes $\Gam 2$ output failure with non-negligible probability, we construct an algorithm $B$ that breaks the binding property of the commitment scheme with non-negligible probability. $B$ works as follows. $B$ receives the parameters of the Pedersen commitment scheme $\paramscom \allowbreak = (\ga, \allowbreak \gamma_1, \allowbreak \gamma_2)$ from the challenger. When running $\Functionality_{\KG}$, $B$ uses those parameters to set the values $(\ga, \allowbreak \gamma_1, \allowbreak \gamma_2)$ in the parameters $\cecparams \allowbreak \gets \allowbreak (\p, \allowbreak \Ga, \allowbreak \Gb, \allowbreak \Gt, \allowbreak \e, \allowbreak \ga, \allowbreak \gb, \allowbreak \gamma_1, \allowbreak \gamma_2, \allowbreak \delta, \allowbreak L)$. When the adversary sends two requests with commitments $\com'$ and $\com$ and proofs $\pi'_s$ and $\pi_s$ such that $\com' \allowbreak = \allowbreak \com$ but, after extraction of the witnesses $(\cecmes'_{1}, \allowbreak \cecmes'_{2}, \allowbreak o', \allowbreak o'_{1}, \allowbreak o'_{2})$ from $\pi'_s$ and $(\cecmes_{1}, \allowbreak \cecmes_{2}, \allowbreak o, \allowbreak o_{1}, \allowbreak o_{2})$ from $\pi_s$, it holds that $(\cecmes'_{1}, \allowbreak \cecmes'_{2}) \allowbreak \neq \allowbreak  (\cecmes_{1}, \allowbreak \cecmes_{2})$, $B$ sends $(\com, \cecmes_{1}, \allowbreak \cecmes_{2}, \allowbreak o, \cecmes'_{1}, \allowbreak \cecmes'_{2}, \allowbreak o')$ to the challenger.
	\end{proof}

	\item[$\Gam 3$:] This game proceeds as $\Gam 2$, except that $\Gam 3$ runs the extractor $\mathcal{E}_v$ for the non-interactive ZK proofs of knowledge $\pi_v$ sent by the adversary. Under the weak simulation extractability property of the proof system (Definition~\ref{def:weakse}), we have that $|\Pr[\Gam 3]-\Pr[\Gam 2]| \allowbreak \leq \allowbreak \Adv_{\Adversary}^{\mathsf{ext}}$.

	\item[$\Gam 4$:] This game proceeds as $\Gam 3$, except that, after extracting the witness $(\ssk_{\fcecUser_j}, \allowbreak \cecsn, \allowbreak r, \allowbreak o_c, \allowbreak \langle l_k, \allowbreak o_{a_k}, \allowbreak \mu_k, \allowbreak o_{\mu_k} \rangle_{k=0}^{V-1})$ from a proof $\pi_v$, $\Gam 4$ takes the payment $\cecpayment \allowbreak = \allowbreak (\kappa, \allowbreak \sigma', \allowbreak \langle S_k, \allowbreak T_k, \allowbreak A_k \rangle_{k\in[0,V-1]}, \allowbreak V, \allowbreak C, \allowbreak \pi_v)$ that contains $\pi_v$ and parses $\sigma'$ as $(\h', \allowbreak s')$. $\Gam 4$ computes $\hat{\sigma} \allowbreak = \allowbreak (\hat{\h}, \allowbreak \hat{s}) \allowbreak = \allowbreak (\h', \allowbreak s' (\h')^{-r})$. Then $\Gam 4$ outputs failure if $\hat{\sigma}$ is not a valid signature. As shown below, the computation $\hat{\sigma} \allowbreak = \allowbreak (\hat{\h}, \allowbreak \hat{s}) \allowbreak = \allowbreak (\h', \allowbreak s' (\h')^{-r})$ always produces a valid signature, and thus $|\Pr[\Gam 4]-\Pr[\Gam 3]| \allowbreak = \allowbreak 0$.
	\begin{proof}
	We follow the proof in~\cite{cryptoeprint:2022:011}. We observe that, after extraction from $\pi_v$ is successful, $\kappa$ is of the form $\kappa \allowbreak \gets \allowbreak  \tilde{\alpha} \tilde{\beta}_1^{\ssk_{\fcecUser_j}} \tilde{\beta}_2^{\cecsn} \gb^{r}$. We also know that the following equality holds
	\begin{equation*}
	    \e(\h', \kappa) = \e(s',\gb)
	\end{equation*}
	If we replace $\kappa$ by $\tilde{\alpha} \tilde{\beta}_1^{\ssk_{\fcecUser_j}} \tilde{\beta}_2^{\cecsn} \gb^{r}$, we have that
	\begin{equation*}
	    \e(\h', \tilde{\alpha} \tilde{\beta}_1^{\ssk_{\fcecUser_j}} \tilde{\beta}_2^{\cecsn} \gb^{r}) = \e(s',\gb)
	\end{equation*}
	If we now multiply the two sides of the equality by $\e(\h',\gb^{-r})$, we have that
	\begin{equation*}
	    \e(\h', \tilde{\alpha} \tilde{\beta}_1^{\ssk_{\fcecUser_j}} \tilde{\beta}_2^{\cecsn} \gb^{r}) \e(\h',\gb^{-r}) = \e(s',\gb) \e(\h',\gb^{-r})
	\end{equation*}
	and this gives us
	\begin{equation*}
	    \e(\h', \tilde{\alpha} \tilde{\beta}_1^{\ssk_{\fcecUser_j}} \tilde{\beta}_2^{\cecsn}) = \e(s' (\h')^{-r},\gb)
	\end{equation*}
	which is the verification equation of the Pointcheval-Sanders signature scheme for the signature $(\h', s' (\h')^{-r})$. Therefore, the computation $\hat{\sigma} \allowbreak = \allowbreak (\hat{\h}, \allowbreak \hat{s}) \allowbreak = \allowbreak (\h', \allowbreak s' (\h')^{-r})$ always produces a valid signature.
	\end{proof}

	\item[$\Gam 5$:] This game proceeds as $\Gam 4$, except that $\Gam 5$ outputs failure if, after computing the signature $\hat{\sigma} \allowbreak = \allowbreak (\hat{\h}, \allowbreak \hat{s}) \allowbreak = \allowbreak (\h', \allowbreak s' (\h')^{-r})$ on $(\ssk_{\fcecUser_j}, \allowbreak \cecsn)$, it is the case that the adversary was not issued at least $t-\tilde{t}$ signatures from $t-\tilde{t}$ different authorities on $(\ssk_{\fcecUser_j}, \allowbreak \cecsn)$. Under the unforgeability property of Pointcheval-Sanders signatures in the random oracle model, we have that $|\Pr[\Gam 5]-\Pr[\Gam 4]| \leq \Adv_{\Adversary}^{\mathsf{unf}}\cdot ((n-\tilde{t})!/((t-1-\tilde{t})!(n-t+1)!))$, where $n$ is the number of authorities, $t$ is the threshold and $\tilde{t}$ is the number of corrupt authorities.
	\begin{proof}
	 We follow the proof in~\cite{cryptoeprint:2022:011}. We construct an algorithm $B$ that interacts with the challenger of the existential unforgeability game in the RO model (Definition~\ref{def:unfro}) and the adversary $\Adversary$ and that shows that, if $\Adversary$ makes $\Gam 5$ output failure with non-negligible probability, then $\Adversary$ can be used by $B$ to break the existential unforgeability property in the RO model of Pointcheval-Sanders signatures.

	 $B$ receives a public key $(\theta, \allowbreak \tilde{\alpha}, \allowbreak \beta_1, \tilde{\beta}_1, \allowbreak \beta_2, \allowbreak \tilde{\beta}_2)$ from the challenger. To set up the keys when running functionality $\Functionality_{\KG}$, $B$ proceeds as follows.
	 \begin{itemize}

        \item $B$ uses the bilinear map setup $\theta$ to set $\cecparams$. $B$ uses the public key received from the challenger to set the verification key $\spk = (\cecparams, \allowbreak \tilde{\alpha}, \allowbreak \beta_{1}, \allowbreak \tilde{\beta}_{1}, \allowbreak \beta_{2}, \tilde{\beta}_{2})$.

        \item Let $\mathbb{T}$ be the set of indices of corrupt authorities. Let $\mathbb{U}$ be a set of indices of size $t-1-|\mathbb{T}|$ picked at random from $[1,n] \allowbreak \setminus \allowbreak \mathbb{T}$. Let $\mathbb{S}' \gets \mathbb{T} \cup \mathbb{U}$. To compute the secret keys and public keys of the authorities $\fcecAuthority_i$ such that $i \allowbreak \in \allowbreak \mathbb{S}'$, $B$ picks random $(x_i, \allowbreak y_{i,1},  \allowbreak y_{i,2}) \allowbreak \gets \allowbreak \Zp$ and computes $\spk_{\fcecAuthority_i} \allowbreak = \allowbreak (\tilde{\alpha}_i, \allowbreak \beta_{i,1}, \allowbreak \tilde{\beta}_{i,1}, \allowbreak \beta_{i,2}, \allowbreak \tilde{\beta}_{i,2}) \allowbreak \gets \allowbreak (\gb^{x_i}, \allowbreak \ga^{y_{i,1}}, \allowbreak \gb^{y_{i,1}}, \allowbreak \ga^{y_{i,2}}, \allowbreak \gb^{y_{i,2}})$.

        \item Let $\mathbb{S} \allowbreak \gets \allowbreak \mathbb{S}' \allowbreak \cup \allowbreak \{0\}$ and let $\mathbb{D} \allowbreak = \allowbreak [1,n] \setminus \mathbb{S}'$. To compute the public keys of the remaining authorities, i.e. the authorities in the set $\mathbb{D}$, $B$ does the following. Let $(\tilde{\alpha}_0, \allowbreak \beta_{0,1}, \allowbreak \tilde{\beta}_{0,1}, \allowbreak \beta_{0,2}, \allowbreak \tilde{\beta}_{0,2}) \gets (\tilde{\alpha}, \allowbreak \beta_{1}, \allowbreak \tilde{\beta}_{1}, \allowbreak \beta_{2}, \tilde{\beta}_{2})$. For all $d \in \mathbb{D}$:
        \begin{itemize}

            \item For all $i \in \mathbb{S}$, evaluate at $d$ the Lagrange basis polynomials
            \begin{equation*}
                l_i = [\prod_{j\in\mathbb{S},j\neq i} (d-j)] [\prod_{j\in\mathbb{S},j\neq i} (i-j)]^{-1}\ \mathrm{mod}\ \p
            \end{equation*}

            \item  For all $i \in \mathbb{S}$, take $(\tilde{\alpha}_i, \allowbreak \beta_{i,1}, \allowbreak \tilde{\beta}_{i,1}, \allowbreak \beta_{i,2}, \allowbreak \tilde{\beta}_{i,2})$ and then do $\spk_{\fcecAuthority_d} = (\tilde{\alpha}_d, \allowbreak \beta_{d,1}, \allowbreak \tilde{\beta}_{d,1}, \allowbreak \beta_{d,2}, \allowbreak \tilde{\beta}_{d,2}) = ( \prod_{i \in \mathbb{S}} \tilde{\alpha}_i^{l_i}, \allowbreak \prod_{i \in \mathbb{S}} \beta_{i,1}^{l_i}, \allowbreak \prod_{i \in \mathbb{S}} \tilde{\beta}_{i,1}^{l_i},  \allowbreak \prod_{i \in \mathbb{S}} \beta_{i,2}^{l_i}, \allowbreak \prod_{i \in \mathbb{S}} \tilde{\beta}_{i,2}^{l_i})$.

        \end{itemize}

    \end{itemize}

    To reply the random oracle queries $H(\com)$ of the adversary $\Adversary$, $B$ forwards the query $\com$ to the random oracle provided by the challenger and sends $\Adversary$ the response $\h$ given by the challenger.

    When $\Adversary$ sends a valid request $\cecrequest \allowbreak \gets \allowbreak (\h, \allowbreak \com, \allowbreak \com_{1}, \allowbreak \com_{2}, \allowbreak \pi_s)$, $B$ runs the extractor $\mathcal{E}_{s}$ to extract the witness $(\cecmes_{1}, \allowbreak \cecmes_{2}, \allowbreak o, \allowbreak o_{1}, \allowbreak o_{2})$ from $\pi_s$. $B$ outputs failure if two request messages were received with commitments $\com'$ and $\com$ and proofs $\pi'_s$ and $\pi_s$ such that $\com' \allowbreak = \allowbreak \com$ but, after extraction of the witnesses from $\pi'_s$ and $\pi_s$, $(m'_1, \allowbreak m'_2) \allowbreak \neq \allowbreak  (m_1, \allowbreak m_2)$. As shown in $\Gam 2$, the probability that $B$ fails is negligible if the commitment scheme is binding. This guarantees that $\com$ is different for each tuple of messages $(m'_1, \allowbreak m'_2)$, which is necessary when querying the signing oracle.

    If the request is sent to an authority $\fcecAuthority_{i}$ such that $i \allowbreak \in \allowbreak \mathbb{U}$, $B$ computes an issuance message by following $\mathrm{\Pi}_{\CEC}$ and stores $(m_1, \allowbreak m_2, \allowbreak \fcecAuthority_i)$. (We note that in this case $B$ knows the secret key of the authority.) If the request is sent to $\fcecAuthority_{d}$ such that $d \allowbreak \in \allowbreak \mathbb{D}$, $B$ proceeds as follows:

    \begin{itemize}

        \item $B$ submits the message tuple $(m_1, \allowbreak m_2)$ that was extracted by $\mathcal{E}_{s}$ and the commitment $\com$ to the signing oracle provided by the challenger. The challenger sends a signature $\sigma_{0} \allowbreak = \allowbreak (\h, \allowbreak s_{0})$ and state information $\sstate'$.

        \item For all $i \allowbreak \in \allowbreak \mathbb{S}'$, $B$ computes a signature $\sigma_{i} = (\h, s_{i})$ by using the secret keys of authorities in $\mathbb{S}'$.

        \item For all $i \allowbreak \in \allowbreak \mathbb{S}$, $B$ evaluates at $d$ the Lagrange basis polynomials
        \begin{equation*}
            l_i = [\prod_{j\in\mathbb{S},j\neq i} (d-j)] [\prod_{j\in \mathbb{S},j\neq i} (i-j)]^{-1}\ \mathrm{mod}\ \p
        \end{equation*}

        \item $B$ computes the signature $\sigma_d = (\h, s_d) \gets (\h, \prod_{i \in \mathbb{S}} s_i^{l_i})$. We note that in this computation the signature sent by the challenger is used.

        \item $B$ computes $\hat{\sigma}_{d} = (\h, s_{d} \beta_{d,1}^{o_1} \beta_{d,2}^{o_2})$ and includes it in the issuance message sent to $\Adversary$. $B$ stores  $(m_1, \allowbreak m_2, \fcecAuthority_d)$.

    \end{itemize}

	After the adversary $\Adversary$ sends a valid payment $\cecpayment \allowbreak \gets \allowbreak (\kappa, \allowbreak \sigma', \allowbreak \langle S_k, \allowbreak T_k, \allowbreak A_k \rangle_{k\in[0,V-1]}, \allowbreak V, \allowbreak C, \allowbreak \pi_v)$, $B$ proceeds as follows.
	\begin{itemize}

        \item $B$ runs the extractor $\mathcal{E}_v$ to extract the witness  $(\ssk_{\fcecUser_j}, \allowbreak \cecsn, \allowbreak r, \allowbreak o_c, \allowbreak \langle l_k, \allowbreak o_{a_k}, \allowbreak \mu_k, \allowbreak o_{\mu_k} \rangle_{k=0}^{V-1})$ from the proof $\pi_v$.

        \item $B$ parses $\sigma'$ as $(\h', \allowbreak s')$ and computes $\hat{\sigma} \allowbreak = \allowbreak (\hat{\h}, \allowbreak \hat{s}) \allowbreak = \allowbreak (\h', \allowbreak s' (\h')^{-r})$. $B$ runs the verification equation $\e(\hat{\h}, \allowbreak \tilde{\alpha} \allowbreak \tilde{\beta}_1^{\ssk_{\fcecUser_j}} \tilde{\beta}_2^{\cecsn}) \allowbreak =  \allowbreak \e(\hat{s}, \allowbreak \gb)$ of the Pointcheval-Sanders signature scheme.  If for any signature $\hat{\sigma}_l$ the verification equation does not hold, $B$ outputs failure. As shown in $\Gam 4$, the probability that $B$ outputs failure is $0$.

        \item $B$ checks that there are at least $t-|\mathbb{T}|$ tuples $(m_{1}, \allowbreak m_{2}, \allowbreak \fcecAuthority_i)$ stored for $t-|\mathbb{T}|$ honest authorities. If that is the case, $B$ does nothing because $\Adversary$ was issued enough signatures to compute the payment. Else, if the adversary $\Adversary$ received less than $t-|\mathbb{T}|$ signatures from honest authorities, but $\Adversary$ did receive a signature from an authority $\fcecAuthority_{d}$ such that $d \allowbreak \in \allowbreak \mathbb{D}$, $B$ fails because $B$ had to the query signing oracle to issue that signature to the adversary and therefore he cannot use $\hat{\sigma}$ as a forgery. However, if $\Adversary$ received less than $t-|\mathbb{T}|$ signatures from honest authorities, and all those authorities $\fcecAuthority_{i}$ are such that $i \allowbreak \in \allowbreak \mathbb{U}$, $B$ sends $\hat{\sigma}$ to the challenger to win the existential unforgeability game.

    \end{itemize}

	Finally, the probability that $B$ fails can be bound as follows. $B$ needs to query the signing oracle of the challenger whenever $\Adversary$ requests a signature from an authority $\fcecAuthority_{d}$ such that $d \in \mathbb{D}$. Therefore, when $\Adversary$ is able to show a signature without receiving $t - |\mathbb{T}|$ signatures shares from $t - |\mathbb{T}|$ different honest authorities, $B$ fails whenever $\Adversary$ did request a signature from an authority $\fcecAuthority_{d}$ such that $d \in \mathbb{D}$. In the worst case, $\Adversary$ received $t-1-|\mathbb{T}|$ signatures from $t-1-|\mathbb{T}|$ honest authorities. In that worst case, $B$ only succeeds when those $t-1-|\mathbb{T}|$ authorities are those authorities $\fcecAuthority_{i}$ such that $i \in \mathbb{U}$.  The probability that $B$ succeeds, i.e. the probability that $\Adversary$ picks those $t-1-|\mathbb{T}|$ authorities from the set of $n-|\mathbb{T}|$ authorities is given by the inverse of the number of $(t-1-|\mathbb{T}|)$-element combinations of $n-|\mathbb{T}|$ objects taken without repetition
    \begin{equation*}
        \frac{(t-1-|\mathbb{T}|)!(n-t+1!)}{(n-|\mathbb{T}|)!}
    \end{equation*}
    We remark that, in the frequent case in which $t=n$, then $B$ succeeds with probability $1/(t-|\mathbb{T}|)$.
	\end{proof}

	\item[$\Gam 6$:] This game proceeds as $\Gam 5$, except that, after extracting the witness $(\ssk_{\fcecUser_j}, \allowbreak \cecsn, \allowbreak r, \allowbreak o_c, \allowbreak \langle l_k, \allowbreak o_{a_k}, \allowbreak \mu_k, \allowbreak o_{\mu_k} \rangle_{k=0}^{V-1})$ from a proof $\pi_v$, $\Gam 6$ takes the payment $\cecpayment \allowbreak = \allowbreak (\kappa, \allowbreak \sigma', \allowbreak \langle S_k, \allowbreak T_k, \allowbreak A_k \rangle_{k\in[0,V-1]}, \allowbreak V, \allowbreak C, \allowbreak \pi_v)$ that contains $\pi_v$ and the associated payment information $\cecpaymentinfo$ and computes the serial numbers $S'_{k} \allowbreak \gets \allowbreak \delta^{1/(\cecsn+l_k+1)}$ and the double spending tags $T'_k \allowbreak \gets \allowbreak \ga^{\ssk_{\fcecUser_j} + R_{k}/(\cecsn+l_k+1)}$, where $R_k \allowbreak = \allowbreak H'(\cecpaymentinfo, \allowbreak k)$. Then $\Gam 6$ outputs failure if, for any $k \allowbreak \in \allowbreak [0,V-1]$, $S'_k \allowbreak \neq \allowbreak S_k$ or $T'_k \allowbreak \neq \allowbreak T_k$. Under the hardness of the discrete logarithm problem, $|\Pr[\Gam 6]-\Pr[\Gam 5]| \allowbreak \leq \allowbreak \Adv_{\Adversary}^{\mathsf{dlog}}$.
    \begin{proof}
     We have an adversary $\Adversary$ that, with non-negligible probability, sends payment information $\cecpaymentinfo$ and a payment   $\cecpayment \allowbreak = \allowbreak (\kappa, \allowbreak \sigma', \allowbreak \langle S_k, \allowbreak T_k, \allowbreak A_k \rangle_{k\in[0,V-1]}, \allowbreak V, \allowbreak C, \allowbreak \pi_v)$ such that, after extracting from $\pi_v$ the witness $(\ssk_{\fcecUser_j}, \allowbreak \cecsn, \allowbreak r, \allowbreak o_c, \allowbreak \langle l_k, \allowbreak o_{a_k}, \allowbreak \mu_k, \allowbreak o_{\mu_k} \rangle_{k=0}^{V-1})$ and computing $S'_{k} \allowbreak \gets \allowbreak \delta^{1/(\cecsn+l_k+1)}$ and $T'_k \allowbreak \gets \allowbreak \ga^{\ssk_{\fcecUser_j} + R_{k}/(\cecsn+l_k+1)}$ for all $k \allowbreak \in \allowbreak [0,V-1]$, we have that $S'_k \allowbreak \neq \allowbreak S_k$ or  $T'_k \allowbreak \neq \allowbreak T_k$ for some $k \allowbreak \in \allowbreak [0,V-1]$. We construct an algorithm $B$ that uses that adversary to solve the discrete logarithm problem.

    $B$ works as follows. Given an instance $(h, \allowbreak h^x)$ of the discrete logarithm problem, when running the functionality $\Functionality_{\KG}$ to set up the parameters $\cecparams \gets (\p,\Ga,\Gb,\allowbreak \Gt,\e,\ga,\gb,\gamma_1,\gamma_2,\delta, L)$, $B$ sets $\ga \allowbreak \gets \allowbreak  h$ and $\gamma_1 \allowbreak \gets \allowbreak h^x$. When the adversary outputs a payment $\cecpayment$ that fulfills the condition described above for a certain $k \allowbreak \in \allowbreak [0,V-1]$, $B$ outputs
    \begin{equation*}
    x \allowbreak \gets \allowbreak \frac{(o_{a_k}+o_c)\mu_k+o_{\mu_k}}{1-((\cecsn+l_k+1)\mu_k)}
    \end{equation*}
    We show that the discrete logarithm $x$ is computed correctly as follows. The non-interactive ZK proof of knowledge $\pi_v$ is described by
    \begin{align*}
        \pi_v =  & \NIZK\{(\ssk_{\fcecUser_j}, \cecsn, r, o_c, \langle l_k, o_{a_k}, \mu_k, o_{\mu_k} \rangle_{k=0}^{V-1}): \\
        &\kappa = \tilde{\alpha} \tilde{\beta}_1^{\ssk_{\fcecUser_j}} \tilde{\beta}_2^{\cecsn} \gb^{r}\ \land\ C = \ga^{o_c} \gamma_1^{\cecsn}\ \land\  \\&
        \langle A_k = \ga^{o_{a_k}} \gamma_1^{l_k}\ \land\ l_k \in [0,L-1]\ \land\ \\&
        S_k=\delta^{\mu_k}\ \land\ \gamma_1 = (A_k C\gamma_1)^{\mu_k} \ga^{o_{\mu_k}}\ \land\ \\&
        T_k=\ga^{\ssk_{\fcecUser_j}} (\ga^{R_{k}})^{\mu_k}\  \rangle_{k\in[0,V-1]}
        \}
    \end{align*}
    After successful extraction of the witness $(\ssk_{\fcecUser_j}, \allowbreak \cecsn, \allowbreak r, \allowbreak o_c, \allowbreak \langle l_k, \allowbreak o_{a_k}, \allowbreak \mu_k, \allowbreak o_{\mu_k} \rangle_{k=0}^{V-1})$, we know that the statements proven by $\pi_v$ hold. In the case that $S'_k \allowbreak \neq \allowbreak S_k$ for some $k \allowbreak \in \allowbreak [0,V-1]$, we have that
    \begin{equation*}
    \delta^{1/(\cecsn+l_k+1)} \neq \delta^{\mu_k}
    \end{equation*}
    and thus $1/(\cecsn+l_k+1) \neq {\mu_k}$.
    In the case that $T'_k \allowbreak \neq \allowbreak T_k$ for some $k \allowbreak \in \allowbreak [0,V-1]$, we have that
    \begin{equation*}
    \ga^{\ssk_{\fcecUser_j} + R_{k}/(\cecsn+l_k+1)} \neq \ga^{\ssk_{\fcecUser_j}} (\ga^{R_{k}})^{\mu_k}
    \end{equation*}
    and thus we can also deduce that $1/(\cecsn+l_k+1) \neq {\mu_k}$.

    We also have the following:
    \begin{align*}
    \gamma_1 &=  (A_k C\gamma_1)^{\mu_k} \ga^{o_{\mu_k}} \\&
    = (\ga^{o_{a_k}} \gamma_1^{l_k} \ga^{o_c} \gamma_1^{\cecsn} \gamma_1)^{\mu_k} \ga^{o_{\mu_k}} \\ &
    = \gamma_1^{(\cecsn+l_k+1)\mu_k} \ga^{(o_{a_k}+o_c)\mu_k+o_{\mu_k}}
    \end{align*}
    and therefore
    \begin{equation*}
    \gamma_1 = \ga^{\frac{(o_{a_k}+o_c)\mu_k+o_{\mu_k}}{1-((\cecsn+l_k+1)\mu_k)}}
    \end{equation*}
    This equation shows that, when $1/(\cecsn+l_k+1) \allowbreak \neq \allowbreak {\mu_k}$, we can compute the discrete logarithm $x$ as described above.
    \end{proof}

	\item[$\Gam 7$:] This game proceeds as $\Gam 6$, except that in $\Gam 7$ the non-interactive ZK proofs of knowledge $\pi_v$ that are sent to the adversary are replaced by simulated proofs computed by the simulator $\mathcal{S}_{v}$. Under the zero-knowledge property of the proof system (see Definition~\ref{def:zero-knowledge}), we have that $|\Pr[\Gam 7]-\Pr[\Gam 6]| \allowbreak \leq \allowbreak \Adv_{\Adversary}^{\mathsf{zk}}$.

	\item[$\Gam 8$:] This game proceeds as $\Gam 7$, except that in $\Gam 8$, for the payments $\cecpayment \allowbreak \gets \allowbreak (\kappa, \allowbreak \sigma', \allowbreak \langle S_k, \allowbreak T_k, \allowbreak A_k \rangle_{k\in[0,V-1]}, \allowbreak V, \allowbreak C, \allowbreak \pi_v)$ that are sent to the adversary, the values $\kappa$ and $\sigma'$ are computed as follows:
	\begin{itemize}

	    \item Pick random $t \gets \Zp$ and $t' \gets \Zp$.

        \item Compute $\sigma' = (\h',s') \gets (\ga^{t'}, \ga^{t t'})$.

        \item Compute $\kappa \gets \gb^{t}$.

	\end{itemize}
	As shown below, $|\Pr[\Gam 8]-\Pr[\Gam 7]| = 0$.
	\begin{proof}
	 This proof follows the proof in~\cite{cryptoeprint:2022:011}. We show that values $\kappa$ and $\sigma'$ follow the same distribution as the ones computed by the honest user in the real-world protocol. Observe that the honest user computes the following:
    \begin{itemize}

        \item Pick random $r \gets \Zp$ and $r' \gets \Zp$.

        \item Set $\sigma' = (\h',s') \gets (\h^{r'}, s^{r'}(\h')^{r})$, where we have that
        \begin{align*}
             (\h^{r'}, s^{r'}(\h')^{r}) & = (\h^{r'}, \allowbreak \h^{(x+\ssk_{\fcecUser_j} y_{1} + \cecsn y_{2} + r)r'}) \\ &
             = (\ga^{u r'}, \ga^{(x+\ssk_{\fcecUser_j} y_{1} + \cecsn y_{2} + r) u r'})
        \end{align*}
        \item Set $\kappa \gets \alpha \beta_1^{\ssk_{\fcecUser_j}} \beta_2^{\cecsn} \gb^{r} = \gb^{x+\ssk_{\fcecUser_j} y_{1} + \cecsn y_{2} + r}$.

    \end{itemize}
    Therefore, $t$ corresponds to $(x+\ssk_{\fcecUser_j} y_{1} + \cecsn y_{2} + r_l)$ and $t'$ corresponds to $u r'$, where $u$ is a random value such that $\h = \ga^{u}$. Both $(x+ \ssk_{\fcecUser_j} y_{1} + \cecsn y_{2} + r)$ and $u r'$ are random. Observe as well that the verification equation $\e(\h', \kappa) = \e(s',\gb)$ still holds because $\e(\ga^{t'}, \gb^{t}) = \e(\ga^{t t'},\gb)$.
	\end{proof}

	\item[$\Gam 9$:] This game proceeds as $\Gam 8$, except that in $\Gam 9$ the non-interactive ZK proofs of knowledge $\pi_s$ that are sent to the adversary are replaced by simulated proofs computed by the simulator $\mathcal{S}_{s}$. Under the zero-knowledge property of the proof system (see Definition~\ref{def:zero-knowledge}), we have that $|\Pr[\Gam 9]-\Pr[\Gam 8]| \allowbreak \leq \allowbreak \Adv_{\Adversary}^{\mathsf{zk}}$.

	\item[$\Gam 10$:] This game proceeds as $\Gam 9$, except that in $\Gam 10$, in each request that is sent to the adversary, the values $(\com_1, \allowbreak \com_2)$ are replaced by $\com_{1} \allowbreak \gets \allowbreak \ga^{o_{1}} \h^{\cecmes_{1}}$ and $\com_{2} \allowbreak \gets \allowbreak \ga^{o_{2}} \h^{\cecmes_{2}}$, where $(o_1, \allowbreak \cecmes_1, \allowbreak o_2, \allowbreak \cecmes_2)$ are random values in $\Zp$. At this point, the non-interactive ZK proofs of knowledge $\pi_s$ are simulated proofs of false statements. We also remark that, after $\Gam 8$, the computation of payment messages does not use the signatures obtained in the issuance phase. Since the values $(\com_1, \allowbreak \com_2)$ are uniformly distributed at random, this change does not alter the view of the environment and we have that $|\Pr[\Gam 10]-\Pr[\Gam 9]| = 0$.

	\item[$\Gam 11$:] This game proceeds as $\Gam 10$, except that in $\Gam 11$, in each request that is sent to the adversary, the value $\com$ is replaced by picking random $\com \allowbreak \gets \allowbreak \Ga$. Under the hiding property of the commitment scheme,  $|\Pr[\Gam 11]-\Pr[\Gam 10]| \allowbreak \leq \allowbreak N \cdot\Adv_{\Adversary}^{\mathsf{hid}}$, where $N$ is the number of commitments $\com$ sent to the adversary. Since the Pedersen commitment scheme is perfectly hiding, $|\Pr[\Gam 11]-\Pr[\Gam 10]| \allowbreak = \allowbreak 0$.

	\begin{proof}
	The proof uses a sequence of games $\Gam 10.i$, for $i=0$ to $N$. $\Gam 10.0$ is equal to $\Gam 10$, whereas $\Gam 10.N$ is equal to $\Gam 11$. In $\Gam 10.i$, the first $i$ commitments sent to the adversary are set to random values, whereas the remaining ones are set as in $\Gam 10$.

	Given an adversary that distinguishes between $\Gam 10.i$ and $\Gam 10.(i+1)$ with non-negligible probability, we construct an algorithm $B$ that breaks the hiding property of the commitment scheme. $B$ works as follows. $B$ receives the parameters of the Pedersen commitment scheme $\paramscom \allowbreak = (\ga, \allowbreak \gamma_1, \allowbreak \gamma_2)$ from the challenger. When running $\Functionality_{\KG}$, $B$ uses those parameters to set the values $(\ga, \allowbreak \gamma_1, \allowbreak \gamma_2)$ in the parameters $\cecparams \allowbreak \gets \allowbreak (\p, \allowbreak \Ga, \allowbreak \Gb, \allowbreak \Gt, \allowbreak \e, \allowbreak \ga, \allowbreak \gb, \allowbreak \gamma_1, \allowbreak \gamma_2, \allowbreak \delta, \allowbreak L)$. To compute the first $i$ commitments, $B$ sets $\com$ to random. To compute the commitment $i+1$, $B$ sends the messages $(\ssk_{\fcecUser_j}, \allowbreak \cecsn)$ to the challenger. $B$ sets $\com$ to the challenge commitment received from the challenger. As can be seen, if the challenge commitment commits to $(\ssk_{\fcecUser_j}, \allowbreak \cecsn)$, then we are in $\Gam 10.i$, whereas if the challenge commitment commits to a random message, then we are in $\Gam 10.(i+1)$. The remaining commitments are computed as in $\Gam 10$. $B$ sends the adversarial guess to distinguish between $\Gam 10.i$ and $\Gam 10.(i+1)$ to the challenger of the hiding game.
	\end{proof}

   	\item[$\Gam 12$:] This game proceeds as $\Gam 11$, except that in $\Gam 12$, for the payments $\cecpayment \allowbreak \gets \allowbreak (\kappa, \allowbreak \sigma', \allowbreak \langle S_k, \allowbreak T_k, \allowbreak A_k \rangle_{k\in[0,V-1]}, \allowbreak V, \allowbreak C, \allowbreak \pi_v)$ that are sent to the adversary, for all $k \allowbreak \in \allowbreak [0,V-1]$, the values $A_k \allowbreak \gets \allowbreak \Ga$ are set to random elements in $\Ga$. The value $C$ is also set to a random element in $\Ga$. Under the hiding property of the commitment scheme, we have that $|\Pr[\Gam 12]-\Pr[\Gam 11]| \allowbreak \leq \allowbreak N \cdot\Adv_{\Adversary}^{\mathsf{hid}}$, where $N$ is the number of commitments $C$ and $A_k$ sent to the adversary. Since the Pedersen commitment scheme is perfectly hiding, $|\Pr[\Gam 12]-\Pr[\Gam 11]| \allowbreak = \allowbreak 0$. We omit the proof, which is similar to the proof of indistinguishability between $\Gam 10$ and $\Gam 11$.

	\item[$\Gam 13$:] This game proceeds as $\Gam 12$, except that in $\Gam 13$, for the payments $\cecpayment \allowbreak \gets \allowbreak (\kappa, \allowbreak \sigma', \allowbreak \langle S_k, \allowbreak T_k, \allowbreak A_k \rangle_{k\in[0,V-1]}, \allowbreak V, \allowbreak C, \allowbreak \pi_v)$ that are sent to the adversary,  for all $k \allowbreak \in \allowbreak [0,V-1]$, the values $S_k$ and $T_k$ are computed by doing $S_k \allowbreak \gets \allowbreak \delta^{r_k}$ and $T_k \allowbreak \gets \allowbreak \ga^{\ssk_{\fcecUser_j}}  (\ga^{r_k})^{R_{k}}$, where $r_k \gets \Zp$ is picked up randomly. Under the pseudorandomness property of the pseudorandom function, we have that $|\Pr[\Gam 13]-\Pr[\Gam 12]| \allowbreak \leq \allowbreak \Adv_{\Adversary}^{\mathsf{pseu}}$.
	\begin{proof}
	 Given an adversary that is able to distinguish $\Gam 12$ from $\Gam 13$ with non-negligible probability, we construct an algorithm $B$ that breaks the pseudorandomness property of the pseudorandom function $f_{\G,\p,\ga,s}(\cdot)$ described in~\S\ref{subsec:pseudorandomfunction}. $B$ receives from the challenger the parameters $(\Ga, \allowbreak \p, \allowbreak \ga)$. When running $\Functionality_{\KG}$, $B$ picks up random $a \allowbreak \in \allowbreak \Zp$, computes $\delta \allowbreak \gets \allowbreak \ga^{1/a}$ and sets the parameters $\cecparams \allowbreak \gets \allowbreak (\p, \allowbreak \Ga, \allowbreak \Gb, \allowbreak \Gt, \allowbreak \e, \allowbreak \ga, \allowbreak \gb, \allowbreak \gamma_1, \allowbreak \gamma_2, \allowbreak \delta, \allowbreak L)$. We remark that, after $\Gam 11$, request messages are computed without requiring knowledge of the coin secret $\cecsn$, and so $B$ does not need to know the secret $s$ of the pseudorandom function to compute them. Similarly, after $\Gam 12$, payment messages are computed without requiring knowledge of the coin secret $\cecsn$.  To compute a payment $\cecpayment \allowbreak \gets \allowbreak (\kappa, \allowbreak \sigma', \allowbreak \langle S_k, \allowbreak T_k, \allowbreak A_k \rangle_{k\in[0,V-1]}, \allowbreak V, \allowbreak C, \allowbreak \pi_v)$ to be sent to the adversary, $B$ follows the changes described up to $\Gam 12$ and, additionally, to compute $\langle S_k, \allowbreak T_k \rangle_{k\in[0,V-1]}$, $B$ does the following. For all $k\in[0,V-1]$, $B$ sends the coin index $l_k$ to the oracle of the challenger, which provides a response $Z_k$. $B$ sets $T_k \allowbreak \gets \allowbreak \ga^{\ssk_{\fcecUser_j}}  (Z_k)^{R_{k}}$ and $S_k \allowbreak \gets \allowbreak Z_k^{1/a}$. As can be seen, if $Z_k = f_{\G,\p,\ga,s}(l_k)$, $T_k$ and $S_k$ are computed as in $\Gam 12$, whereas if $Z_k$ is random, $Z_k$ are computed as in $\Gam 13$. Therefore, $B$ uses the guess of the adversary to distinguish between $\Gam 12$ and $\Gam 13$ in order to break the pseudorandomness property of the pseudorandom function.
	\end{proof}

	\item[$\Gam 14$:] This game proceeds as $\Gam 13$, except that in $\Gam 14$, for the payments $\cecpayment \allowbreak \gets \allowbreak (\kappa, \allowbreak \sigma', \allowbreak \langle S_k, \allowbreak T_k, \allowbreak A_k \rangle_{k\in[0,V-1]}, \allowbreak V, \allowbreak C, \allowbreak \pi_v)$ that are sent to the adversary,  for all $k \allowbreak \in \allowbreak [0,V-1]$, the values $S_k$ and $T_k$ are computed by picking random $S_k \allowbreak \gets \Ga$ and $T_k \allowbreak \gets \allowbreak \Ga$. Under the external Diffie-Hellman (XDH)  assumption in $\Ga$, we have that $|\Pr[\Gam 14]-\Pr[\Gam 13]| \allowbreak \leq \allowbreak N_{s} \cdot \Adv_{\Adversary}^{\mathsf{xdh}}$, where $N_{s}$ is the number of serial numbers and double spending tags sent to the adversary.
	\begin{proof}
	The proof uses a sequence of games $\Gam 13.i$, for $i=0$ to $N_{s}$. $\Gam 13.0$ is equal to $\Gam 13$, whereas $\Gam 13.(N_{s})$ is equal to $\Gam 14$. In $\Gam 13.i$, the values $S_k$ and $T_k$ of the first $i$ payments sent to the adversary are set to random values, whereas in the remaining payments they are set as in $\Gam 13$.

	Given an adversary that distinguishes between $\Gam 13.i$ and $\Gam 13.(i+1)$ with non-negligible probability, we construct an algorithm $B$ that uses that adversary to solve the XDH problem with non-negligible probability. $B$ works as follows. Given an instance $(h, \allowbreak h^a, \allowbreak h^b, \allowbreak Z)$ of the XDH problem in $\Ga$, when running $\Functionality_{\KG}$ to set up the parameters $\cecparams \allowbreak \gets \allowbreak (\p, \allowbreak \Ga, \allowbreak \Gb, \allowbreak \Gt, \allowbreak \e,\allowbreak \ga,\allowbreak \gb,\allowbreak \gamma_1,\allowbreak \gamma_2,\allowbreak \delta, \allowbreak L)$, $B$ sets $\ga \allowbreak \gets \allowbreak \h^a$ and $\delta \allowbreak \gets \allowbreak h$. When setting the $i+1$ serial number and double spending tag sent to the adversary, $B$ sets $S_k \allowbreak \gets \allowbreak \h^b$ and $T_k \allowbreak \gets \allowbreak \ga^{\ssk_{\fcecUser_j}}  (Z)^{R_{k}}$. As can be seen, when $Z$ is random, $S_k$ and $T_k$ are random values and we are thus in $\Gam 13.(i+1)$. In contrast, when $Z \allowbreak = \allowbreak h^{ab}$, we have that $S_k \allowbreak = \allowbreak h^b \allowbreak = \allowbreak \delta^b$ and $T_k \allowbreak = \allowbreak (h^a)^{\ssk_{\fcecUser_j}} (h^{ab})^{R_{k}} = \allowbreak (\ga)^{\ssk_{\fcecUser_j}} (\ga^{b})^{R_{k}}$, and thus the distribution is equal to that of $\Gam 13.i$. Therefore, $B$ can use the guess of the adversary to distinguish between $\Gam 13.i$ and $\Gam 13.(i+1)$ in order to solve the XDH problem in $\Ga$ with non-negligible probability.
	\end{proof}

    \item[$\Gam 15$:] This game proceeds as $\Gam 14$, except that $\Gam 15$ checks that algorithm $\cecIdentify$ identifies the double spender when there is double spending. To do that, when the adversary sends two valid payments $(\cecpayment, \allowbreak \cecpaymentinfo)$ and $(\cecpayment', \allowbreak \cecpaymentinfo')$, after extracting the witness $\langle \ssk_{\fcecUser_j}, \allowbreak \cecsn, \allowbreak r, \allowbreak o_c, \allowbreak \langle l_k, \allowbreak o_{a_k}, \allowbreak \mu_k, \allowbreak o_{\mu_k} \rangle_{k=0}^{V-1} \rangle)$ from the proofs $\pi_v \allowbreak \in \allowbreak \cecpayment$ and $\langle \ssk'_{\fcecUser_j}, \allowbreak \cecsn', \allowbreak r', \allowbreak o'_c, \allowbreak \langle l'_k, \allowbreak o'_{a_k}, \allowbreak \mu'_k, \allowbreak o'_{\mu_k} \rangle_{k=0}^{V'-1} \rangle)$  from the proof $\pi'_v \allowbreak \in \allowbreak \cecpayment'$, $\Gam 15$ checks whether $\ssk'_{\fcecUser_j} \allowbreak = \allowbreak \ssk_{\fcecUser_j}$, $\cecsn' \allowbreak = \allowbreak \cecsn$ and $\langle l'_k \rangle_{k\in[0,V'-1]} \cap \langle l_k \rangle_{k\in[0,V-1]} \allowbreak \neq \allowbreak \emptyset$, which means that there is a double spending. In that case $\Gam 15$ sets $\cecPK$ to contain all the registered public keys and runs the algorithm $c \allowbreak \gets \allowbreak \cecIdentify(\cecparams, \allowbreak \cecPK, \allowbreak \cecpayment, \allowbreak \cecpayment', \allowbreak \cecpaymentinfo, \allowbreak \cecpaymentinfo')$. Then $\Gam 15$ does the following:
    \begin{itemize}

                \item If $\cecpaymentinfo \allowbreak = \allowbreak \cecpaymentinfo'$ and $c \allowbreak \neq \allowbreak \cecpaymentinfo$, $\Gam 15$ outputs failure.

                \item If $\cecpaymentinfo \allowbreak \neq \allowbreak \cecpaymentinfo'$ and $c \allowbreak \neq \allowbreak \spk_{\fcecUser_j}$, where $\spk_{\fcecUser_j}$ is the public key associated with secret key $\ssk_{\fcecUser_j}$, $\Gam 15$ outputs failure.

    \end{itemize}
    The probability that $\Gam 15$ fails is negligible under the collision-resistance property of the hash function $H'$, i.e. we have that $|\Pr[\Gam 15]-\Pr[\Gam 14]| \allowbreak \leq \allowbreak \Adv_{\Adversary}^{\mathsf{col-res}}$.
    \begin{proof}
    In $\Gam 6$, we have shown that, if a payment is valid, under the hardness of the discrete logarithm assumption, the serial numbers $S_k$ and the double-spending tags $T_k$ are correctly computed. Hence, if there are two payments with witnesses  $\langle \ssk_{\fcecUser_j}, \allowbreak \cecsn, \allowbreak r, \allowbreak o_c, \allowbreak \langle l_k, \allowbreak o_{a_k}, \allowbreak \mu_k, \allowbreak o_{\mu_k} \rangle_{k=0}^{V-1} \rangle)$ and $\langle \ssk'_{\fcecUser_j}, \allowbreak \cecsn', \allowbreak r', \allowbreak o'_c, \allowbreak \langle l'_k, \allowbreak o'_{a_k}, \allowbreak \mu'_k, \allowbreak o'_{\mu_k} \rangle_{k=0}^{V'-1} \rangle)$ such that $\ssk'_{\fcecUser_j} \allowbreak = \allowbreak \ssk_{\fcecUser_j}$, $\cecsn' \allowbreak = \allowbreak \cecsn$ and $\langle l'_k \rangle_{k\in[0,V'-1]} \cap \langle l_k \rangle_{k\in[0,V-1]} \allowbreak \neq \allowbreak \emptyset$, the proof in $\Gam 6$ guarantees that, for those coin indices such that $l'_{k'} \allowbreak = \allowbreak l_k$ (where $k' \allowbreak \in \allowbreak [0,V'-1]$ and $k \allowbreak \in \allowbreak [0,V-1]$), it is the case that $S'_{k'} \allowbreak = \allowbreak S_k = \allowbreak \delta^{1/(\cecsn+l_k+1)}$. Therefore, the algorithm $\cecIdentify$ always detects double spending.

    After detecting double spending, the algorithm $\cecIdentify$ checks if $\cecpaymentinfo \allowbreak = \allowbreak \cecpaymentinfo'$ and in that case sets $c \allowbreak = \allowbreak \cecpaymentinfo$. Therefore, the first condition under which $\Gam 15$ fails never happens.

    If $\cecpaymentinfo \allowbreak \neq \allowbreak \cecpaymentinfo'$, algorithm $\cecIdentify$ computes
    \begin{equation*}
    \spk_{\fcecUser_j} \gets ((T'_{k'})^{R_{k}}/T_{k}^{R'_{k'}})^{(R_{k}-R'_{k'})^{-1}}
    \end{equation*}
    We have that $R_k \allowbreak = \allowbreak H'(\cecpaymentinfo,k)$ and $R'_{k'} \allowbreak = \allowbreak H'(\cecpaymentinfo',k')$. The proof in $\Gam 6$ also guarantees that the double spending tags are correctly computed. Hence, we know that $T_k=\ga^{\ssk_{\fcecUser_j}} (\ga^{1/(\cecsn+l_k+1)})^{R_{k}}$ and $T'_{k'}=\ga^{\ssk_{\fcecUser_j}} (\ga^{1/(\cecsn+l_{k}+1)})^{R'_{k'}}$. Let $Z \allowbreak = \allowbreak \ga^{1/(\cecsn+l_k+1)}$. We have that
    \begin{align*}
    \spk_{\fcecUser_j} & = ((T'_{k'})^{R_{k}}/T_{k}^{R'_{k'}})^{(R_{k}-R'_{k'})^{-1}} \\ & = ((\ga^{\ssk_{\fcecUser_j}} (Z)^{R'_{k'}})^{R_k} / (\ga^{\ssk_{\fcecUser_j}} (Z)^{R_{k}})^{R'_{k'}})^{(R_{k}-R'_{k'})^{-1}} \\&
    = ((\ga^{\ssk_{\fcecUser_j} R_k} (Z)^{R'_{k'}R_k}) / (\ga^{\ssk_{\fcecUser_j}R'_{k'}} (Z)^{R_{k}R'_{k'}}))^{(R_{k}-R'_{k'})^{-1}} \\&
    = (\ga^{\ssk_{\fcecUser_j} R_k}  / \ga^{\ssk_{\fcecUser_j}R'_{k'}} )^{(R_{k}-R'_{k'})^{-1}} \\ &
    = (\ga^{\ssk_{\fcecUser_j} (R_k-R'_{k'})}  )^{(R_{k}-R'_{k'})^{-1}} \\ &
    = \ga^{\ssk_{\fcecUser_j}}
    \end{align*}
    Therefore, algorithm $\cecIdentify$ outputs the public key of the double spender, except when $R_{k} \allowbreak = \allowbreak R'_{k'}$. Given that $R_k \allowbreak = \allowbreak H'(\cecpaymentinfo,k)$ and $R'_{k'} \allowbreak = \allowbreak H'(\cecpaymentinfo',k')$, if $R_{k} \allowbreak = \allowbreak R'_{k'}$, a collision for the hash function $H'$ has been found.
    \end{proof}

    \item[$\Gam 16$:] This game proceeds as $\Gam 15$, except that, even if there is not double spending, $\Gam 16$ sets $\cecPK$ to contain all the registered public keys and runs the algorithm $c \allowbreak \gets \allowbreak \cecIdentify(\cecparams, \allowbreak \cecPK, \allowbreak \cecpayment, \allowbreak \cecpayment', \allowbreak \cecpaymentinfo, \allowbreak \cecpaymentinfo')$. If it is the case that $c \allowbreak = \allowbreak \spk_{\fcecUser_j}$, where $\spk_{\fcecUser_j}$ is a public key associated with an honest user, $\Gam 16$ outputs failure. We show that $\Gam 16$ outputs failure with negligible probability under the hardness of the discrete logarithm problem, i.e. $|\Pr[\Gam 16]-\Pr[\Gam 15]| \allowbreak \leq \allowbreak N_{u} \allowbreak \cdot \allowbreak \Adv_{\Adversary}^{\mathsf{dlog}}$, where $N_u$ is the number of public keys of honest users.
    \begin{proof}
    Given an adversary that makes $\Gam 16$ fail with non-negligible probability, we construct an algorithm $B$ that solves the discrete logarithm problem with non-negligible probability. $B$ works as follows. $B$ receives an instance $(h, \allowbreak h^x)$ of the discrete logarithm problem from the challenger. When running $\Functionality_{\KG}$, $B$ sets $\ga \allowbreak \gets h$ and sets the parameters $\cecparams \allowbreak \gets \allowbreak (\p, \allowbreak \Ga, \allowbreak \Gb, \allowbreak \Gt, \allowbreak \e, \allowbreak \ga, \allowbreak \gb, \allowbreak \gamma_1, \allowbreak \gamma_2, \allowbreak \delta, \allowbreak L)$. $B$ picks randomly an honest user $\fcecUser_j$ and, when that user registers her public key, $B$ sets $\spk_{\fcecUser_j} \allowbreak \gets \allowbreak \h^x$. We remark that, since $\Gam 9$, knowledge of $\ssk_{\fcecUser_j}$ is not needed to compute request messages. We also remark that, since $\Gam 8$, knowledge of $\ssk_{\fcecUser_j}$ is not needed to compute payment messages. Therefore, $B$ can simulate those messages without knowledge of $\ssk_{\fcecUser_j}$.

    At some point $B$ receives from the adversary two payments $(\cecpayment, \allowbreak \cecpaymentinfo)$ and $(\cecpayment', \allowbreak \cecpaymentinfo')$ that make $\Gam 16$ fail. If the public key $\spk_{\fcecUser_j}$ obtained after running $\cecIdentify$ is different from the value $h^x$ received from the challenger, $B$ fails. Otherwise $B$ computes $\ssk_{\fcecUser_j}$ as follows. First, $B$ extracts the witness $\langle \ssk, \allowbreak \cecsn, \allowbreak r, \allowbreak o_c, \allowbreak \langle l_k, \allowbreak o_{a_k}, \allowbreak \mu_k, \allowbreak o_{\mu_k} \rangle_{k=0}^{V-1} \rangle)$ from the proofs $\pi_v \allowbreak \in \allowbreak \cecpayment$ and $\langle \ssk', \allowbreak \cecsn', \allowbreak r', \allowbreak o'_c, \allowbreak \langle l'_k, \allowbreak o'_{a_k}, \allowbreak \mu'_k, \allowbreak o'_{\mu_k} \rangle_{k=0}^{V'-1} \rangle)$ from the proof $\pi'_v \allowbreak \in \allowbreak \cecpayment'$. Given that double spending has been detected, and that in $\Gam 6$ we proved that serial numbers and double spending tags are correctly computed, we know that there is $S_k \in \cecpayment$ and $S'_{k'} \in \cecpayment'$ such that $S_k = S'_{k'} = \delta^b$, where $b= (1/(\cecsn+l_k+1)) = (1/(\cecsn'+l'_{k'}+1))$. Therefore, we also know that the double spending tags are of the form $T_k \allowbreak = \allowbreak \ga^{\ssk + b R_{k}}$ and $T'_{k'} \allowbreak = \allowbreak \ga^{\ssk' + b R'_{k'}}$. From the computation of algorithm $\cecIdentify$, we have that
    \begin{align*}
    \spk_{\fcecUser_j} & = ((T'_{k'})^{R_{k}}/T_{k}^{R'_{k'}})^{(R_{k}-R'_{k'})^{-1}} \\ &
    = ((\ga^{\ssk' + b R'_{k'}})^{R_{k}}/(\ga^{\ssk + b R_{k}})^{R'_{k'}})^{(R_{k}-R'_{k'})^{-1}} \\ &
    = ((\ga^{R_{k} \ssk' + b R_{k} R'_{k'}})/(\ga^{R'_{k'} \ssk + b R_{k} R'_{k'}}))^{(R_{k}-R'_{k'})^{-1}} \\ &
    = (\ga^{R_{k} \ssk' - R'_{k'} \ssk })^{(R_{k}-R'_{k'})^{-1}} \\ &
    = (\ga^{(R_{k} \ssk' - R'_{k'} \ssk)/ (R_{k}-R'_{k'})}) \\ &
    \end{align*}
    Therefore, $B$ computes $x \gets (R_{k} \ssk' - R'_{k'} \ssk)/ (R_{k}-R'_{k'})$ to solve the discrete logarithm problem. If the adversary succeeds with probability $\alpha$, $B$ succeeds with probability $\alpha/N_{u}$, where $N_u$ is the number of public keys of honest users.
    \end{proof}

    \item[$\Gam 17$:] This game proceeds as $\Gam 16$, except that $\Gam 17$ outputs failure when the serial number of a payment computed by an honest user is equal to a serial number of another payment. The probability that two serial numbers have the same value is bounded by $N_s/|\Ga|$, where $N_s$ is the number of serial numbers and $|\Ga|$ is the size of $\Ga$. Additionally, we show below that the probability that an adversarial user computes a payment with a serial number that is equal to a serial number in a payment computed by an honest user is negligible thanks to the hardness of the discrete logarithm problem. Therefore, we have that $|\Pr[\Gam 17]-\Pr[\Gam 16]| \allowbreak \leq \allowbreak N_s/|\Ga| + N_{s} \allowbreak \cdot \allowbreak \Adv_{\Adversary}^{\mathsf{dlog}}$.
    \begin{proof}
    Given an adversary that makes $\Gam 17$ fail with non-negligible probability, we construct an algorithm $B$ that solves the discrete logarithm problem with non-negligible probability. $B$ works as follows. $B$ receives an instance $(h, \allowbreak h^x)$ of the discrete logarithm problem from the challenger. When running $\Functionality_{\KG}$, $B$ sets $\delta \allowbreak \gets h$ and sets the parameters $\cecparams \allowbreak \gets \allowbreak (\p, \allowbreak \Ga, \allowbreak \Gb, \allowbreak \Gt, \allowbreak \e, \allowbreak \ga, \allowbreak \gb, \allowbreak \gamma_1, \allowbreak \gamma_2, \allowbreak \delta, \allowbreak L)$. $B$ picks randomly a serial number in a payment computed by an honest user and sets $S_k \allowbreak \gets \allowbreak h^x$. We recall that, since $\Gam 14$, serial numbers in payments computed by honest users are random values in $\Ga$.

    At some point, $B$ finds that a payment received from the adversary has a serial number that is equal to a serial number in a payment computed by an honest user. If that serial number is not equal to $h^x$, $B$ fails. Otherwise $B$ extracts the witness $\langle \ssk, \allowbreak \cecsn, \allowbreak r, \allowbreak o_c, \allowbreak \langle l_k, \allowbreak o_{a_k}, \allowbreak \mu_k, \allowbreak o_{\mu_k} \rangle_{k=0}^{V-1} \rangle)$ from the proof $\pi_v$ in the payment $\cecpayment$ sent by the adversary. Thanks to the proof in $\Gam 6$, we know that the serial number $S_k \allowbreak = \allowbreak h^x$ in $\cecpayment$ is of the form $S_k \allowbreak = \allowbreak \delta^{1/(\cecsn+l_k+1)}$. Therefore, $B$ outputs $x \allowbreak \gets \allowbreak 1/(\cecsn+l_k+1)$. If the adversary succeeds with probability $\alpha$, $B$ succeeds with probability $\alpha/N_{s}$, where $N_s$ is the number of serial numbers.
    \end{proof}

\end{description}
The distribution of $\Gam 17$ is identical to that of our simulation. In $\Gam 17$, the request message is computed without knowledge of the values $\ssk_{\fcecUser_j}$ and $\cecsn$. The payment is computed without knowledge of the signatures and without knowledge of the signed messages $\ssk_{\fcecUser_j}$ and $\cecsn$. Additionally, it is guaranteed that the adversary cannot compute a payment on a wallet defined by $\ssk_{\fcecUser_j}$ and $\cecsn$ unless the adversary obtained enough signatures from honest authorities. It is also guaranteed that, if the adversary double spends coins, then an adversarial user or provider will be identified. Moreover, it is guaranteed that an honest user or provider will not be found guilty of double spending. The overall advantage of the environment to distinguish between the real and the ideal protocol is $|\Pr[\Gam 17]-\Pr[\Gam 0]| \leq \Adv_{\Adversary}^{\mathsf{ext}} + \Adv_{\Adversary}^{\mathsf{bin}} + \Adv_{\Adversary}^{\mathsf{unf}}\cdot ((n-\tilde{t})!/((t-1-\tilde{t})!(n-t+1)!)) + \Adv_{\Adversary}^{\mathsf{zk}} + (N_u+N_s+1) \cdot\Adv_{\Adversary}^{\mathsf{dlog}} + N_s/|\Ga| + \Adv_{\Adversary}^{\mathsf{pseu}} + N_s \cdot \Adv_{\Adversary}^{\mathsf{xdh}} + \Adv_{\Adversary}^{\mathsf{col-res}}$, where $N_u$ is the number of users and $N_s$ is the number of serial numbers. The discrete logarithm assumption are implied by the DDHI assumption, which is used to prove that the function in~\S\ref{subsec:pseudorandomfunction} is pseudorandom. This concludes the proof of Theorem~\ref{th:all}.

%% file: CSecurityAnalysisDivisible.tex
\section{Security Proof for Our Divisible E-Cash Scheme}
\label{sec:securityProofDivisible}

In~\S\ref{sec:securityProofCompact}, we provide a detailed security proof for our compact $\CEC$ scheme. In this section, we analyze the security of our divisible $\CEC$ scheme, but we omit a full proof. Instead, we discuss the points where the security analysis of our divisible $\CEC$ scheme differs from the analysis of our compact $\CEC$ scheme.

For the interfaces $\fcecsetup$, $\fcecregister$, $\fcecrequest$, and $\fcecissue$, the simulator for the divisible $\CEC$ scheme is equal to that of our compact $\CEC$ scheme. We recall that both schemes use the same algorithms for the $\fcecregister$, $\fcecrequest$ and $\fcecissue$ interfaces. For the $\fcecsetup$ interface, although algorithm $\cecSetup$ is different in our schemes, the algorithm is run in both cases by $\Functionality_{\KG}$, as specified in our simulation. 

In the spending phase, the simulator for the divisible $\CEC$ scheme, like in the compact $\CEC$ scheme, outputs failure when the adversary submits a payment such that the simulator fails to extract a valid PS signature, or when the adversary did not receive signatures from $t - \tilde{t}$ authorities on the signed messages. Additionally, the simulator for the divisible $\CEC$ scheme outputs failure if it is unable to extract a SPS signature on $(\varsigma_{l+V-1}, \allowbreak \theta_{l+V-1})$, where $(\varsigma_{l+V-1}, \allowbreak \theta_{l+V-1})$ are part of the public parameters. It also outputs failure if the extracted $(\varsigma_{l}, \allowbreak \theta_{l})$ are not part of the public parameters. In the security proof, the probability that extraction fails is negligible thanks to the weak simulation extractability property of the ZK argument $\pi_v$, the existential unforgeability of the SPS signature scheme, and the BDHI assumption. All these arguments ensure that $\phi_{V,l}$ and $\varphi_{V,l}$ are computed correctly.

In the spending phase, to simulate a payment $(\kappa, \allowbreak \sigma', \allowbreak \phi_{V,l}, \allowbreak \varphi_{V,l}, \allowbreak R, \allowbreak \pi_v, \allowbreak  V)$, the values $\kappa$ and $\sigma'$ for the proof of possession of a PS signature are simulated like in the simulator for our compact $\CEC$ scheme (see Figure~\ref{fig:proc2}). The proof $\pi_v$ is also simulated by using the simulator $S_v$. The values $\phi_{V,l}$ and $\varphi_{V,l}$ are set to random. In the security proof, it can be shown that a simulated payment is indistinguishable from an honestly computed payment under the $N$-MXDH' assumption. The proof is similar to one given in~\cite{DBLP:conf/pkc/PointchevalST17}. We remark that, in the proof in~\cite{DBLP:conf/pkc/PointchevalST17}, an element of the $N$-MXDH' instance is needed to simulate the request message because the element $U_2 = u_2^\cecsn$ is revealed to the bank, whereas in our case that is not needed thanks to the hiding property of the commitment scheme.

In the deposit phase, the behavior of both simulators is similar, taking into account that the serial numbers and double spending tags are computed differently. The simulator also outputs failure if double spending happened but the identification algorithm does not identify the double spender. In the security proof, after it is ensured, as described above, that the ElGamal encryptions $\phi_{V,l}$ and $\varphi_{V,l}$ sent by the adversary are computed correctly, we know that serial numbers and double spending tags can be retrieved. Then we can prove, as in the case of the compact $\CEC$ scheme, that the user guilty of double spending can be identified under the collision resistance property of the hash function $H'$.

The simulator also outputs failure when there is not double spending, but an honest user is found guilty. In the security proof, like in our $\CEC$ scheme, we can show that the simulator fails with negligible probability under the hardness of the discrete logarithm problem.

%% file: DCompactECashRangeProof.tex
\section{Compact E-Cash with Range Proof Instantiation}
\label{sec:compactECashRangeProof}

The zero-knowledge argument of knowledge in algorithm $\cecSpend$ involves a statement $l \allowbreak \in \allowbreak [0, L-1]$ to prove the validity of the index of the spent coin. To implement it, one option is to use the set membership proof described in~\cite{DBLP:conf/asiacrypt/CamenischCS08}. This proof consists in proving possession of a signature that signs the index $l$. We use the Pointcheval-Sanders signature scheme to instantiate it. Algorithm $\cecSetup$ is extended to compute a signing key pair and signatures on all the values in the range $[0, \allowbreak L-1]$. Algorithm $\cecSpend$ is extended to include a zero-knowledge argument of knowledge of the signature that signs $l$, and algorithm $\cecSpendVf$ is extended to verify it. The modified algorithms work as follows:
\begin{description}[leftmargin=*]

\item[$\cecSetup(1^{\securityparameter}, L)$.] Execute the following steps:

\begin{itemize}

    \item Run $(\p,\Ga,\Gb,\allowbreak \Gt,\e,\ga,\gb) \gets \BilinearSetup(1^\securityparameter)$.

    \item Pick $3$ random generators $(\gamma_1, \gamma_2, \delta) \gets \Ga$.

    \item Run $(\ssk,\spk) \allowbreak \gets \allowbreak \SKeygen(1^\securityparameter, \allowbreak 1)$, i.e., pick random secret key $(x, \allowbreak y) \gets \mathbb{Z}_p^{2}$ and output the secret key $\ssk = (x, \allowbreak y)$ and the public key $\spk = (\tilde{\alpha}_{sm}, \tilde{\beta}_{sm}) \gets (\gb^x, \allowbreak \gb^{y})$.

    \item For all $l \allowbreak \in \allowbreak [0, L-1]$, compute $\ssig_l \gets \SSign(\ssk, \allowbreak l)$, i.e., pick random $r_l \gets \Zp$, set $h_l \gets \ga^{r_l}$ and output the signature $\ssig_l = (h_l, s_l) \gets (h_l, h_l^{x+yl})$.

    \item Set the parameters $\cecparams \gets (\p,\Ga,\Gb,\allowbreak \Gt,\e,\ga,\gb,\gamma_1, \gamma_2, \allowbreak \delta, \allowbreak \tilde{\alpha}_{sm}, \allowbreak \tilde{\beta}_{sm}, \allowbreak \ssig_0, \allowbreak \ldots, \allowbreak \ssig_{L-1}, \allowbreak L)$.

    \item Output $\cecparams$.

\end{itemize}

\item[$\cecSpend(\spk, \ssk_{\fcecUser_j}, \cecwallet, \cecpaymentinfo, V)$.] Execute the following steps:

\begin{itemize}

    \item Parse $\cecwallet$ as $(\sigma, \allowbreak \cecsn, \allowbreak l)$. If $l + V -1 \geq L$, output $0$.

    \item Parse $\sigma$ as $(\h, \allowbreak s)$.

    \item Parse $\spk$ as $(\cecparams, \allowbreak \tilde{\alpha}, \allowbreak \beta_{1}, \allowbreak \tilde{\beta}_{1}, \beta_{2}, \tilde{\beta}_{2})$.

    \item Pick random $r \gets \Zp$ and $r' \gets \Zp$.

    \item Compute $\sigma' = (\h',s') \gets (\h^{r'}, s^{r'}(\h')^{r})$.

    \item Compute $\kappa \gets \tilde{\alpha} \tilde{\beta}_1^{\ssk_{\fcecUser_j}} \tilde{\beta}_2^{\cecsn} \gb^{r}$.

    \item Pick random $o_c \gets \Zp$ and compute the commitment $C \gets \ga^{o_c} \gamma_1^{\cecsn}$.

    \item For $k \in [0,V-1]$, compute $R_k \gets H'(\cecpaymentinfo,k)$, where $\cecpaymentinfo$ must contain the identifier of the merchant, and $H'$ is a collision-resistant hash function.

    \item For $k \in [0,V-1]$, set $l_k \gets l+k$, pick random $o_{a_k}$ and compute $A_k = \ga^{o_{a_k}} \gamma_1^{l_k}$.

    \item For $k \in [0,V-1]$, evaluate the pseudorandom functions $S_{k} \gets f_{\delta,\cecsn}(l_k) = \delta^{1/(\cecsn+l_k+1)}$ and $T_k \gets \ga^{\ssk_{\fcecUser_j}} (f_{\ga,\cecsn}(l_k))^{R_{k}} = \ga^{\ssk_{\fcecUser_j} + R_{k}/(\cecsn+l_k+1)}$.

    \item For $k \in [0,V-1]$, compute the values $\mu_k \gets 1/(\cecsn+l_k+1)$ and $o_{\mu_k} \gets -(o_{a_k} + o_c) \mu_k$.

    \item Parse $\cecparams$ as the tuple $(\p,\Ga,\Gb,\allowbreak \Gt,\e,\ga,\gb,\gamma_1, \allowbreak \gamma_2, \allowbreak \delta, \allowbreak \tilde{\alpha}_{sm}, \allowbreak  \tilde{\beta}_{sm}, \allowbreak \ssig_0, \allowbreak \ldots, \allowbreak \ssig_{L-1}, \allowbreak L)$.

    \item For $k \in [0,V-1]$, pick random $r_{k} \gets \Zp$ and $r'_{k} \gets \Zp$.

    \item For $k \in [0,V-1]$, parse $\ssig_{l_k}$ as $(\h_{l_k}, \allowbreak s_{l_k})$. Compute $\sigma'_{l_k} = (\h'_{l_k},s'_{l_k}) \allowbreak \gets \allowbreak (\h_{l_k}^{r'_{k}}, s_{l_k}^{r'_{k}}(\h'_{l_k})^{r_{k}})$.

    \item For $k \in [0,V-1]$, compute $\kappa_{k} \gets \tilde{\alpha}_{sm} \tilde{\beta}_{sm}^{l_k} \gb^{r_{k}}$.

    \item Compute a ZK argument of knowledge $\pi_v$ via the Fiat-Shamir heuristic for the following relation:
    \begin{align*}
        \pi_v =  & \NIZK\{(\ssk_{\fcecUser_j}, \cecsn, r, o_c, \langle l_k, r_{k}, o_{a_k}, \mu_k, o_{\mu_k} \rangle_{k=0}^{V-1}): \\
        &\kappa = \tilde{\alpha} \tilde{\beta}_1^{\ssk_{\fcecUser_j}} \tilde{\beta}_2^{\cecsn} \gb^{r}\ \land\ C = \ga^{o_c} \gamma_1^{\cecsn}\ \land\  \\&
        \langle A_k = \ga^{o_{a_k}} \gamma_1^{l_k}\ \land\ \kappa_{k} = \tilde{\alpha}_{sm} \tilde{\beta}_{sm}^{l_k} \gb^{r_{k}} \land\ \\&
        S_k=\delta^{\mu_k}\ \land\ \gamma_1 = (A_k C\gamma_1)^{\mu_k} \ga^{o_{\mu_k}}\ \land\ \\&
        T_k=\ga^{\ssk_{\fcecUser_j}} (\ga^{R_{k}})^{\mu_k}\  \rangle_{k\in[0,V-1]}
        \}
    \end{align*}
    The equation $\kappa = \tilde{\alpha} \tilde{\beta}_1^{\ssk_{\fcecUser_j}} \tilde{\beta}_2^{\cecsn} \gb^{r}$ proves that $\kappa$ commits to the signed messages. ($\kappa$ is used as part of the proof of signature possession.) The equation $C = \ga^{o_c} \gamma_1^{\cecsn}$ proves that $C$ commits to the same value $\cecsn$ committed in $\kappa$. The equations $A_k = \ga^{o_{a_k}} \gamma_1^{l_k}$ and $\kappa_{k} = \tilde{\alpha}_{sm} \tilde{\beta}_{sm}^{l_k} \gb^{r_{k}}$ prove that the value $l_k$ committed in $A_k$ is in the valid range $[0,L-1]$ for a wallet with $L$ coins. The equations $S_k=\delta^{\mu_k}$, $\gamma_1 = (A_k C\gamma_1)^{\mu_k} \ga^{o_{\mu_k}}$ and $T_k=\ga^{\ssk_{\fcecUser_j}} (\ga^{R_{k}})^{\mu_k}$ prove that the serial numbers $S_k$ and the security tags $T_k$ are correctly computed. This non-interactive argument signs the payment information $\cecpaymentinfo$.

    \item Output $\cecpayment \gets (\kappa, \sigma', \langle S_k, T_k, A_k, \kappa_{k}, \sigma'_{l+k}  \rangle_{k\in[0,V-1]}, V, C, \allowbreak \pi_v)$    and an updated wallet $\cecwallet' \gets (\sigma, \allowbreak \cecsn, \allowbreak l+V)$.

\end{itemize}

\item[$\cecSpendVf(\spk, \cecpayment, \cecpaymentinfo)$.] Execute the following steps:

\begin{itemize}

    \item Parse $\spk$ as $(\cecparams, \allowbreak \tilde{\alpha}, \allowbreak \beta_{1}, \allowbreak \tilde{\beta}_{1}, \beta_{2}, \tilde{\beta}_{2})$.

    \item Parse $\cecpayment$ as $(\kappa, \sigma', \langle S_k, T_k, A_k, \kappa_{k}, \sigma'_{l_k} \rangle_{k\in[0,V-1]}, V, C,  \pi_v)$.

    \item Parse $\sigma'$ as $(\h', \allowbreak s')$ and output $0$ if $\h' = 1$ or if $\e(\h', \kappa) \allowbreak = \allowbreak \e(s',\gb)$ does not hold.

    \item For $k \in [0,V-1]$, parse $\sigma'_{l_k}$ as $(\h'_{l_k}, \allowbreak s'_{l_k})$ and output $0$ if $\h'_{l_k} = 1$ or if $\e(\h'_{l_k}, \kappa_{k}) \allowbreak = \allowbreak \e(s'_{l_k},\gb)$ does not hold.

    \item Output $0$ if not all the serial numbers $\langle S_k \rangle_{k\in[0,V-1]}$ are different from each other.

    \item For $k \in [0,V-1]$, compute $R_k \gets H'(\cecpaymentinfo,k)$.

    \item Verify $\pi_v$ by using $\cecpaymentinfo$, $\spk$, $\langle S_k, T_k, A_k, R_{k}, \kappa_{k} \rangle_{k\in[0,V-1]}$, $C$ and $\kappa$. Output $0$ if the proof is not correct, else output $V$.







\end{itemize}

\paragraph{Cost reduction by using one secret in the wallet.} We quantify the reduction of costs attained by having a wallet that signs one secret instead of two (as done in the compact e-cash scheme in~\cite{DBLP:conf/eurocrypt/CamenischHL05}),  in addition to the user secret key. First, we analyze the communication and storage costs. Let $|\Ga|$, $|\Gb|$ and $|\Zp|$ denote the bit size of elements in $\Ga$, $\Gb$ and $\Zp$ respectively. The size of the public parameters does not change. The size of public keys of authorities is $2|\Ga|+3|\Gb|$ with one secret and $3|\Ga|+4|\Gb|$ with two secrets. The size of the secret key of authorities is $3|\Zp|$ with one secret and $4|\Zp|$ with two secrets. In the withdrawal phase, the size of a response does not change, but the size of a request is $8|\Ga|+6|\Zp|$ with one secret and $10|\Ga|+8|\Zp|$ with two secrets. The wallet size is $2|\Ga|+1|\Zp|$ with one secret and $2|\Ga|+2|\Zp|$ with two secrets. The size of a payment of $V$ coins is $(2+5V+1)|\Ga|+(1+V)|\Gb|+(5+5V)|\Zp|$ with one secret and $(2+5V+2)|\Ga|+(1+V)|\Gb|+(7+7V)|\Zp|$ with two secrets. 

Second, we analyze the computation cost. We remark that the number of bilinear map computations does not change. Let $|M|$ and $|E|$ denote the cost of a multi-exponentiation and of a exponentiation respectively. In a withdrawal phase in which the user contacts $t$ authorities, the total cost is $9|M|+(4+7t)|E|$ with one secret and $12|M|+(5+10t)|E|$ with two secrets. In a spending phase in which $V$ coins are spent, the total cost is $(6+11V)|M|+(4+5V)|E|$ with one secret and $(9+13V)|M|+(5+5V)|E|$ with two secrets. The cost of the deposit phase does not change.
\end{description}

%% file: EDivisibleECashSpendProof.tex
\section{Complete Description of the Divisible E-Cash Scheme}
\label{sec:divisibleEcashSpend}

The zero-knowledge argument of knowledge in algorithm $\cecSpend$ in~\S\ref{sec:divisibleecash} involves proving knowledge of secret bases. For this purpose, the transformation described in~\S\ref{subsec:zkpk} needs to be applied. In algorithm $\cecSetup$, the generators $\psi \allowbreak \in \allowbreak \Ga$ and $\tilde{\psi} \allowbreak \in \allowbreak \Gb$ are added. In algorithm $\cecSpend$, the secret bases are blinded, and the zero-knowledge argument is modified accordingly. We describe below the modified algorithms $\cecSetup$, $\cecSpend$ and $\cecSpendVf$. Moreover, we also describe the algorithms  $\cecKeyGenA$, $\cecKeyGenU$, $\cecRequest$, $\cecRequestVf$, $\cecWithdraw$, $\cecWithdrawVf$ and $\cecCreateWallet$, which were not depicted in~\S\ref{sec:divisibleecash}.

\begin{description}[leftmargin=*]

\item[$\cecSetup(1^{\securityparameter}, L)$.] Execute the following steps:

\begin{itemize}

    \item Run $\grp = (\p,\Ga,\Gb,\allowbreak \Gt,\e,\ga,\gb) \gets \BilinearSetup(1^\securityparameter)$.

    \item Pick random generators $\eta, \allowbreak \gamma_1, \allowbreak \gamma_2, \psi \in \Ga$ and $\tilde{\psi} \in \Gb$.

    \item Generate random scalars $(z,y) \allowbreak \gets \allowbreak \Zp$ and, for $l \allowbreak \in \allowbreak [1,L]$, $a_l \gets \Zp$.

    \item Compute $(\varsigma, \theta) \gets (\ga^z, \eta^z)$.

    \item For $l \allowbreak \in \allowbreak [1,L]$, compute $(\varsigma_l, \theta_l) \allowbreak \gets \allowbreak (\varsigma^{y^l}, \allowbreak \theta^{y^l})$.

    \item For $k \in [0,L-1]$, compute $\tilde{\delta}_k \gets \gb^{y^k}$.

    \item For $l \allowbreak \in \allowbreak [1,L]$, compute $\eta_l \gets \ga^{a_l}$.

    \item For $l \allowbreak \in \allowbreak [1,L]$, for $k \allowbreak \in \allowbreak [0, \allowbreak l-1]$, compute $\tilde{\eta}_{l,k} \allowbreak \gets \allowbreak \gb^{-a_l \cdot y^k}$.

    \item Run the algorithm $(\spk_{sps}, \allowbreak \ssk_{sps}) \allowbreak \gets \allowbreak \SKeygen(\grp,\allowbreak 2,\allowbreak 0)$ of the structure-preserving signature scheme in~\S\ref{subsec:signatureSchemes}.

    \item For $l \allowbreak \in \allowbreak [1,L]$, compute $\tau_l \gets \SSign(\ssk_{sps},\langle \varsigma_l, \theta_l \rangle)$.

    \item Set the parameters for users $\cecparams_u \gets (\p,\Ga,\Gb,\allowbreak \Gt,\e,\ga,\gb, \allowbreak \eta, \allowbreak \gamma_1, \allowbreak \gamma_2, \allowbreak \{\eta_l, \allowbreak \varsigma_l, \allowbreak \theta_l, \allowbreak \tau_l\}_{l=1}^L, \allowbreak \spk_{sps}, \allowbreak \psi, \allowbreak \tilde{\psi})$. Set the additional parameters for authorities $\cecparams_a \allowbreak \gets \allowbreak (\{\tilde{\delta}_k\}_{k=0}^{L-1}, \allowbreak \{\langle \tilde{\eta}_{l,k} \rangle_{k=0}^{l-1}\}_{l=1}^{L-1})$.

    \item Set the parameters $\cecparams \gets (\cecparams_u, \cecparams_a)$.

    \item Output $\cecparams$.

\end{itemize}

\item[$\cecKeyGenA(\cecparams_u, t, n)$.] Execute the following steps:
\begin{itemize}

    \item Choose $(1 + 2)$ polynomials $(v, w_1, w_2)$ of degree $(t - 1)$ with random coefficients in $\Zp$.

    \item Set $(x, y_1, y_2) \gets (v(0), w_1(0), w_2(0))$.

    \item For $i = 1$ to $n$, set the secret key $\ssk_{\fcecAuthority_i}$ of each authority $\fcecAuthority_i$ as $\ssk_{\fcecAuthority_i} \allowbreak = \allowbreak (x_i, \allowbreak y_{i,1}, \allowbreak y_{i,2}) \allowbreak \gets \allowbreak (v(i),\allowbreak w_1(i), \allowbreak w_2(i))$.

    \item For $i = 1$ to $n$, set the verification key $\spk_{\fcecAuthority_i}$ of each authority $\fcecAuthority_i$ as $\spk_{\fcecAuthority_i} \allowbreak = \allowbreak (\tilde{\alpha}_i, \allowbreak \beta_{i,1}, \allowbreak \tilde{\beta}_{i,1},  \allowbreak \beta_{i,2}, \allowbreak \tilde{\beta}_{i,2}) \allowbreak \gets \allowbreak (\gb^{x_i}, \allowbreak \ga^{y_{i,1}}, \allowbreak \gb^{y_{i,1}},  \allowbreak \ga^{y_{i,2}}, \allowbreak \gb^{y_{i,2}})$.

    \item Compute the verification key $\spk = (\cecparams_u, \allowbreak \tilde{\alpha}, \allowbreak \beta_{1}, \allowbreak \tilde{\beta}_{1}, \allowbreak \beta_{2}, \tilde{\beta}_{2}) \allowbreak \gets \allowbreak (\cecparams_u, \allowbreak \gb^{x}, \allowbreak \ga^{y_{1}}, \allowbreak \gb^{y_{1}}, \allowbreak \ga^{y_{2}}, \allowbreak \gb^{y_{2}})$.

    \item Output $(\spk, \langle \spk_{\fcecAuthority_i}, \ssk_{\fcecAuthority_i} \rangle_{i =1}^{n})$.

\end{itemize}

\item[$\cecKeyGenU(\cecparams_u)$.] Execute the following steps:
\begin{itemize}

    \item Pick random $\ssk_{\fcecUser_j} \gets \Zp$ and compute $\spk_{\fcecUser_j} \gets \ga^{\ssk_{\fcecUser_j}}$.

    \item Output $(\ssk_{\fcecUser_j}, \allowbreak \spk_{\fcecUser_j})$.

\end{itemize}

\item[$\cecRequest(\cecparams_u, \ssk_{\fcecUser_j})$.] Execute the following steps:

\begin{itemize}

    \item Pick random $\cecsn \gets \Zp$ and set $(\cecmes_{1}, \allowbreak \cecmes_{2}) \allowbreak = (\ssk_{\fcecUser_j}, \allowbreak \cecsn)$.

    \item Pick random $o \gets \Zp$ and compute  $\com = \ga^{o} \prod_{j=1}^{2} \gamma_j^{\cecmes_{j}}$.

    \item Compute $\h \gets H(\com)$, where $H$ is modeled as a random oracle.

    \item Compute commitments to each of the messages. For $j = 1$ to $2$, pick random $o_{j} \gets \Zp$ and set $\com_{j} = \ga^{o_{j}} \h^{\cecmes_{j}}$.

    \item Compute a ZK argument of knowledge $\pi_s$ via the Fiat-Shamir heuristic for the following relation:
    \begin{align*}
        \pi_s =  & \NIZK\{(\cecmes_{1}, \cecmes_{2}, o, o_{1}, o_{2}): \\ &
        \com = \ga^{o} \prod_{j=1}^{2} \gamma_j^{\cecmes_{j}}\ \land\  \spk_{\fcecUser_j} \gets \ga^{\cecmes_{1}} \land\ \\ &
        \{\com_{j}= \ga^{o_{j}} \h^{\cecmes_{j}} \}_{\forall j \in [1,2]}  \}
    \end{align*}

    \item Set $\cecrequestinfo \allowbreak \gets \allowbreak (\h, \allowbreak o_{1}, \allowbreak o_{2}, \allowbreak \cecsn)$.

    \item Set $\cecrequest \allowbreak \gets \allowbreak (\h, \allowbreak \com, \allowbreak \com_{1}, \allowbreak \com_{2}, \allowbreak \pi_s)$.

    \item Output $\cecrequest$ and $\cecrequestinfo$.

\end{itemize}

\item[$\cecRequestVf(\cecparams_u, \cecrequest, \spk_{\fcecUser_j})$.] Execute the following steps:
\begin{itemize}

    \item Parse $\cecrequest$ as $(\h, \allowbreak \com, \allowbreak \com_{1}, \allowbreak \com_{2}, \allowbreak \pi_s)$.

    \item Compute $\h' \gets H(\com)$, where $H$ is modeled as a random oracle. Output $0$ if $\h \neq \h'$.

    \item Verify $\pi_s$ by using the tuple $(\cecparams_u, \allowbreak \h, \allowbreak \com, \allowbreak \com_{1}, \allowbreak \com_{2})$. Output $0$ if the proof $\pi_s$ is not correct, else output $1$.

\end{itemize}

\item[$\cecWithdraw(\cecparams_u, \ssk_{\fcecAuthority_i}, \cecrequest)$.] Execute the following steps:
\begin{itemize}

    \item Parse $\cecrequest$ as $(\h, \allowbreak \com, \allowbreak \com_{1}, \allowbreak \com_{2}, \allowbreak \pi_s)$.

    \item Parse $\ssk_{\fcecAuthority_i}$ as $(x_i, \allowbreak y_{i,1}, \allowbreak y_{i,2})$.

    \item Compute $c = \h^{x_i} \prod_{j=1}^{2} \com_j^{y_{i,j}}$.

    \item Set the blinded signature share $\hat{\sigma}_{i} \allowbreak \gets \allowbreak (\h, \allowbreak c)$.

    \item Output $\cecresponse \gets \hat{\sigma}_{i}$.

\end{itemize}

\item[$\cecWithdrawVf(\cecparams_u, \spk_{\fcecAuthority_i}, \ssk_{\fcecUser_j}, \cecresponse, \cecrequestinfo)$.] Execute the following steps:

\begin{itemize}

    \item Parse $\cecrequestinfo$ as $(\h', \allowbreak o_{1},  \allowbreak o_{2}, \allowbreak \cecsn)$.

    \item Parse $\cecresponse$ as $\hat{\sigma}_{i} \allowbreak = \allowbreak (\h, \allowbreak c)$. Output $0$ if $\h \allowbreak \neq \allowbreak \h'$.

    \item Parse $\spk_{\fcecAuthority_i}$ as $(\tilde{\alpha}_i, \allowbreak \beta_{i,1}, \allowbreak \tilde{\beta}_{i,1}, \allowbreak \beta_{i,2}, \allowbreak \tilde{\beta}_{i,2})$.

    \item Compute $\sigma_{i} \allowbreak = \allowbreak (\h, \allowbreak s) \allowbreak \gets \allowbreak (\h, \allowbreak c \prod_{j=1}^{2} \beta_{i,j}^{-o_{j}})$.

    \item Set $(\cecmes_{1}, \cecmes_{2}) = (\ssk_{\fcecUser_j}, \cecsn)$. Output $0$ if $\e(\h, \tilde{\alpha}_i \prod_{j=1}^{2} \tilde{\beta}_{i,j}^{\cecmes_{j}}) = \e(s, \gb)$ does not hold.

    \item Output $\cecwallet_i \gets (i,\sigma_{i},\cecsn)$.

\end{itemize}

\item[$\cecCreateWallet(\spk, \ssk_{\fcecUser_j}, \cecset, \langle \cecwallet_i \rangle_{i \in \cecset})$.] Execute the following steps:
\begin{itemize}

    \item If $|\cecset| \neq t$, output $0$.

    \item For all $i \in \cecset$, evaluate at 0 the Lagrange basis polynomials
    \begin{equation*}
        l_i = [\prod_{j \in \cecset,j\neq i} (0-j)] [\prod_{j\in\cecset,j\neq i} (i-j)]^{-1}\ \mathrm{mod}\ \p
    \end{equation*}

    \item For all $i \in \cecset$, parse $\cecwallet_{i}$ as  $(i,\sigma_{i},\cecsn)$ and $\sigma_{i}$ as $(\h, \allowbreak s_i)$.

    \item Compute the signature $\sigma = (\h, s) \gets (\h, \allowbreak \prod_{i \in \cecset} s_{i}^{l_i})$.

    \item Parse $\spk$ as $(\cecparams_u, \allowbreak \tilde{\alpha}, \allowbreak \beta_{1}, \allowbreak \tilde{\beta}_{1}, \beta_{2}, \tilde{\beta}_{2})$.

    \item Set $(\cecmes_{1}, \allowbreak \cecmes_{2}) \allowbreak = \allowbreak (\ssk_{\fcecUser_j}, \allowbreak \cecsn)$ and output $0$ if $\e(\h, \tilde{\alpha} \prod_{j=1}^{2} \tilde{\beta}_j^{\cecmes_{j}}) = \e(s, \gb)$ does not hold, else output $\cecwallet \allowbreak \gets \allowbreak (\sigma, \allowbreak \cecsn, l)$, where $l$ is a counter from $1$ to $L$ initialized to $1$.

\end{itemize}

\item[$\cecSpend(\spk, \ssk_{\fcecUser_j}, \cecwallet, \cecpaymentinfo, V)$.] Execute the following steps:

\begin{itemize}

    \item Parse $\cecwallet$ as $(\sigma, \allowbreak \cecsn, \allowbreak l)$. If $l+V \geq L$, output $0$.

    \item Parse $\sigma$ as $(\h, \allowbreak s)$.

    \item Parse $\spk$ as $(\cecparams_u, \allowbreak \tilde{\alpha}, \allowbreak \beta_{1}, \allowbreak \tilde{\beta}_{1}, \beta_{2}, \tilde{\beta}_{2})$.

    \item Pick random scalars $r \gets \Zp$ and $r' \gets \Zp$.

    \item Compute $\sigma' = (\h',s') \gets (\h^{r'}, s^{r'}(\h')^{r})$.

    \item Compute $\kappa \gets \tilde{\alpha} \tilde{\beta}_1^{\ssk_{\fcecUser_j}} \tilde{\beta}_2^{\cecsn}  \gb^{r}$.

    \item Pick random scalars $r_1,r_2 \gets \Zp$.

    \item Compute $\phi_{V,l} = (\phi_{V,l}[1],\phi_{V,l}[2]) \gets (\ga^{r_1}, \varsigma_l^\cecsn \eta_{V}^{r_1})$.

    \item Set $R \gets H(\cecpaymentinfo)$, where $H$ is a collision-resistant hash function, and set $\varphi_{V,l} = (\varphi_{V,l}[1],\varphi_{V,l}[2]) \gets (\ga^{r_2}, (\ga^R)^{\ssk_{\fcecUser_j}} \theta_l^{\cecsn} \eta_{V}^{r_2})$.

    \item Take $\cecparams_u$ from $\spk$. Take the public key $\spk_{sps} \allowbreak = \allowbreak (Y, \allowbreak W_1, \allowbreak W_2, \allowbreak Z)$ and the signature $\tau_{l+V-1} \allowbreak = \allowbreak (R_{l+V-1}, \allowbreak S_{l+V-1}, \allowbreak T_{l+V-1})$.

    \item Pick random $(\rho_{\varsigma_l}, \rho_{\theta_l}, \rho_{\varsigma_{l+V-1}}, \rho_{\theta_{l+V-1}}, \rho_{R_{l+V-1}}, \rho_{S_{l+V-1}}, \rho_{T_{l+V-1}}) \allowbreak \gets \allowbreak \Zp$.

    \item Compute the blinded bases $\varsigma'_l \allowbreak \gets \allowbreak  \varsigma_l \psi^{\rho_{\varsigma_l}}$, $\theta'_l \allowbreak \gets \allowbreak  \theta_l \psi^{\rho_{\theta_l}}$, $\varsigma'_{l+V-1} \allowbreak \gets \allowbreak \varsigma_{l+V-1}  \psi^{\rho_{\varsigma_{l+V-1}}}$ and $\theta'_{l+V-1} \allowbreak \gets \allowbreak \theta_{l+V-1}  \psi^{\rho_{\theta_{l+V-1}}}$, and the blinded bases for the signature $R'_{l+V-1} \allowbreak \gets \allowbreak R_{l+V-1}  \psi^{\rho_{R_{l+V-1}}}$, $S'_{l+V-1} \allowbreak \gets \allowbreak S_{l+V-1}  \psi^{\rho_{S_{l+V-1}}}$ and $T'_{l+V-1} \allowbreak \gets \allowbreak T_{l+V-1}  \tilde{\psi}^{\rho_{T_{l+V-1}}}$.

    \item Compute $\rho_1 \gets -\cecsn \rho_{\varsigma_l}$, $\rho_2 \gets -\cecsn \rho_{\theta_l}$ and $\rho_3 \gets \rho_{R_{l+V-1}}\rho_{T_{l+V-1}}$.

    \item Compute a ZK argument of knowledge $\pi_v$ via the Fiat-Shamir heuristic for the following relation:
    \begin{align}
        \pi_v =  & \NIZK\{(\ssk_{\fcecUser_j}, \cecsn, r, r_1, r_2, \rho_{\varsigma_l}, \rho_{\theta_l}, \rho_{\varsigma_{l+V-1}}, \rho_{\theta_{l+V-1}}, \nonumber \\ & \rho_{R_{l+V-1}}, \rho_{S_{l+V-1}},  \rho_{T_{l+V-1}}, \rho_1, \rho_2, \rho_3): \nonumber \\ &
        \kappa = \tilde{\alpha} \tilde{\beta}_1^{\ssk_{\fcecUser_j}} \tilde{\beta}_2^{\cecsn} \gb^{r}\ \land\ \label{equ1}\\&
        \phi_{V,l}[1] = \ga^{r_1}\ \land\ \phi_{V,l}[2] = (\varsigma'_l)^{\cecsn} \psi^{\rho_1} \eta_{V}^{r_1}\ \land\ \label{equ2} \\&
        \varphi_{V,l}[1] = \ga^{r_2}\ \land\ \varphi_{V,l}[2] = (\ga^R)^{\ssk_{\fcecUser_j}} (\theta'_l)^{\cecsn} \psi^{\rho_2} \eta_{V}^{r_2}\ \land\ \label{equ3} \\&
        \e(\varsigma'_l, \tilde{\delta}_{V-1})  \e(\psi, \tilde{\delta}_{V-1})^{-\rho_{\varsigma_l}} = \e(\varsigma'_{l+V-1},\tilde{g}) \e(\psi,\tilde{g})^{-\rho_{\varsigma_{l+V-1}}}\ \land\ \label{equ4}\\& \e(\theta'_l, \tilde{\delta}_{V-1}) \e(\psi, \tilde{\delta}_{V-1})^{-\rho_{\theta_l}} = \e(\theta'_{l+V-1}, \tilde{g}) \e(\psi,\tilde{g})^{-\rho_{\theta_{l+V-1}}}\ \land\ \label{equ5} \\&
        \e(R'_{l+V-1},Y) \e(\psi,Y)^{-\rho_{R_{l+V-1}}} \e(S'_{l+V-1}, \gb) \e(\psi,\gb)^{-\rho_{S_{l+V-1}}} \cdot \nonumber \\& e(\varsigma'_{l+V-1}, W_1)  e(\psi, W_1)^{-\rho_{\varsigma_{l+V-1}}} e(\theta'_{l+V-1}, W_2) e(\psi, W_2)^{-\rho_{\theta_{l+V-1}}} \cdot \nonumber \\& \e(\ga, Z)^{-1} =1\ \land \label{equ6} \\&
        \e(R'_{l+V-1},T'_{l+V-1}) \e(R'_{l+V-1},\tilde{\psi})^{-\rho_{T_{l+V-1}}} \cdot \nonumber \\& \e(\psi,T'_{l+V-1})^{-\rho_{R_{l+V-1}}} \e(\psi,\tilde{\psi})^{\rho_3} \e(\ga, \gb)^{-1} =1 \label{equ7}
    \end{align}
     Equation~\ref{equ1} is part of the proof of possession of the wallet signature that signs $(\ssk_{\fcecUser_j}, \allowbreak \cecsn)$. Equation~\ref{equ2} and \ref{equ3} prove that $\phi_{V,l}$ and the security tag $\varphi_{V,l}$ are well-formed and are computed on input $(\ssk_{\fcecUser_j}, \allowbreak \cecsn)$. Equation~\ref{equ4}, \ref{equ5}, \ref{equ6} and \ref{equ7}  prove that the values $\varsigma_l$ and $\theta_l$, which were used to compute  $\phi_{V,l}$ and $\varphi_{V,l}$ respectively, are part of the public parameters and fulfill $l \leq L-V+1$. This is accomplished by proving possession of a signature on $\varsigma_{l+V-1}$ and $\theta_{l+V-1}$ in equations~\ref{equ6} and \ref{equ7}, and proving that the indices of $\varsigma_{l+V-1}$ and $\theta_{l+V-1}$ and $\varsigma_l$ and $\theta_l$ are related by a difference of $V-1$ is equation~\ref{equ4} and \ref{equ5} respectively. This non-interactive argument signs the payment information $\cecpaymentinfo$.

    \item Output a payment $\cecpayment \gets (\kappa, \sigma', \phi_{V,l}, \varphi_{V,l}, \allowbreak \varsigma'_l, \allowbreak \theta'_l, \allowbreak \varsigma'_{l+V-1}, \allowbreak \theta'_{l+V-1}, \allowbreak R'_{l+V-1}, \allowbreak S'_{l+V-1}, \allowbreak T'_{l+V-1}, \allowbreak R, \allowbreak \pi_v, \allowbreak V)$ and an updated wallet $\cecwallet' \gets (\sigma, \allowbreak \cecsn, \allowbreak l+V)$.

\end{itemize}

\item[$\cecSpendVf(\spk, \cecpayment, \cecpaymentinfo)$.] Execute the following steps:

\begin{itemize}

    \item Parse $\cecpayment$ as $(\kappa, \sigma', \phi_{V,l}, \varphi_{V,l}, \varsigma'_l, \allowbreak \theta'_l, \allowbreak \varsigma'_{l+V-1}, \allowbreak \theta'_{l+V-1}, \allowbreak R'_{l+V-1}, \allowbreak S'_{l+V-1}, \allowbreak T'_{l+V-1}, \allowbreak R, \allowbreak \pi_v, \allowbreak V)$.

    \item Parse $\sigma'$ as $(\h', \allowbreak s')$ and output $0$ if $\h' = 1$ or if $\e(\h', \kappa) \allowbreak = \allowbreak \e(s',\gb)$ does not hold.

    \item Output $0$ if $R \neq H(\cecpaymentinfo)$ or if $\cecpaymentinfo$ does not contain the identifier of the merchant.

    \item Verify $\pi_v$ by using $\cecpaymentinfo$, $\spk$,  $\phi_{V,l}$, $\varphi_{V,l}$, $\varsigma'_l$, $\theta'_l$, $\varsigma'_{l+V-1}$, $\theta'_{l+V-1}$, $R'_{l+V-1}$, $S'_{l+V-1}$, $T'_{l+V-1}$, $V$, $R$ and $\kappa$. Output $0$ if the proof is not correct, else output $V$.

\end{itemize}

\item[$\cecIdentify(\cecparams, \cecPK, \cecpayment_1, \cecpayment_2, \cecpaymentinfo_1, \cecpaymentinfo_2)$.] Execute these steps:
\begin{itemize}

    \item Parse $\cecpayment_1$ as  $(\kappa_1, \allowbreak \sigma'_1, \allowbreak \phi_{V_1,l_1,1}, \allowbreak \varphi_{V_1,l_1,1}, \varsigma'_{l_1,1}, \allowbreak \theta'_{l_1,1}, \allowbreak \varsigma'_{l_1+V_1-1,1}, \allowbreak \theta'_{l_1+V_1-1,1}, \allowbreak \\ R'_{l_1+V_1-1,1}, \allowbreak S'_{l_1+V_1-1,1}, \allowbreak T'_{l_1+V_1-1,1}, \allowbreak R_1, \allowbreak \pi_{v,1}, \allowbreak V_1)$.

    \item Parse $\cecpayment_2$ as $(\kappa_2, \allowbreak \sigma'_2, \allowbreak \phi_{V_2,l_2,2}, \allowbreak \varphi_{V_2,l_2,2}, \varsigma'_{l_2,2}, \allowbreak \theta'_{l_2,2}, \allowbreak \varsigma'_{l_2+V_2-1,2}, \allowbreak \theta'_{l_2+V_2-1,2}, \allowbreak \\ R'_{l_2+V_2-1,2}, \allowbreak S'_{l_2+V_2-1,2}, \allowbreak T'_{l_2+V_2-1,2}, \allowbreak R_2, \allowbreak \pi_{v,2}, \allowbreak V_2)$.

    \item For $k \in [0,V_1-1]$, compute the serial numbers $\mathrm{SN}_{k,1} \allowbreak \gets \allowbreak \e(\phi_{V_1,l_1,1}[2], \tilde{\delta}_k) \allowbreak \e(\phi_{V_1,l_1,1}[1], \tilde{\eta}_{V_1,k})$.  For $k \in [0,V_2-1]$, compute $\mathrm{SN}_{k,2} \allowbreak \gets \allowbreak \e(\phi_{V_2,l_2,2}[2], \tilde{\delta}_k) \allowbreak \e(\phi_{V_2,l_2,2}[1], \tilde{\eta}_{V_2,k})$.

    \item Output $1$ if none of the serial numbers $\mathrm{SN}_{k_1,1}$, for $k_1 \in [0,V_1-1]$, is equal to $\mathrm{SN}_{k_2,2}$, for $k_2 \in [0,V_2-1]$.

    \item Else, output $\cecpaymentinfo_1$ if $\cecpaymentinfo_1 = \cecpaymentinfo_2$.

    \item Else, let $k_1 \in [0,V_1-1]$ and $k_2 \in [0,V_2-1]$ be two indices such that $\mathrm{SN}_{k_1,1}=\mathrm{SN}_{k_2,2}$. Compute
    \begin{align*}
        T_1 \gets \e(\varphi_{V_1,l_1,1}[2],\tilde{\delta}_{k_1}) \e(\varphi_{V_1,l_1,1}[1],\tilde{\eta}_{V_1,k_1}),
    \end{align*}
    and
    \begin{align*}
        T_2 \gets \e(\varphi_{V_2,l_2,2}[2],\tilde{\delta}_{k_2}) \e(\varphi_{V_2,l_2,2}[1],\tilde{\eta}_{V_2,k_2}).
    \end{align*}
    For each $\spk_{\fcecUser_j} \in \mathrm{PK}$, check whether $T_1T_2^{-1} = \e(\spk_{\fcecUser_j},\tilde{\delta}_{k_1}^{R_1}\tilde{\delta}_{k_2}^{-R_2})$ and output $\spk_{\fcecUser_j}$ if the equality holds. Output $\bot$ if the equality does not hold for any $\spk_{\fcecUser_j} \in \mathrm{PK}$.

\end{itemize}

\end{description}